\documentclass{aa}  
\usepackage{arydshln}
\usepackage{scrextend}
\usepackage[dvipsnames]{xcolor}
\usepackage{rotating}
\usepackage{multirow}
\usepackage{ulem}
\usepackage{natbib,twoopt}
\usepackage{anyfontsize}
\usepackage{ccaption}
\usepackage[breaklinks=true]{hyperref} 
\bibpunct{(}{)}{;}{a}{}{,}             
\makeatletter
  \newcommandtwoopt{\citeads}[3][][]{\href{http://adsabs.harvard.edu/abs/#3}%
    {\def\hyper@linkstart##1##2{}%
     \let\hyper@linkend\@empty\citealp[#1][#2]{#3}}}
  \newcommandtwoopt{\citepads}[3][][]{\href{http://adsabs.harvard.edu/abs/#3}%
    {\def\hyper@linkstart##1##2{}%
     \let\hyper@linkend\@empty\citep[#1][#2]{#3}}}
  \newcommandtwoopt{\citetads}[3][][]{\href{http://adsabs.harvard.edu/abs/#3}%
    {\def\hyper@linkstart##1##2{}%
     \let\hyper@linkend\@empty\citet[#1][#2]{#3}}}
  \newcommandtwoopt{\citeyearads}[3][][]%
    {\href{http://adsabs.harvard.edu/abs/#3}
    {\def\hyper@linkstart##1##2{}%
     \let\hyper@linkend\@empty\citeyear[#1][#2]{#3}}}
\makeatother

\usepackage{xcolor}
\usepackage{pdflscape}
\usepackage{multirow}
\usepackage{hyperref}
\usepackage{amsmath}
\usepackage{txfonts}

\usepackage{xintcore}
\usepackage{xintfrac}

\usepackage{cleveref}
\crefdefaultlabelformat{\textsubscript{\textbf{(#2#1#3.)}}}
\usepackage[inline]{enumitem}

\usepackage{tikz}
\usetikzlibrary{shapes, arrows.meta, positioning}
\usepackage{longtable}
\usepackage{orcidlink}

\newcommand{\FeKav}{\ion{Fe}{xxv} K$\alpha$}
\newcommand{\FeKavi}{\ion{Fe}{xxvi} K$\alpha$}

\newcommand{\FeKbv}{\ion{Fe}{xxv} K$\beta$}
\newcommand{\FeKbvi}{\ion{Fe}{xxvi} K$\beta$}

\newcommand{\FeKgvi}{\ion{Fe}{xxv} K$\gamma$}

\newcommand{\msun}{$M_\odot$}

\newcommand{\nustar}{\textit{NuSTAR}}
\newcommand{\chandra}{\textit{Chandra}}
\newcommand{\xmm}{XMM-\textit{Newton}}
\newcommand{\nicer}{\textit{NICER}}
\newcommand{\swift}{\textit{Swift}}
\newcommand{\suzaku}{\textit{Suzaku}}
\newcommand{\integral}{\textit{INTEGRAL}}
\newcommand{\maxi}{\textit{MAXI}}
\newcommand{\rxte}{\textit{RXTE}}
\newcommand{\ixpe}{\textit{IXPE}}
\newcommand{\hxmt}{\textit{hxmt}}

\newcommand\T{\rule{0pt}{2.6ex}}       
\newcommand\B{\rule[-1.2ex]{0pt}{0pt}} 

\begin{document} 

   \title{20 years of disk winds in 4U~1630$-$47  - I.\\ Long-term behavior and influence of hard X-rays}
\titlerunning{20 years of disk winds in 4U~1630$-$47  - I: Long-term behavior and influence of hard X-rays}
   \author{M. Parra\inst{1}\orcidlink{0009-0003-8610-853X}, S. Bianchi \inst{2}\orcidlink{0000-0002-4622-4240}, P.-O. Petrucci \inst{3}\orcidlink{0000-0001-6061-3480}, T. Bouchet\inst{4}\orcidlink{0009-0003-5878-0978}, M. Shidatsu\inst{1}\orcidlink{0000-0001-8195-6546}, F. Capitanio\inst{5}\orcidlink{0000-0002-6384-3027}, Michal Dov\v{c}iak\inst{6}\orcidlink{0000-0003-0079-1239}, T. D. Russell\inst{7}\orcidlink{0000-0002-7930-2276}, V. E. Gianolli\inst{8}\orcidlink{0000-0002-9719-8740}, F. Carotenuto\inst{9}\orcidlink{0000-0002-0426-3276}}
          \authorrunning{M. Parra et al.} 

   \institute{
   Department of Physics, Ehime University, 2-5, Bunkyocho, Matsuyama, Ehime 790-8577, Japan\\
   e-mail:\href{mailto:maxime.parrastro@gmail.com}{maxime.parrastro@gmail.com}
   \and
   Dipartimento di Matematica e Fisica, Università degli Studi Roma Tre, via della Vasca Navale 84, 00146 Roma, Italy
   \and
   Univ. Grenoble Alpes, CNRS, IPAG, 38000 Grenoble, France
   \and
   Université Paris Cité, Université Paris-Saclay, CEA, CNRS, AIM, 91191 Gif-sur-Yvette, France
   \and
   INAF, Istituto di Astrofisica e Planetologia Spaziali, via del fosso del Cavaliere 100, I-00133 Roma, Italy
   \and
   Astronomical Institute of the Czech Academy of Sciences, Bo\v{c} n\'{i} II 1401, 14100 Praha, Czech Republic
   \and
    INAF, Istituto di Astrofisica Spaziale e Fisica Cosmica, Via U. La Malfa 153, I-90146 Palermo, Italy
   \and
   Dep. of Physics and Astronomy, Clemson University, Kinard Lab of Physics, 140 Delta Epsilon Ct, Clemson, SC 29634, USA 
   \and
   INAF, Osservatorio Astronomico di Roma, Via Frascati 33, I-00078 Monte Porzio Catone, Italy
}

   \date{}

 
  \abstract
   {Highly ionized X-ray wind signatures have been found in the soft states of high-inclination Black Hole Low Mass X-ray Binaries (BHLMXBs) for more than two decades. Yet signs of a systematic evolution of the outflow itself along the outburst remain elusive, due to the limited sampling of individual sources and the necessity to consider the broad-band evolution of the Spectral Energy Distribution (SED). We perform an holistic analysis of archival X-ray wind signatures in the most observed wind-emitting transient BHLMXB to date, 4U~1630$-$47 . The combination of \chandra{}, \nicer{}, \nustar{}, \suzaku{}, and \xmm{}, complemented in hard X-rays by \swift{}/BAT and \integral{}, spans more than 200 individual days over 9 individual outbursts, and provides a near complete broad-band coverage of the brighter portion of the outburst. Our results show that the hard X-rays allow to define "soft" states with ubiquitous wind detections, and their contribution is strongly correlated with the Equivalent Width (EW) of the lines. We then constrain the evolution of the outflow in a set of representative observations, using thermal stability curves and photoionization modeling. The former confirms that the switch to unstable SEDs occurs well after the wind signatures disappear, to the point where the last canonical hard states are thermally stable. The latter shows that intrinsic changes in the outflow are required to explain the main correlations of the line EWs, be it with luminosity or the hard X-rays. These behaviors are seen systematically over all outbursts and confirm individual links between the wind properties, the thermal disk, and the corona.}

   \maketitle

\section{Introduction} \label{sec:intro}

X-ray binaries are compact binary stellar systems emitting primarily at X-ray energies, due to the energy liberated during the mass transfer from a main sequence star (or companion) to any type of compact object, be it a black hole (BH), a neutron star (NS), or a white dwarf (WD). Systems hosting a BH or a NS share many similarities and can be further divided into high-  and low-mass  X-ray binaries. For high-mass X-ray binaries (HMXB; \citealt{Fornasini2023_HMXB_review}) the companion is either an O or B-type supergiant feeding the compact object directly via powerful stellar winds. For low-mass X-ray binaries (LMXB; \citealt{Bahramian2023_LMXB_review}) the companion is a low-mass (typically $\lesssim$ 1 M$_\odot$) with mass transfer occurring via Roche Lobe overflow, where the infalling matter forms an accretion disk around the compact object. The majority of black hole LMXBs are transients, exhibiting rare, erratically repeating outburst patterns lasting from several months to years, in between much longer periods of quiescence. 

During these outbursts, the accretion rate increases by more than 5 orders of magnitude, as the accretion structure around the BH goes through several major transitions. These transitions can be traced over the entire electromagnetic spectrum and exhibit a wealth of spectral-timing states and signatures. The most drastic changes occur in X-rays, where the emission switches between a "hard" and "soft" spectral state (see, e.g., \citealt{Done2007_BHXRB_accretion} for a review) and then back, with the return occuring at precise luminosity thresholds \citep{VahdatMotlagh2019_soft-hard_transition_lum,Wang2023_soft-hard_transition_lum}. The hard state is dominated by a comptonized component of $\Gamma\sim1.5$ and a cutoff at $\sim100$ keV, whose origin is thought to be an optically thin, hot plasma region close to the BH, dubbed the corona. On the other hand, the soft state is dominated by a thermal emission from a geometrically thin, optically thick accretion disc extending close to the innermost stable circular orbit (ISCO). This rich diversity of configurations, combined with the repetition of outbursts on human timescales producing some of the brightest X-ray signatures in the sky, makes BH LMXBs prime candidates to understand the physics of accretion and the long-lasting effects of mass transfer on stellar evolution.

Some of the most remarkable features of LMXBs, found in BH and NS alike, are distinct ejection processes that occur during precise spectral-timing states. Jets (see, e.g., \citealt{fender2003_jet_review} for a review) are collimated, relativistic ejections of matter that release significant amounts of energy and angular momentum while exhibiting  a weak mass loss rate. They primarily emit via a synchrotron component which extends from the radio to the infrared band. Although this emission is observed only during the hard state, around the hard-to-soft transition, discrete ejecta can be launched from the system and travel outwards at relativistic speeds on scales of several parsecs. Their presence is ubiquitous among accreting com8
pact objects. Meanwhile, winds (see \citealt{Ponti2016_winds_XRB_review,DiazTrigo2016_winds_XRB_review} for reviews) are slow, equatorial ejections of matter with a much higher mass outflow rate, seen primarily via the continuum or atomic absorption features they imprint in different wavebands. In BH systems, the detection of X-ray wind signatures has been traditionally restricted to highly-inclined, jet-less soft states (\citealt{Ponti2012_ubhw,Parra2023_winds_global_BHLMXBs}, hereafter P24). However, recent observations at higher wavelengths have revealed a plethora of "cold" wind detections, observed in the optical band during the hard state, and in the infrared throughout the entire outburst (see \citealt{Munoz-Darias2019_MAXIJ1820+070_winds_hard_visible,Sanchez-Sierras2020_MAXIJ1820+070_wind_emission_infrared_soft_hard,Panizo-Espinar2022_MAXIJ1348-630-winds_optical} and references therein). 

The lack of visibility of the "hot" X-ray winds is a natural consequence of the illumination by the hard state spectral energy distribution (SED), which makes the range of ionization primarily seen in X-rays unstable \citep{Chakravorty2013_thermal_stability,Bianchi2017_stability_NS,Petrucci2021_outburst_wind_stability} and prevents the formation of absorption lines. However, several other elements, such as the lack of ubiquitous detection in soft states, the lack of any X-ray wind detections in many highly-inclined BHLMXBs showing optical and infrared (OIR) wind signatures, and velocities barely detectable by charge-coupled devices (CCDs), greatly hamper our ability to understand the structure of these outflows (P24). However, the lack of ability to detect and resolve line profiles over a wide range of ionization parameters, combined with the very limited sampling of BH outbursts with instruments able to detect absorption lines, means that the nature of the outflow remains for now out of reach. The two wind launching mechanisms most relevant to X-ray binaries: thermal driving \citep{Begelman1983_wind_thermal_init_1,Woods1996_wind_thermal_init_2,Done2018_thermal_winds_modeling_H1743_GROJ1655} and magnetic driving \citep{Blandford1982_Blandford-Payne,Fukumura2010_MHD_wind_AGNs,Jacquemin-Ide2019_wind_weak_magnetic_JEDSAD_modeling}. Both of these mechanisms show promise, reproducing at least some of the features seen in the current generation of instruments (see e.g. \citealt{Tomaru2020_H1743-322_wind_model_thermalradiative_2,Fukumura2017GROJ1655-40_wind_magnetic}), but remain indistinguishable without micro-calorimeter-level spectral resolution \citep{Gandhi2022}.

4U~1630$-$47  is one of the first identified transient X-ray sources \citep{Jones1976_4U1630-47_discovery}, and known for its pattern of recurring outbursts over a 600-700 days period \citep{Kuulkers1997_4U1630-47_outburst_patterns}. Despite no dynamical mass measurements or a proper distance estimate \citep{Kalemci2018_4U1630-47_d}, the source has been classified as a BH due to its spectral and timing properties \citep[see e.g.][and references therein]{Seifina2014_4U1630-47_properties}, and the detection of a K=16.1 mag infrared counterpart \citep{Augusteijn2001_4U1630-47_counterpart_infrared} cements it as a solid candidate BHLMXB. Finally, recurring dips in the X-ray lightcurve constrain its inclination to $\sim60-75$° \citep{Tomsick1998_4U1630-47_i,Kuulkers1998_GROJ1655-40_and_4U1630-47_dips}.

Nevertheless, its spectral and timing behavior rarely match the standard outburst patterns seen in typical black hole X-ray binaries. First, in addition of its recurring standard outbursts, the source occasionally enters "super-outbursts" of much longer duration \citep{Kuulkers1997_4U1630-47_outburst_patterns} and with varying spectral and timing evolution \citep{Abe2005_4U1630-47_long-term_outburst,Tomsick2005_4U1630-47_2002-2004_outburst_integral}.
Second, the source rarely exhibits proper hard states with very small disk contribution and standard $\Gamma<2$ high-energy component. Instead, a large fraction of its outbursts are spent alternating between Soft/thermal dominated states, intermediate (sometimes flaring) states, and Steep Powerlaw states (SPL, also called Very High State or VHS), the latter 2 showing an increasing contribution of a $\Gamma\sim2.5-3.5$ high-energy component and very distinct timing properties \citep{Tomsick2005_4U1630-47_2002-2004_outburst_integral}. Finally, 4U~1630$-$47 has both a history of mostly/completely soft outbursts \citep{Capitanio2015_4U1630-47_outbursts_hard_states}, and of decays in the soft state down to extremely low luminosities ($<10^{-4}L_{Edd}$)  before transitioning back to hard states (\citealt{Tomsick2014_4U1630-47_decay_2010_anomalous}, see also \citealt{Kalemci2018_4U1630-47_d}).

The unusual behavior of this source has prompted many observational campaigns and studies in the last decades, and it has become one of the archetypal wind-producing, high-inclination BHLMXBs. The first report of a wind detection from 4U~1630$-$47 came from a set of \suzaku{} observations in 2006 \citep{Kubota2007_4U1630-47_winds}, although a 2004 \chandra{}-HETG observation already exhibited an absorption line feature, but was later reported by \citet{Trueba2019_4U1630-47_wind_2012-13Chandra}.
Afterward, extensive monitoring campaigns were performed during the 2012--2013 outburst, using \xmm{} \citep{DiazTrigo2014_4U1630-47_wind_2012-13XMM}, \chandra{} HETG \citep{Neilsen2014_4U1630-47_2012-13emjetdebate}, and \suzaku{} \citep{Hori2014_4U1630-47_Suzaku_high_2012}. In parallel, three \nustar{} observations were performed over this period due to a planned survey of the Norma Cluster \citep{Fornasini2017_Norma_NuSTAR_survey}. One of the \nustar{} observations was too off axis to be analyzed, while another was studied in detail and reported by \cite{King2014_4U1630-47_wind_2014Nustar}. A single \xmm{} detection of a relativistic emission line during this outburst has been interpreted as a baryonic jet \citep{Trigo2013_4U1630-47_2012-13emjet}, but this conclusion remains debated \citep{Neilsen2014_4U1630-47_2012-13emjetdebate}.

In the following years, few individual observations with a variety of instruments continued to exhibit wind signatures, such as \suzaku{} and \nustar{} in 2015 \citep{Hori2018_4U1630-47_2015SuzakuNustar,Connors2021_4U1630-47_reflection_wind_soft_NuSTAR_2012_2015}, Astrosat and HETG in 2016 \citep{Pahari2018_4U1630-47_wind_2016Astrosat,Trueba2019_4U1630-47_wind_2012-13Chandra}, and \nicer{} in 2018 \citep{Neilsen2018_4U16300-47_wind_NICER}. \nicer{} coverage has continued during every subsequent outburst until now. We note that a serendipitous \xmm{} pointing of a nearby source performed during a bright portion of the 2018 outburst provided a completely over-saturated and pile-uped spectrum, which could be analyzed but requires a careful analysis. 

4U~1630$-$47's most recent and longest recorded outburst, which started in the second half of 2022, lasting until April 2024. This outburst  was extensively monitored with multiple instruments in order to study the source's X-ray polarisation properties. Wind signatures have been reported in observations taken with \ixpe{}, \nustar{}, and \nicer{} \citep{Ratheesh2023_4U1630-47_pola_soft,Cavero2023_4U1630-47_pola_SPL}.
We note that this latest outburst indicates a near-decadal recurrence of "super-outbursts" of this source in the last $\sim$30 years, as the very bright outburst of 2018-2019 seen in \maxi{}light curves (see Fig.~\ref{fig:4U_monit_full}) is in fact from the nearby binary MAXI J1631-479, as indicated in the \maxi{} webpage for this source \footnote{\href{http://maxi.riken.jp/star\_data/J1634-473/J1634-473.html}{http://maxi.riken.jp/star\_data/J1634-473/J1634-473.html}}.

The sampling of 4U~1630$-$47's outburst evolution is one of the best among BH LMXBs, and even more so for the (currently very limited) population of wind-emitting sources. However, the tens of high-quality exposures with multiple highly sensitive instruments have for now exclusively been studied for a single outburst (e.g., 2012-2013 in  \citealt{Gatuzz2019_4U1630-47_wind_2012-13Chandra}) or instruments (e.g., \chandra{} in \citealt{Trueba2019_4U1630-47_wind_2012-13Chandra}), limiting potential interpretations. Moreover, they are now backed up by hundreds of \nicer{} observations performed in the last few years, and other observations with e.g. \suzaku{} or \nustar{}, which provide additional understanding to previous outbursts, remained to be studied.

We thus perform an exhaustive study of archival observations of 4U~1630$-$47 until the end of 2023. The combination of the five main X-ray telescopes generally used for X-ray wind studies, namely \chandra{}\nicer{}, \nustar{}, \suzaku{}, and \xmm{}, totals more than 200 observations spanned over 9 separate outbursts and two decades. We complement our high-energy coverage using the daily \swift{}/BAT transient monitoring \citep{Krimm2013_Swift_BAT_transient_monitor}, as well as the entirety of the \integral{} archives for this source, which together allow us to derive the high-energy flux behavior of this source for a large fraction of the soft X-ray coverage.

In this first paper, we detail our global results and new diagnostics made possible with the high-energy coverage of this source. We detail the data reduction in Sect.~\ref{sec:data_red} and our spectral analysis procedure in Sect.~\ref{sec:spectral_analysis}. In Sect.~\ref{sec:global_behavior}, we study the global behavior of the absorption lines in the source. In Sect.~\ref{sec:wind_evol}, we disentangle the different effects of the changes in illuminating SED to probe the true evolution of the outflow. We discuss the physical interpretation of these changes and compare our result with the literature in Sect.~\ref{sec:discussion}, and conclude in Sec.~\ref{sec:conclusion}. 
We provide additional details about the NICER filtering procedure in App.~\ref{app:NICER_filtering}, and on the extension of the high-energy coverage from \swift{}/BAT and \integral{} monitoring in App.~\ref{app:highE_coverage}. App.~\ref{app:2D_corner} provides the full list of 2D projections of the comparison between the wind parameters of couples of observations performed in Sect.\ref{sub:photo_mod}. We provide summarized tables of the main spectral and line properties derived in our analysis in App.~\ref{app:tables}. Finally, similarly to P24, except for the photionization modeling, the entirety of our results and all of our figures, monitoring and line correlation properties are reproducible and downloadable in the online tool visual-line\footnote{\href{https://visual-line.streamlit.app/}{https://visual-line.streamlit.app/}}. 

\section{Observations \& Data Reduction} \label{sec:data_red}

We refer the reader to P24  for details on the data reduction methodology used with \chandra{} and \xmm{}, which remains identical to that used in that paper, as no additional observations for this source were performed with these telescopes since then. For our analysis, we use \texttt{Heasoft} v6.32.1 \citep{Blackburn1999_ftools}. The following paragraphs highlight our procedures for each telescope not included in our previous study.

\subsection{\nicer{}}

The Neutron Star Interior Composition Explorer (NICER, \citealt{Gendreau2016_NICER}) has observed extensively every single outburst of 4U~1630$-$47 since its launch in 2017. We analyze every observation listed in the \nicer{} Master HEASARC catalog\footnote{\href{https://heasarc.gsfc.nasa.gov/db-perl/W3Browse/w3table.pl?tablehead=name\%3Dnicermastr\&Action=More+Options}{https://heasarc.gsfc.nasa.gov/db-perl/W3Browse/w3table.pl\\?tablehead=name\%3Dnicermastr\&Action=More+Options}} as of 2023-12-01, for a total of 224 ObsIDs with non-zero exposure. Our data reduction procedure mainly relies on the simplified pipeline tasks of the NICERDAS software\footnote{\href{https://heasarc.gsfc.nasa.gov/docs/nicer/nicer_analysis.html}{https://heasarc.gsfc.nasa.gov/docs/nicer/nicer\_analysis.html}} version 11, and uses the CALDB calibration files xti202221001, the latest as of the writing of this paper. We downloaded up-to-date geomagnetic data, necessary for our choice of background model, using the \texttt{nigeodown} task of NICERDAS.

The Good Time Intervals (GTIs) are first split into continuous periods, filtered to remove background flares (see Appendix~\ref{app:NICER_filtering}), then used as input for the \texttt{nicerl3-spect} and \texttt{nicerl3-lc} tasks, which respectively create all spectral and lightcurve products. We used the xspec-model version of the scorpeon background model, and created lightcurves with a 1s binning in different bands. This last choice allowed us to manually inspect the continuous GTIs remaining after the filtering procedure, to confirm that any flare had been correctly screened out. Finally, the spectra were grouped according to the \cite{Kaastra2016_binning_opt} optimized binning.

The procedure resulted in 618 individual orbit spectra, taken on 189 days (totaling 189 ObsIDs). For the remaining 35 days/ObsIDs, all of the \nicer{} orbits were entirely discarded by the \nicer{} data reduction pipeline.

\subsection{\nustar{}}

The Nuclear Spectroscopic Telescope Array (\nustar{}, \citealt{Harrison2013_NuSTAR}) has observed 4U~1630$-$47  in several different outbursts since its launch in 2012. On top of its ability to detect iron lines, it provides the most precise view of the 10-80 keV band of any flying instrument, which makes it particularly suited to model the broad band X-ray SED of the source. We analyzed every public observation in the \nustar{} Master HEASARC catalog \footnote{\href{https://heasarc.gsfc.nasa.gov/db-perl/W3Browse/w3table.pl?tablehead=name\%3Dnumaster\&Action=More+Options}{https://heasarc.gsfc.nasa.gov/db-perl/W3Browse/w3table.pl\\?tablehead=name\%3Dnumaster\&Action=More+Options}} as of 2023-12-01. We discarded ObsID 40014006001, which was too off-axis to be usable, and ObsID 30001016002, which was performed in quiescence, and where the source is not detected. For the data reduction, we used the standard \texttt{NuSTARDAS} tasks, the \nustar{} CALDB v20230613, and applied a fully automated procedure to compute spectral and temporal products.

For each ObsID, we first reprocess the data using the \texttt{nupipeline} task, using standard parameters and filter criteria. We extract an image in the [3-79] keV using \texttt{xselect}, then extract a background region from the largest circular region not intersecting with the brightest 2$\sigma$ regions in the field of view, with a radius up to 120". In parallel, we fit a Point Spread Function (PSF) starting on the theoretical source position to optimize its localisation and compute the source region radius that optimizes the signal-to-noise ratio (SNR) of the source region \citep{Piconcelli2004_SNR_opti}, considering the background, up to 120''.  

Once a suitable source and background regions have been defined, we extract a 1 second binned lightcurve of the source in the [3-79] keV band using the \texttt{nuproducts} task. If any part of the lightcurve exceeds 100 counts/sec, following the recommendations of the standard threads\footnote{\href{https://heasarc.gsfc.nasa.gov/docs/nustar/nustar\_faq.html\#bright}{https://heasarc.gsfc.nasa.gov/docs/nustar/nustar\_faq.html\#bright}}, we re-run the previous steps of data analysis (both nupipeline and region definition), this time having added the "\texttt{(STATUS==b0000xxx00xxxx000)\&\&(SHIELD==0)}" keyword in \texttt{nupipeline} to mitigate the mismatching of noise events.  

We then extract the final spectral and temporal products of each focal plane independently, using the \texttt{nuproducts} task, and group the spectra according to the \cite{Kaastra2016_binning_opt} optimized binning.

\begin{table*}[h!]
    \begin{center}
    \begin{tabular}{c|c|c|c|c|c|c|c|c}
    \multirow{3}{*}{Outburst period} & \multirow{3}{*}{Outburst type} & \multicolumn{4}{c}{Observations} &  & \multicolumn{2}{c}{High-E monitoring} \T \B \\
    
    &   & \chandra{} & \nicer{}       & \nustar{}    & \suzaku{}    & \xmm{} & \swift{}/BAT & \integral{} \T \B \\
    \hline
    \hline 
        2002-2004 & super (triple)               
        & \textbf{1}& 0         & 0         & 0         & 0     & 0 & 1    \T \B \\\hdashline
    
        2006 & standard                              
        & 0         & 0         & 0         &\textbf{6} & 0   & / & / \T \B \\\hdashline

        2010 & standard  (double)                           
        & 0         & 0         & 0         &\textbf{1} & 0   & / & / \T \B \\\hdashline
        
        2012-2013 & super (triple)               
        &\textbf{7} & 0         & \textbf{2}  &\textbf{2} &\textbf{8} & 15 & / \T \B \\\hdashline
        
        2015 & standard                                
        & 0         & 0         & \textbf{2}$^\star$ &\textbf{3} & 0    & 1 & /     \T \B \\\hdashline
        
        2016-2017 & standard                           
        &\textbf{1} & 0         & 0         & 0         & 0    & 1 & /     \T \B \\\hdashline
        
        2018 & standard                
        & 0         &\textbf{33}& 0         & 0         & 0      & 32 & 0  \T \B \\\hdashline
        
        2020 & standard                                
        &\textbf{3} &\textbf{39}& 0         & 0         & 0      & 42 &  /  \T \B \\\hdashline

        2021-2022 & standard (double)                          
        & 0         &\textbf{29}& 0         & 0         & 0     & 29 & / \T \B \\\hdashline
        
        2022-2024 $\dagger$ & super (double)     
        & 0         &\textbf{71}& \textbf{7}$^\star$ & 0 & 0 & 61     & 1    \T \B \\\hline
        
        Total &                                     
        & 12        & 172       & 11        & 11        & 8    & 180 & 2     \T \B \\ 
    \end{tabular}
    \end{center}
    \vspace{-1.5em}
    \caption{List of outbursts covered in our sample, number of observations suitable for line detection with each instrument as of 12-2023, and number of observations with additional high-energy telescope from monitoring (see Sec.~\ref{sub:SA_highE_coverage}). $\star$ All the 2015 and 2022-2023 \nustar{} exposures are simultaneous to \suzaku{} and \nicer{} observations, and only the four 2023 observations are considered independently (see section 3.2) and highlighted in the figure below.}
    \label{tab:outburst_list}
\end{table*}

\subsection{\suzaku{}}

Suzaku \citep{Mitsuda2007_Suzaku} observed the source 12 times in 2006, 2010, 2012, and 2015 at different luminosities. Among them, 11 observations were performed at bright phases in outbursts when the source was in the high/soft or intermediate state \citep{Hori2018_4U1630-47_2015SuzakuNustar} and the other one was at the end of the 2010 outburst when the source was in the low/hard state \citep{Tomsick2014_4U1630-47_decay_2010_anomalous}. 

The observations were made with two detectors: the X-ray Imaging Spectrometer (XIS) and Hard X-ray Detector (HXD). The HXD stopped its operation before the observations in 2015 (OBSID=409007010, 409007020, and 409007030) due to the power shortage of the spacecraft, and therefore only the XIS data are available in these epochs. The XIS are composed of three frontside-illuminated (FI) CCDs (XIS-0, XIS-2, and XIS-3) and a backside-illuminated (BI) CCD (XIS-1). The XIS-2 stopped working from the end of 2006 so the data of it are only available in 2006 observations (OBSID=400010010 through 400010060).

We adopted all the available XIS and HXD PIN data and conducted data reduction for the individual observations, using the latest \suzaku{} CALDB (version 20160607). We utilized the cleaned event files produced by the final version (v3.0.22.43 or v3.0.22.44) of pipeline processing. For the 2010 data (OBSID=405051010), which did not suffer from pile-up effects, we extracted the source spectra from circular regions of $1'.3$ radii centered at and $\sim 7'$ apart from the source position as the source and background regions, respectively. For the data taken in 2012 February (OBSID=906008010), we adopted an annular region with inner and outer radii of $0'.7$ and $1'.8$, respectively. The background subtraction was not conducted for this observation, because its contribution was negligible, less than 0.1\% at all energies, and the inner radius is chosen to limit the pile-up fraction to below 1\% , estimated in the same way as in \citet{Hori2018_4U1630-47_2015SuzakuNustar} using the tool {\tt aepileupcheckup.py}\footnote{\url{http://www-x.phys.se.tmu.ac.jp/~syamada/ana/suzaku/XISPileupDoc_20120221/XIS_PileupDoc_20120220_ver1.1.html}}. 
For all other observation, which were already analyzed in \citet{Hori2018_4U1630-47_2015SuzakuNustar}, we employed the same source and background regions as those in that work. The response matrix files and ancillary response files were made with the ftools {\tt xisrmfgen} and {\tt xissimarfgen}, respectively. We merged the FI CCD data taken in the same observations. For the HXD PIN, we created the background data by merging the ``tuned'' Non-X-ray background files provided by the \suzaku{} team\footnote{\url{https://darts.isas.jaxa.jp/astro/suzaku/analysis/hxd/pinnxb/tuned/}} and the modeled cosmic X-ray background\footnote{\url{https://heasarc.gsfc.nasa.gov/docs/suzaku/analysis/pin_cxb.html}}. We used the appropriate versions of the PIN response files\footnote{\url{https://darts.isas.jaxa.jp/astro/suzaku/analysis/hxd/pinnxb/quick/}} included in the CALDB.

\begin{figure*}[h!]
\centering
\includegraphics[width=0.99\textwidth]{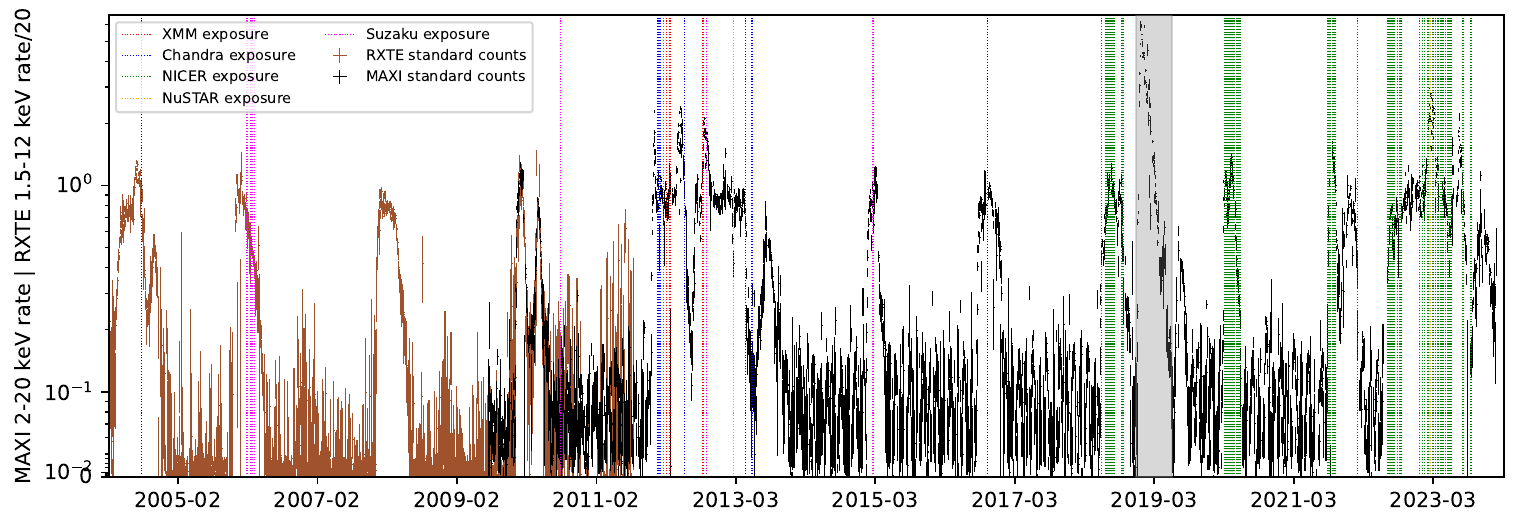}
\vspace{-1.5em}
\caption{Long-term \maxi{}and \rxte{} (normalized to MAXI) lightcurves of 4U~1630$-$47. with exposures used for line detections highlighted by dashed vertical lines. The gray band highlights contamination by the 2018 outburst of the nearby BHB MAXI J1631-479 \citep{Miyasaka2018_MAXIJ1631-479_discovery} 
}\label{fig:4U_monit_full}
\end{figure*}

\subsection{\integral{}}

The INTErnational Gamma-Ray Astrophysical Laboratory (\integral{}) satellite was launched in 2002 and made observations for most outbursts of 4U1630$-$47. We used data from the Imager on-Board the \integral{} Satellite (IBIS, \cite{Ubertini2003_IBIS}), and more specifically from the IBIS Soft Gamma Ray Imager (ISGRI, \cite{Lebrun2003_ISGRI}), which is sensitive in the 30 - 500 keV range and has a 12' angular resolution thanks to its coded aperture. For data reduction we used the Off-line Scientific Analysis (OSA) v11.2 software \footnote{\url{https://www.isdc.unige.ch/integral/download/osa/doc/11.2/osa_um_ibis/man_html.html}}, which allowed us to produce lightcurves and spectra on satellite revolution basis ($\sim$ 2.5 days).

We used 20 logarithmically spaced energy bins between 30-200 keV for every spectra, which we fitted with a simple powerlaw model, allowing us to derive fluxes.

\subsection{\swift{}/BAT}

In order to assess the long-term evolution of the source, and to complement our high-energy coverage, we use the daily BAT lightcurve products available via the BAT Transient Monitor\footnote{\href{https://swift.gsfc.nasa.gov/results/transients/weak/4U1630-472/}{https://swift.gsfc.nasa.gov/results/transients/weak/4U1630-472/}} \citep{Krimm2013_Swift_BAT_transient_monitor}. 

\section{Spectral Analysis} \label{sec:spectral_analysis}

Our spectral analysis, line detection, and line significance methodology remains similar to our previous study. The procedure is explained in detail in Section 3 of P24. In this work, we use Xspec version 12.13.1 \citep{Arnaud1996_xspec}, via \texttt{Pyxspec} version 2.1.2. 

For 4U~1630$-$47, we choose a continuum composed of a \texttt{diskbb} and/or \texttt{nthcomp} \citep{Zdziarski1996_nthcomp_1,Zycki1999_nthcomp_2}, multiplied by  a \texttt{TBabs} for the ISM absorption. The high-energy cutoff of the Comptonized component is unconstrained in all observations and thus kept frozen at 100 keV, and its seed photon temperature is set to the \texttt{diskbb} temperature when a disk component is present, or otherwise fixed at 0.5 keV. In most soft observations without high-energy coverage, the photon index cannot be constrained. Such spectra are overwhelmingly disk dominated and the value of $\Gamma$ value is inconsequential, but for the sake of consistency, it is frozen at a fiducial $\Gamma=2$ (in accordance with the low values of gamma expected in soft states, see Sec.~\ref{sub:SA_highE_coverage} and Sec.~\ref{sub:high_E_comp_evol}). In addition, the procedure can add up to five absorption lines according to the main transitions in the iron complex (\FeKav{}, \FeKavi{}, \FeKbv{}, \FeKbvi{} and \FeKgvi{}), and two broad (width up to 0.7 keV) emission lines for neutral Fe at 6.4 and 7.06 keV.

This list of components, while still very basic, works well with 4U1630$-$47's very simple evolution in the soft states, SPL, and hard state, and allows for a sufficient estimate of the hard X-rays continuum of the source for the sake of photoionization computations. We note that 4U~1630$-$47 only shows weak reflection features, which we found could be sufficiently well modeled with the two empirical emission line components, and detailed reflection modeling remains beyond the scope of our analysis.

We note that 4U~1630$-$47 has a well-known dust scattering halo \citep{Kalemci2018_4U1630-47_d}, which has a known effect on the spectral shape \citep{Gatuzz2019_4U1630-47_wind_2012-13Chandra}, but as we found no significant improvements in the fits using the \texttt{xscat} model \citep{Smith2016_xscat_dust_scattering_model}, we choose not to include this component in our analysis.  

The following paragraphs highlight the specifics of the procedure with each telescope not included in P24, and in the case of simultaneous observations. For \xmm{} and \chandra{} we refer the reader to Section 3 of P24. 

\subsection{Individual satellites}

In\nicer{}-only epochs, in order to limit the effect of instrumental features at low energy, we restrict the broadband fit to 2.5-10 keV for this instrument. This also allows us to keep the very simple list of continuum components used in P24. We group individual GTIs less than one day apart and analyze them together as daily ``epochs''. Similarly to \xmm{} and \chandra{} we apply a predefined count threshold to restrict the analysis to \nicer{} GTIs with sufficient data quality. We use a threshold of 5000 net counts (subtracting the \texttt{scorpeon} model rate of each GTI with default parameters) in the 4-10 keV band. 172 (out of 189) \nicer{} epochs remain after this quality cut. To account for small differences between individual GTIs, we let a \texttt{constant} multiplicative component free to vary in \texttt{Xspec} for each GTI datagroup in a \nicer{} epoch. The background \texttt{scorpeon} model of each GTI/datagroup is left free to vary during the broad band fit, and remains fixed during the remaining part of the spectral analysis. \\
Despite these actions, a small number (9) of days had to be manually re-split due to significant variability in the spectral shape of the continuum between the different GTIs of a given day. These obsids are highlighted in Appendix~\ref{app:tables}. Each daily epoch is then considered as an individual observation for the remainder of this work. \\

\begin{figure*}[h]
\includegraphics[width=0.49\textwidth]{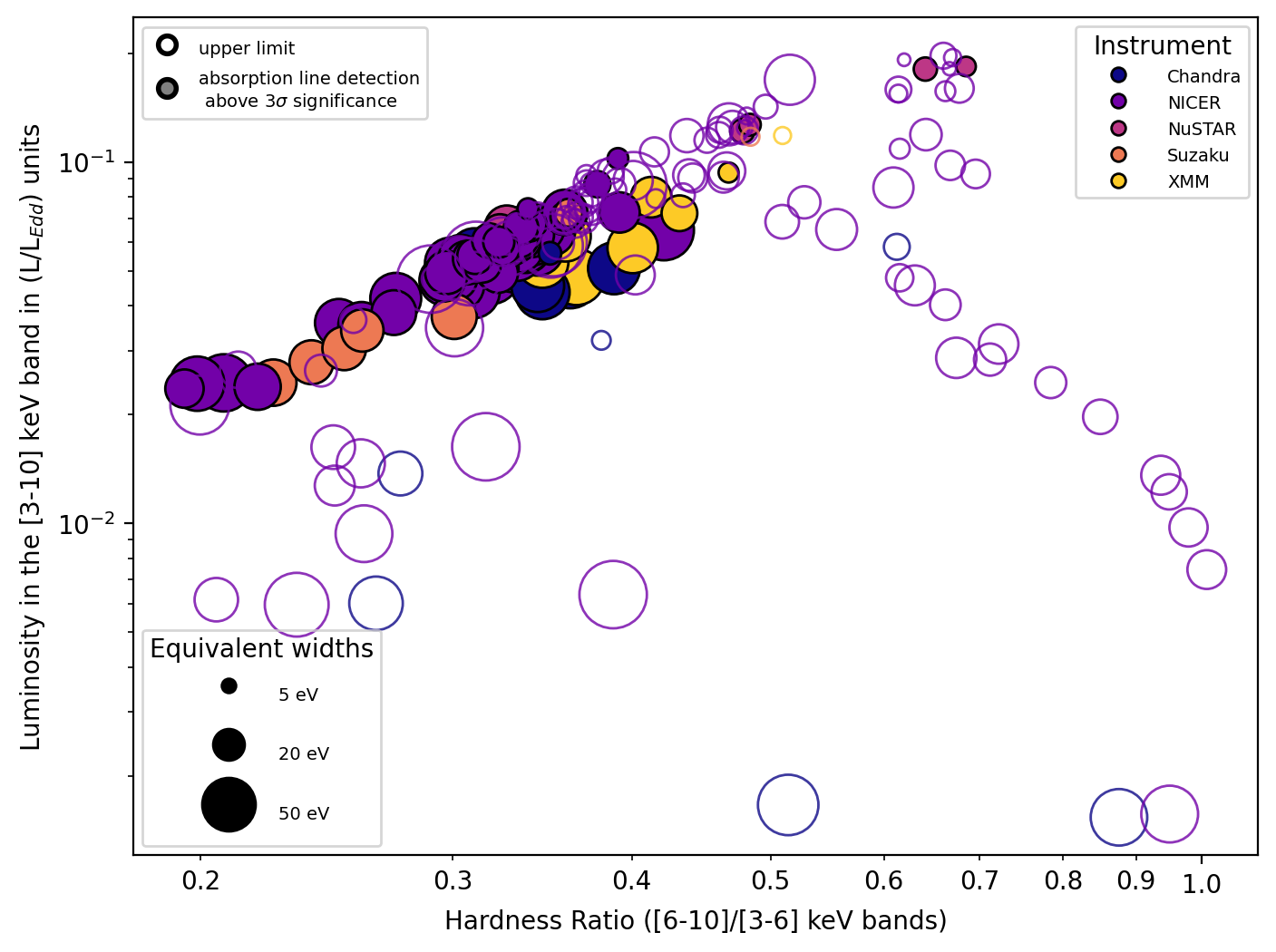}
\includegraphics[width=0.49\textwidth]{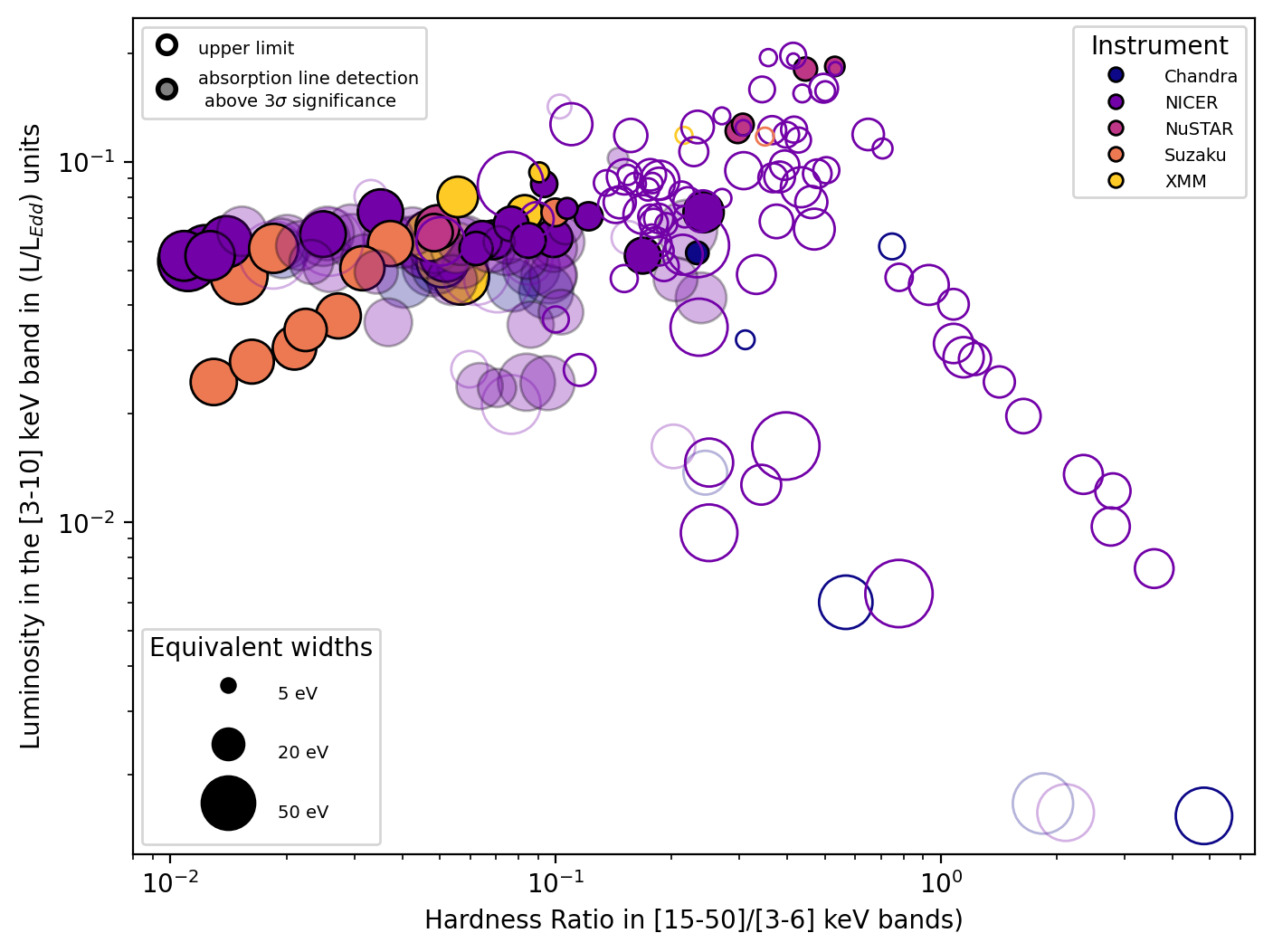}

\vspace{-1.em}
\caption{Multi instrument "soft" (left) and  "hard" (right) HLDs of 4U~1630$-$47, colored according to instruments. In the right panel, transparent markers indicate the position of 1$\sigma$ HR upper limits in non-significant detections (see Appendix \ref{app:highE_coverage})}\label{fig:4U_HID_full}
\end{figure*}

For \suzaku{}, we analyze the spectra of the XIS and PIN detectors together, restricting their energy range to [1.9-9.0] keV and [12-40] keV respectively. 
In addition, we ignore the [2.1-2.3] keV and [3.0-3.4] keV intervals in XIS spectra, due to known calibration uncertainties (see e.g. \citealt{Hori2018_4U1630-47_2015SuzakuNustar}).
We analyze the summed FI XIS, BI XIS and (when available) PIN spectra of individual observations together, and account for known discrepancies between the detectors (notably the FI and BI detectors of XIS, see e.g. \citealt{Shidatsu2013_MAXIJ1305-704_wind_soft_hard_Swift_Suzaku} \citealt{Hori2018_4U1630-47_2015SuzakuNustar}), by applying a global \texttt{crabcorr} correction \citep{Steiner2010_crabcorr} to each datagroup. This model multiplies the rest of the components by a powerlaw with two parameters, the normalization and a $\Delta\Gamma$ between datagroups. Here, we left the normalization free to vary except for the BI-XIS, which serves as a reference and is thus frozen at 1. The $\Delta\Gamma$ is only left free to vary for the FI-XIS datagroup, and kept frozen at 0 for the others. We also verified that all PIN spectra remain above the systematic uncertainty (3\%) of the background modeling of the instrument. As the 2010 low-hard state \suzaku{} observation (ObsID 405051010) ends up with very poorly constrained upper limits for the presence of lines (due to a very low luminosity of $\sim3\cdot10^{-5} L_{Edd}$), we discard this observation (ObsID 400010050) from the main body of the paper, but the results of the line detection are still accessible in the Appendix tables and the online tool.

For \nustar{}, we analyze the data of each focal plane together, allowing a multiplicative \texttt{constant} to vary between the two components. We found no significant discrepancy between the FPMA and FPMB that would warrant the use of the MLI correction model\footnote{\href{https://nustarsoc.caltech.edu/NuSTAR\_Public/NuSTAROperationSite/mli.php}{https://nustarsoc.caltech.edu/NuSTAR\_Public/\\NuSTAROperationSite/mli.php}} in any of the observations. We restrict the lower bound of our energy band to 4 keV, as the residuals strongly deviate from the continuum below this value in every observation. For the upper bound, a dynamical restriction is preferable, since the spectral shape strongly affects the energies at which the spectrum remains significantly above the background. The limit is fixed to the energy where the SNR of the source\footnote{computed according to\\ \href{https://xmm-tools.cosmos.esa.int/external/sas/current/doc/specgroup.pdf}{https://xmm-tools.cosmos.esa.int/external/sas/current/doc/specgroup.pdf}} passes below 3, up to a maximum of 79 keV. We also add a common empirical 9.51 keV \texttt{edge} component to both detectors to account for an instrumental feature, in accordance to previous studies of this source \citep{Ratheesh2023_4U1630-47_pola_soft,Cavero2023_4U1630-47_pola_SPL,Podgorny2023_NuSTAR_calfeature}.

\subsection{Simultaneous observations}

In a number of bright epochs, the soft X-ray coverage provided by \nicer{} or \suzaku{} is complemented by simultaneous \nustar{} coverage. While fitting the different instruments together is beneficial, the low-energies (below $\sim10$ keV) of \nustar{} remains typically very inconsistent with the coverage of the other instruments, and can only be broadly reconciled by applying a very significant $\Delta\Gamma$ (beyond 0.15) to the entire spectrum. However, the low-energy spectrum of \nustar{} barely provides additional constraints on the lines because of the better spectral resolution and already very high SNR of the spectra of other instruments. Thus, whenever there is good agreement between the characteristics of the lines measured by \nustar{} and other instruments, we fit the different instruments together while ignoring \nustar{} energies below 8 keV, which allows us to reach good agreement without the need of a slope correction for the \nustar{} spectrum. 

We make an exception for all observations between 09-03-2023 and 13-03-2023. During this period,  several \nicer{} and \nustar{} observations were triggered to complement the first SPL \ixpe{} exposure of this source \cite{Cavero2023_4U1630-47_pola_SPL}. A \FeKavi{} line is detected with very high significance in all \nustar{} exposures but not in any of the partially simultaneous \nicer{} exposures. This might be due to a combination of high variability and lower SNR in the (much shorter) \nicer{} exposures, and will be studied in more detail in a subsequent paper. For now, the results of the line detection for these \nustar{} exposures are kept independent from the \nicer{} exposures.

\begin{figure*}[h]

\includegraphics[width=0.49\textwidth]{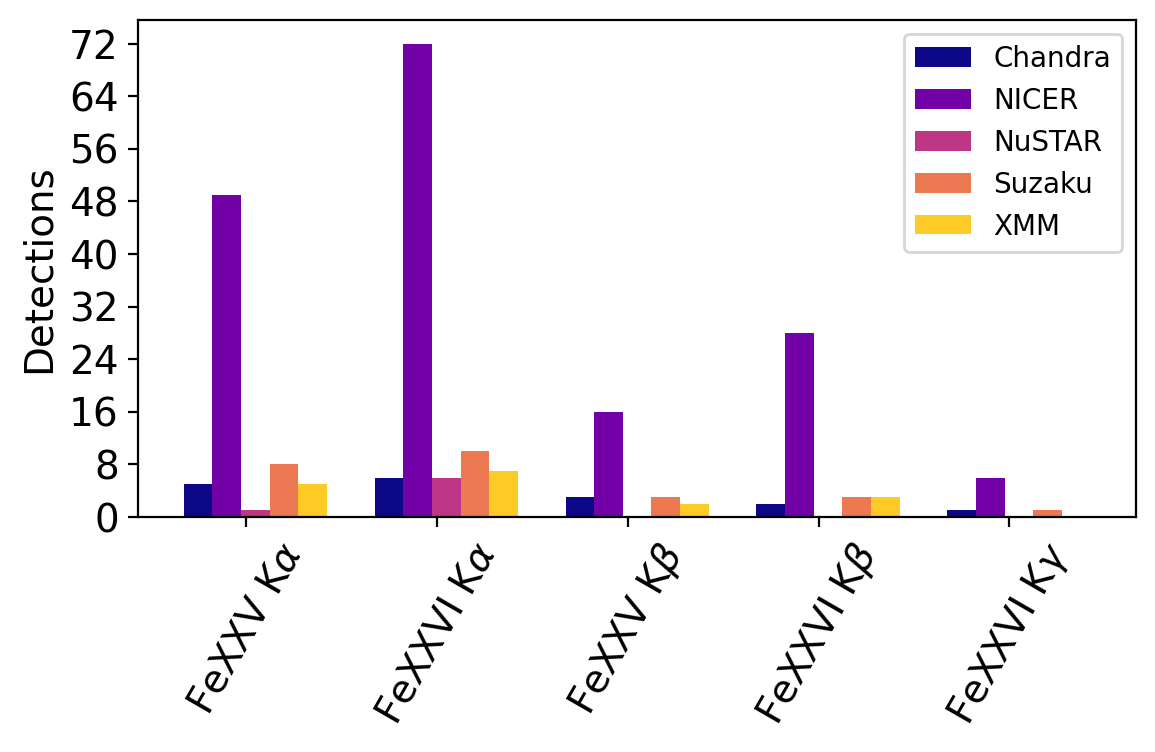}
\includegraphics[width=0.49\textwidth]{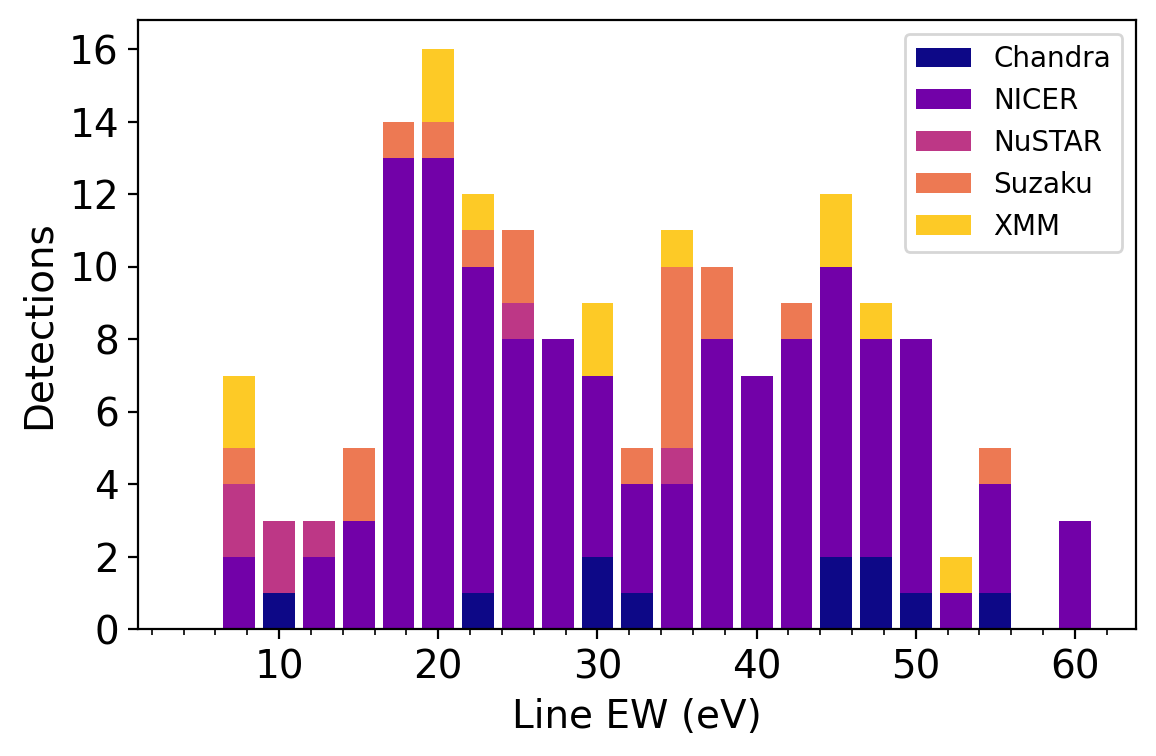}
\includegraphics[width=0.49\textwidth]{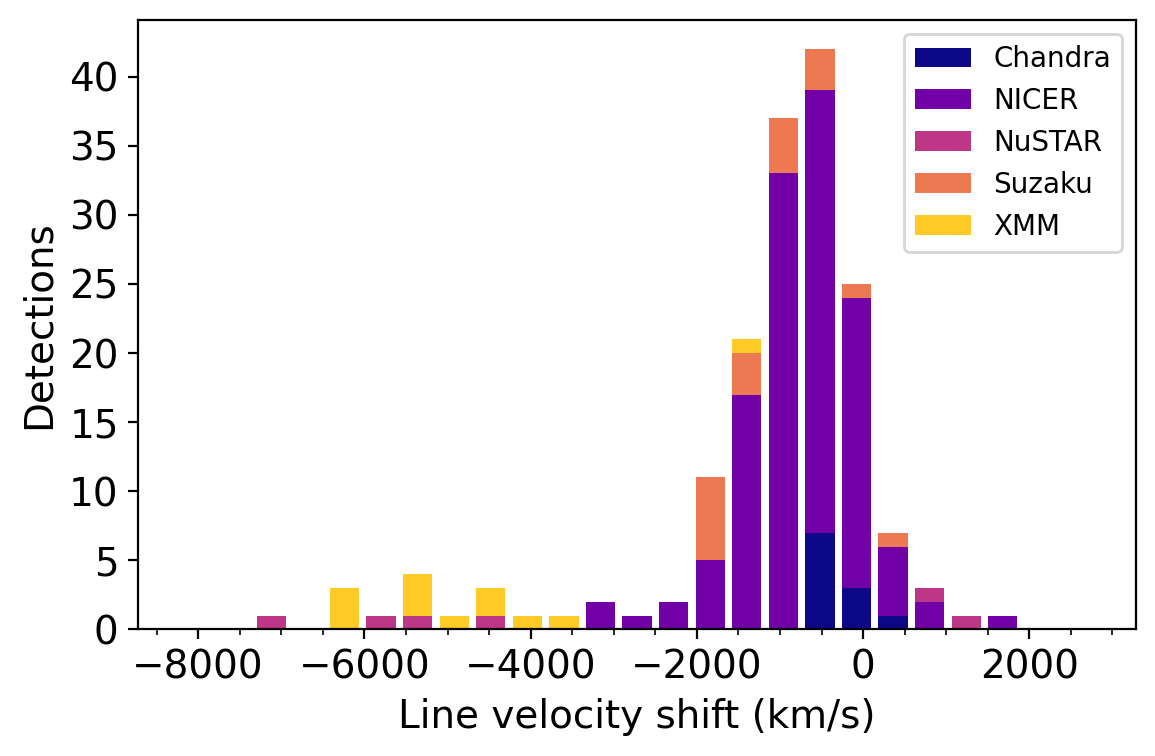}
\includegraphics[width=0.49\textwidth]{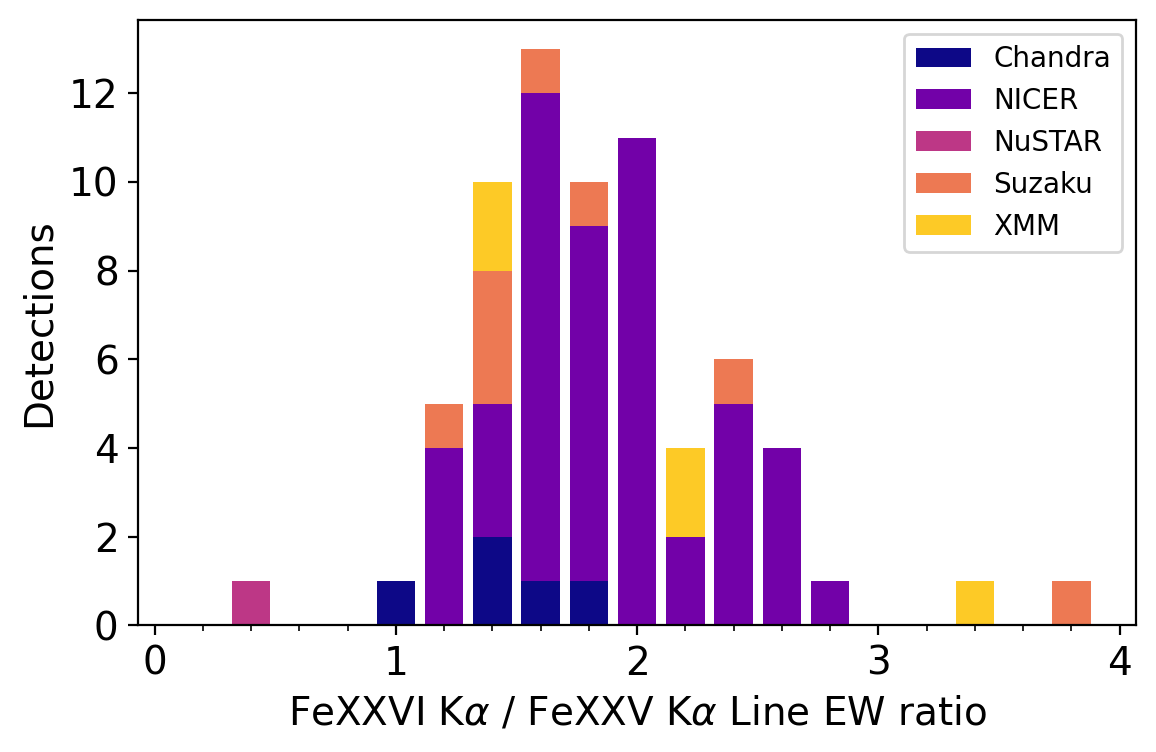}
\vspace{-1.5em}
\caption{Distribution of intrinsic line parameters (detections of each line, EW, K$\alpha$ complex blueshift and K$\alpha$ EW ratio) for the entire sample, split by instrument.}\label{fig:glob_distrib}

\end{figure*}

\subsection{Secondary coverage from \swift{}/BAT and \integral{}}\label{sub:SA_highE_coverage}

The coverage of the high-energy band with \suzaku{}-PIN and \nustar{} exposures remains limited to $\sim20$ daily epochs. While \integral{} has observed the source for a significant number of revolutions (93), the vast majority were performed before the recent increase in the monitoring of this source thanks to\nicer{}, and they are typically not simultaneous with soft X-ray instruments or made in conjunction with \suzaku{} and \nustar{} (and thus do not provide additional information for our purposes). In addition, a significant portion of \integral{} soft-state exposures are too short to derive a proper spectrum. On the other side, the \swift{}/BAT monitoring provides almost daily coverage of the source, but is lacking in sensitivity and provides very limited spectral information. 

However, we can still compute flux estimates using the count rate of both instruments. For this, we take advantage of the very strong correlations which exist between the source high-energy flux, its photon index, and the count rate of \swift{} and \integral{} observations. The full procedure is detailed in Appendix \ref{app:highE_coverage}, and provides first order estimates of the 15-50 keV flux, or its upper limit when the source is observed but not detected with BAT and/or \integral{}.

\section{Global behavior}\label{sec:global_behavior}

In Table~\ref{tab:outburst_list}, we list the numbers of observations analyzed in each of the outbursts covered in this work, and how many of them use the BAT or \integral{} coverage. To highlight the long-term evolution of the source, we also show a long-term monitoring lightcurve of the last 20 years in Fig.~\ref{fig:4U_monit_full}. Due to the lack of precise mass measurements for the mass and distance of the source, for the luminosity estimates we assume a fiducial mass of 8 \msun, and a distance of 8 kpc.

\subsection{HLD evolution at low and high energies}\label{sub:HLD_evol}

We plot in the left panels of Fig.~\ref{fig:4U_HID_full} the unabsorbed Hardness-Luminosity Diagram (HLD) of 4U1630$-$47, using the ratio of the intrinsic luminosities in the [6-10] and [3-6] keV bands for the Hardness Ratio, and the [3-10] keV band luminosity in Eddington units. The full sample provides a near complete coverage of the typical evolution of the source above luminosities of $\sim10^{-2} L_{Edd}$, although spread over different outbursts. The vast majority of soft state observations follow a very narrow diagonal, as expected for highly-inclined binaries \citep{Munoz-Darias2013_HID_i}, and the recent observations (notably from the high cadence \nicer{} monitorings of the outbursts after 2017) confirm the already reported disappearance (or at least strong decrease) of the absorption lines above a HR value of $\sim$0.4-0.45 and $L_{3-10}\sim10^{-1} L_{Edd}$ (see e.g. \citealt{DiazTrigo2014_4U1630-47_wind_2012-13XMM} and Sect.~\ref{sub:lit_compa}). Nevertheless, a significant part of the \nicer{} observations in softer states are non-detections with upper limits far too low to be compatible with the detections seen in other observations at very similar HR and luminosity. 

Since the standard HLD lacks information about the hard X-rays above 10 keV, which can affect the properties of the plasma producing the absorption lines, 
we construct a new "hard" HLD, replacing the 3-6 keV Hardness Ratio (hereafter $HR_{soft}$) by the [15-50]/[3-6] keV Hardness Ratio (hereafter $HR_{hard}$). The [15-50] keV luminosity is the most direct way to use the BAT monitoring. On the other hand, while the [3-6] flux matches the peak of the \texttt{diskbb} component, it remains less affected by uncertainties in interstellar absorption and calibration issues than a wider, softer band. For the y axis, we keep the [3-10] keV Eddington ratio, for an easier comparison with the soft HLD. The hard HLD is presented in the right panel of Fig.~\ref{fig:4U_HID_full}, using shaded markers for epochs where the [15-50] keV estimates are less than 2 $\sigma$ significant. 

\begin{figure*}[t!]
\includegraphics[width=0.5\textwidth]{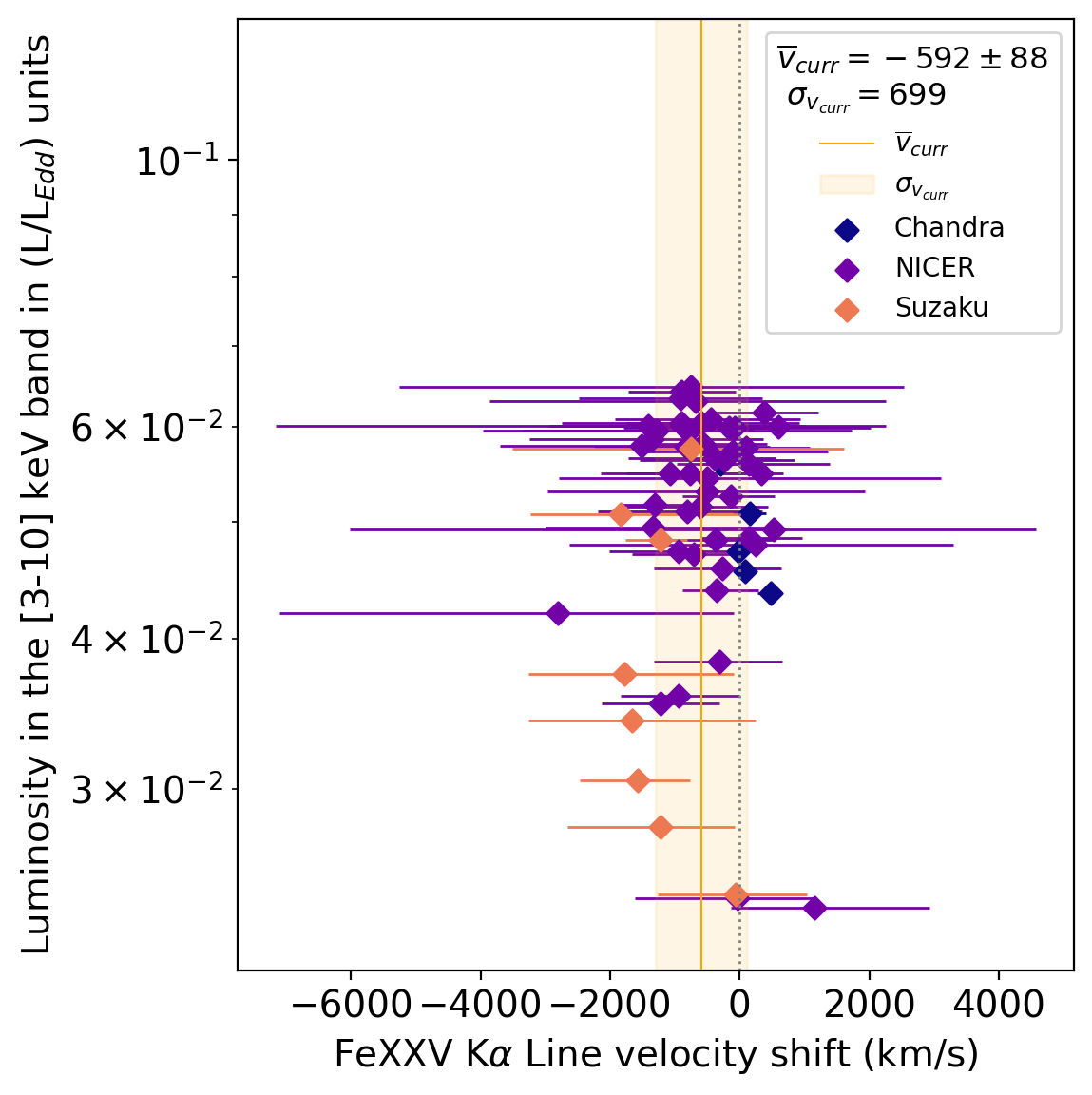}
\includegraphics[width=0.5\textwidth]{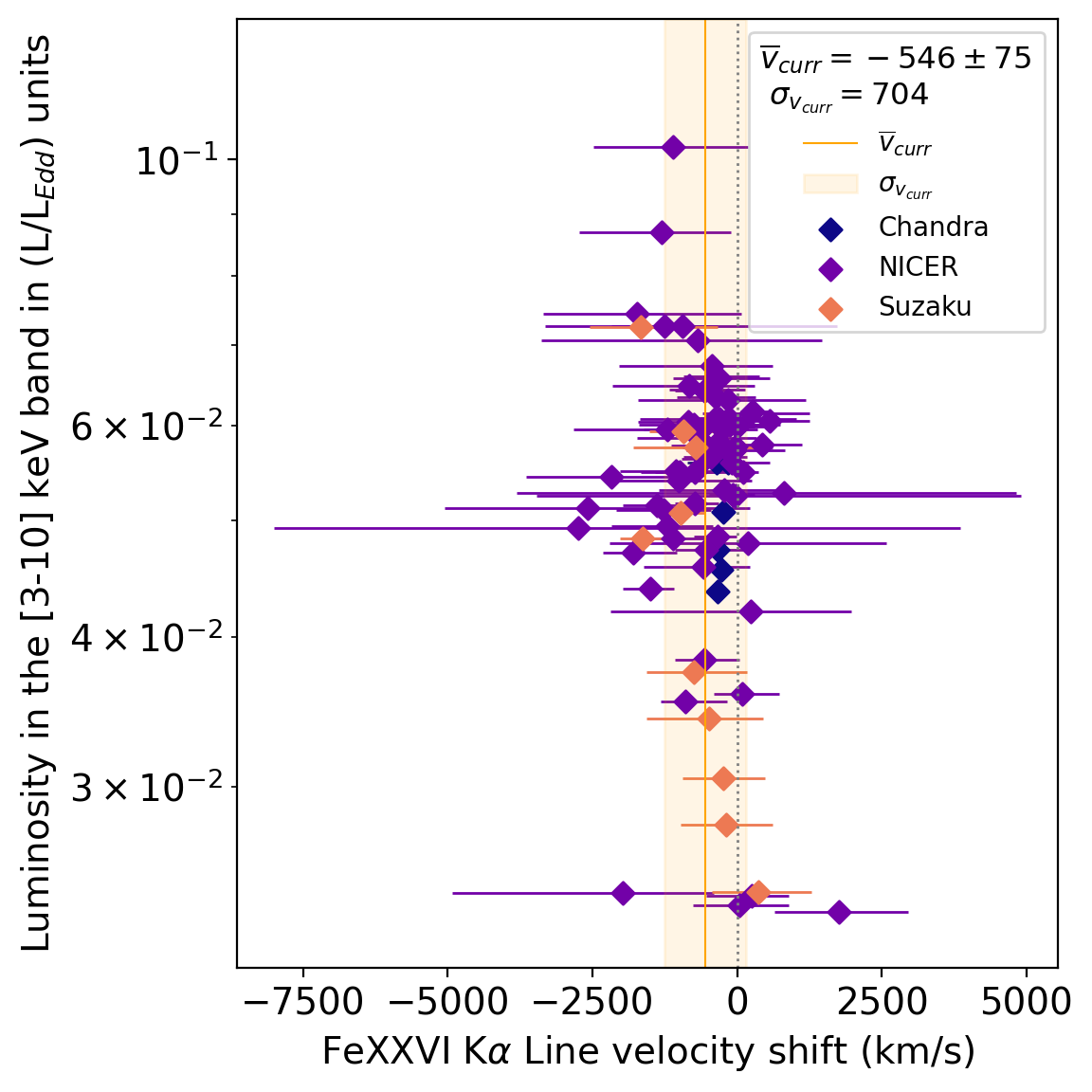}
\vspace{-2.5em}
\caption{Global scatter plots of the blueshift of each of the K $\alpha$ lines for the three instruments with the best line calibration. In each plot, the orange line and region highlight the mean and variance of the current distribution}\label{fig:4U_correl_bshift}
\end{figure*}

This new diagram appears to better separate the states with lines and the states without (or with much weaker) lines. The EW of the absorption lines is clearly anti-correlated with $HR_{hard}$ (which we confirm quantitatively in Sec.~\ref{sub:correlations}), which explains the lack of detections in harder states, as above $HR_{hard}\sim0.1$ the expected EWs of the lines becomes too low to be detected in most observations.
Among the few detections with high EWs ($\geq20$ eV) above $HR_{hard}\sim0.1$, most are associated to $HR_{hard}$ upper limits,  and thus are compatible with much lower $HR_{hard}$ values. The remaining  the remaining 2 are still compatible with sufficiently low $HR_{hard}$ values to match the rest of the observations within errors.
Meanwhile, the few constraining upper limits with low $HR_{hard}$ values systematically have high uncertainties and are compatible with $HR_{hard}\geq0.1$. The majority of them also show hints of weak lines, although not significant enough to be reported.

\subsection{Parameter distribution and correlations}\label{sub:distrib_correl}

We can study the behavior of the absorption lines and quantify the influence of the continuum SED by computing the distribution of absorption lines parameters, as well as statistically significant correlations between line parameters and continuum parameters at low and high energy. We identify correlations with the Spearman coefficient, which traces any monotonic relation between two parameters, considering the uncertainties of each parameter via MC simulations of both the correlation coefficients and their p-values. For this, we follow the perturbation method of \citealt{Curran2014_correl_MC}, via the Python library \texttt{pymccorrelation} \citep{Privon2020_pymmccorrelation}. We consider a correlation "significant" for $p<10^{-3}$. The same perturbation method is applied to consider uncertainties when computing and displaying linear regression between parameters.

\subsubsection{Parameter distribution}\label{sub:distrib}

The main properties of the absorption lines in our sample are the number of individual lines detected, their EWs, and in the case of the better constrained K$\alpha$ complex, the line blueshifts and the ratio between the EW of the lines. We show the corresponding distributions in Fig.~\ref{fig:glob_distrib}. We split them among the different instruments mainly for the sake of visualization, as except for the velocity shifts, the differences between individual distributions are too small to be significant, considering the limited number of detections for all instruments except\nicer{}.\\

The distribution of main parameters follows both trends previously established for wind-emitting sources in general, and the individual results obtained for 4U~1630$-$47 with \chandra{} and \xmm{} specifically (see e.g. \citealt{DiazTrigo2014_4U1630-47_wind_2012-13XMM},\citealt{Gatuzz2019_4U1630-47_wind_2012-13Chandra}, P24). The proportion of detections aligns with the differences in strength between the iron lines, and the sampling of \nicer{} confirms that the EWs of all lines remains below $\sim60$ eV. This suggests ionic column densities and FWHMs similar to those measured by P24 for the few lines resolved in \chandra{}-HETG. Meanwhile, the resolving power of the different instruments allows line detections down to 8-10eV.

\begin{figure*}[h!]
\includegraphics[width=0.33\textwidth]{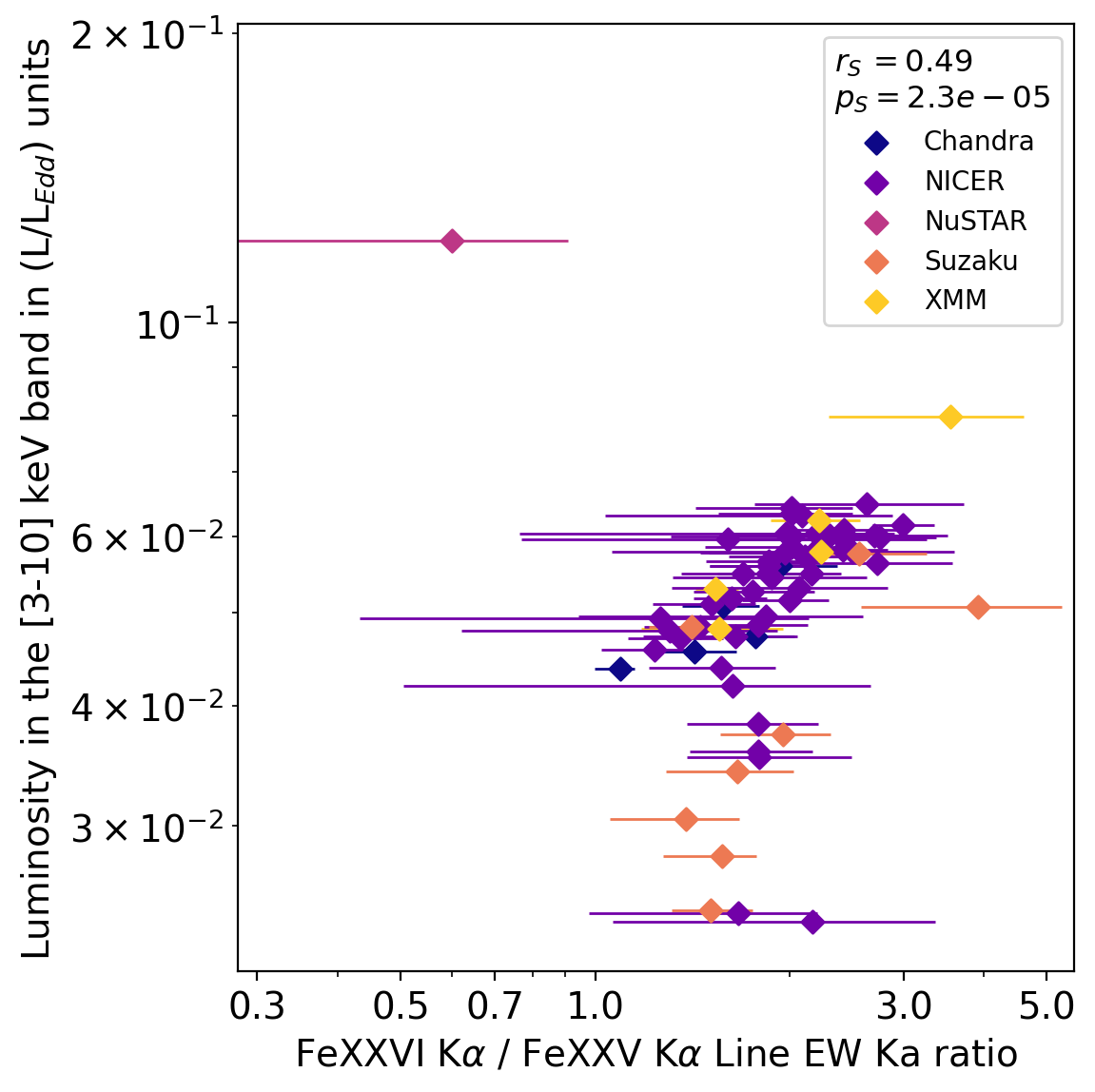}
\includegraphics[width=0.33\textwidth]{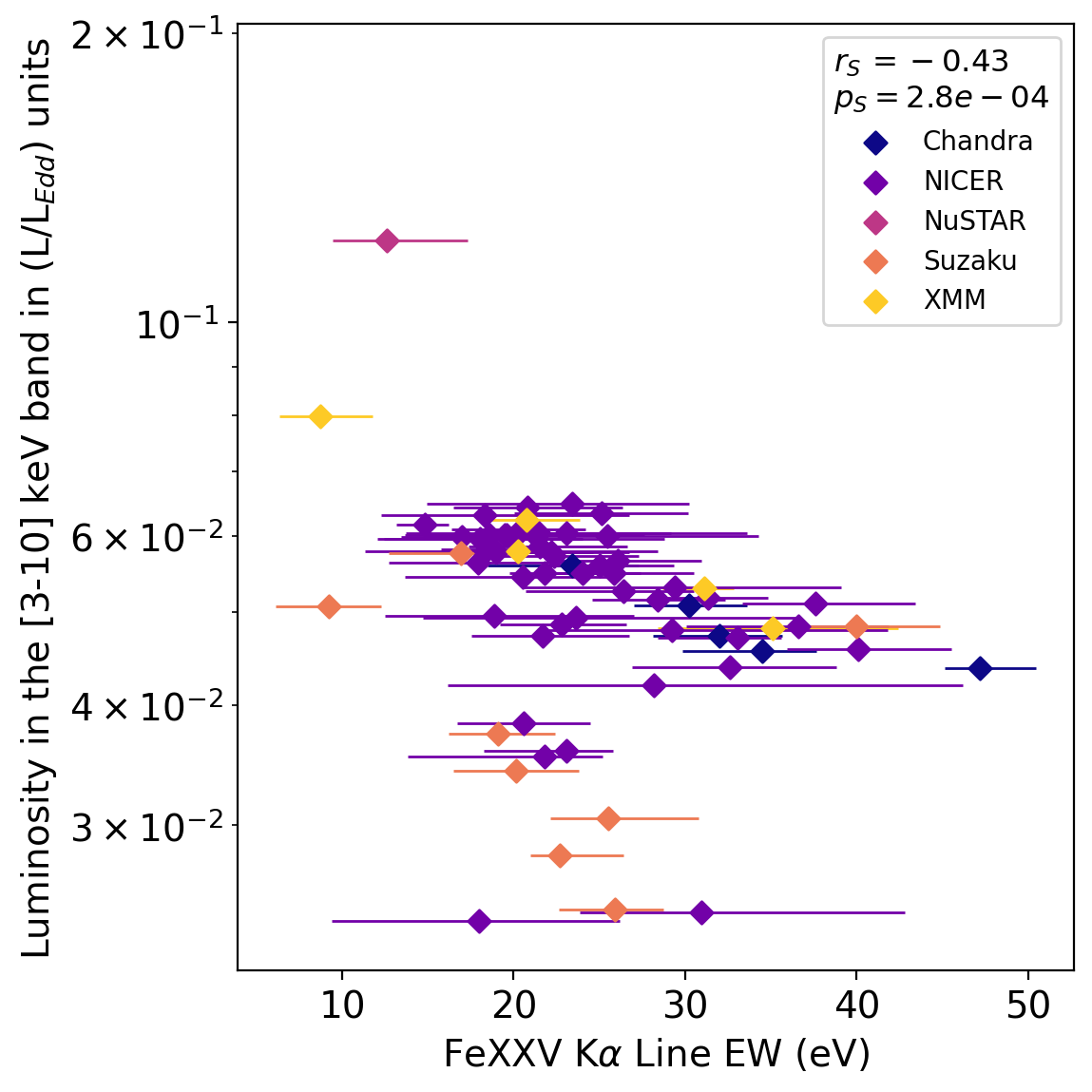}
\includegraphics[width=0.33\textwidth]{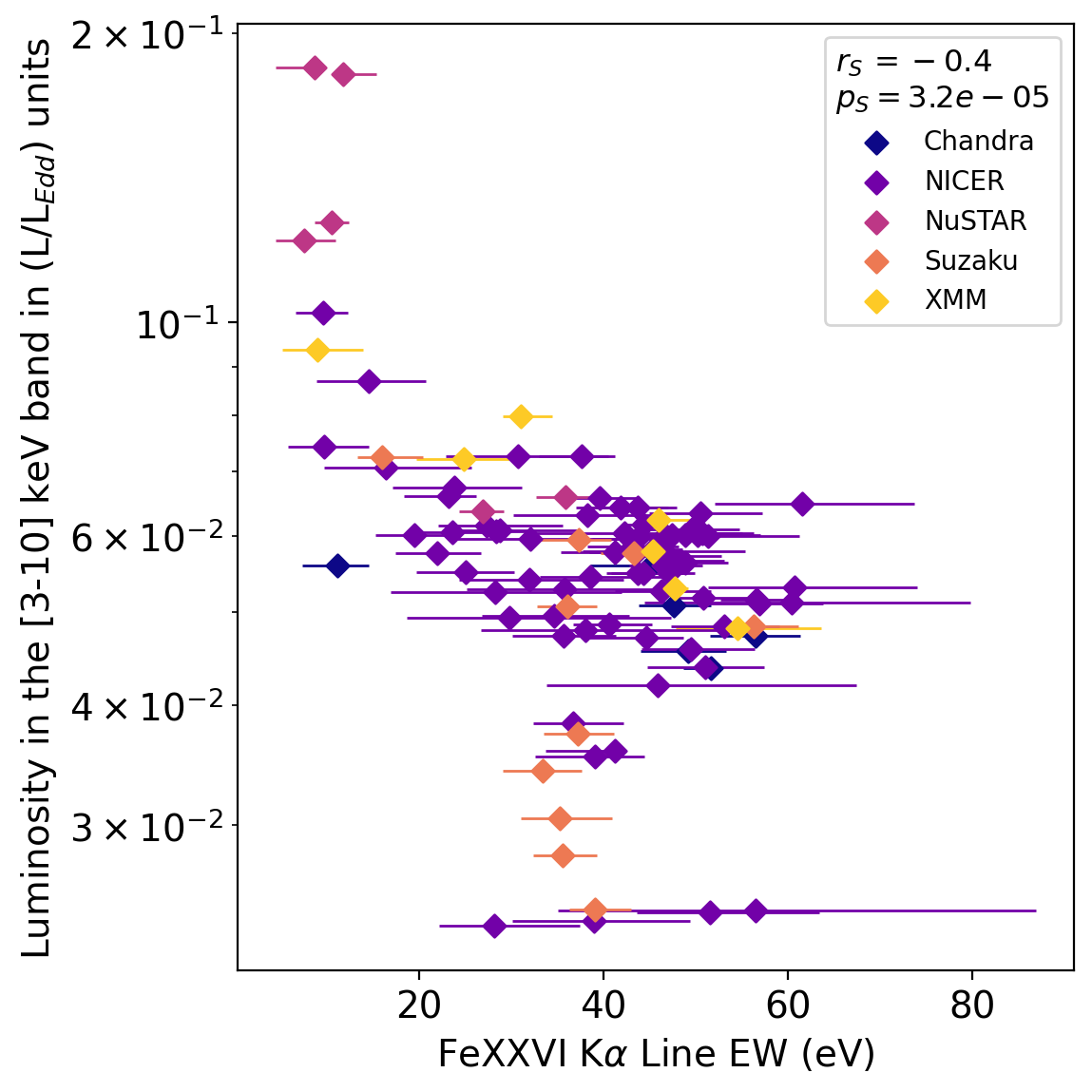}
\includegraphics[width=0.33\textwidth]{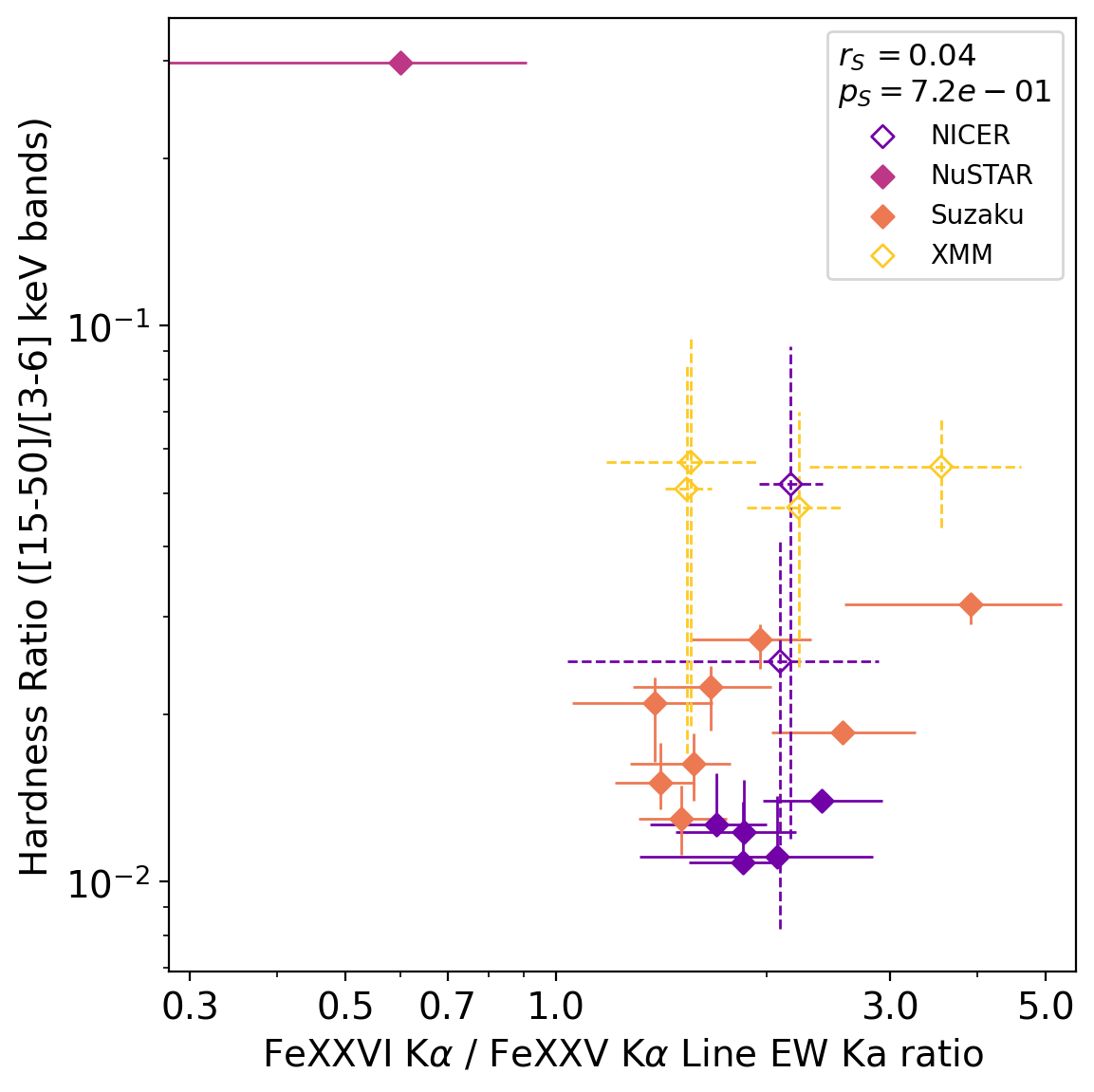}
\includegraphics[width=0.33\textwidth]{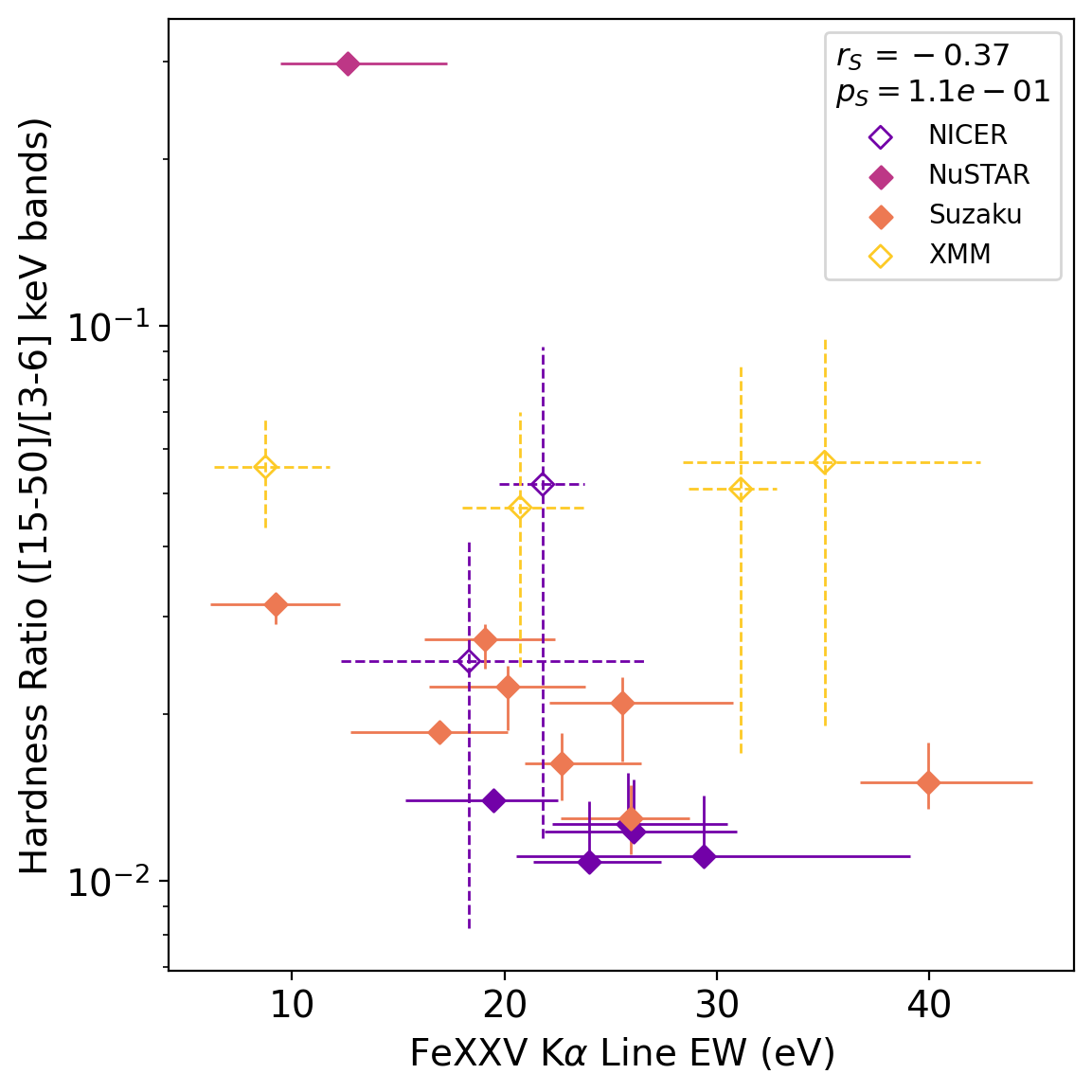}
\includegraphics[width=0.33\textwidth]{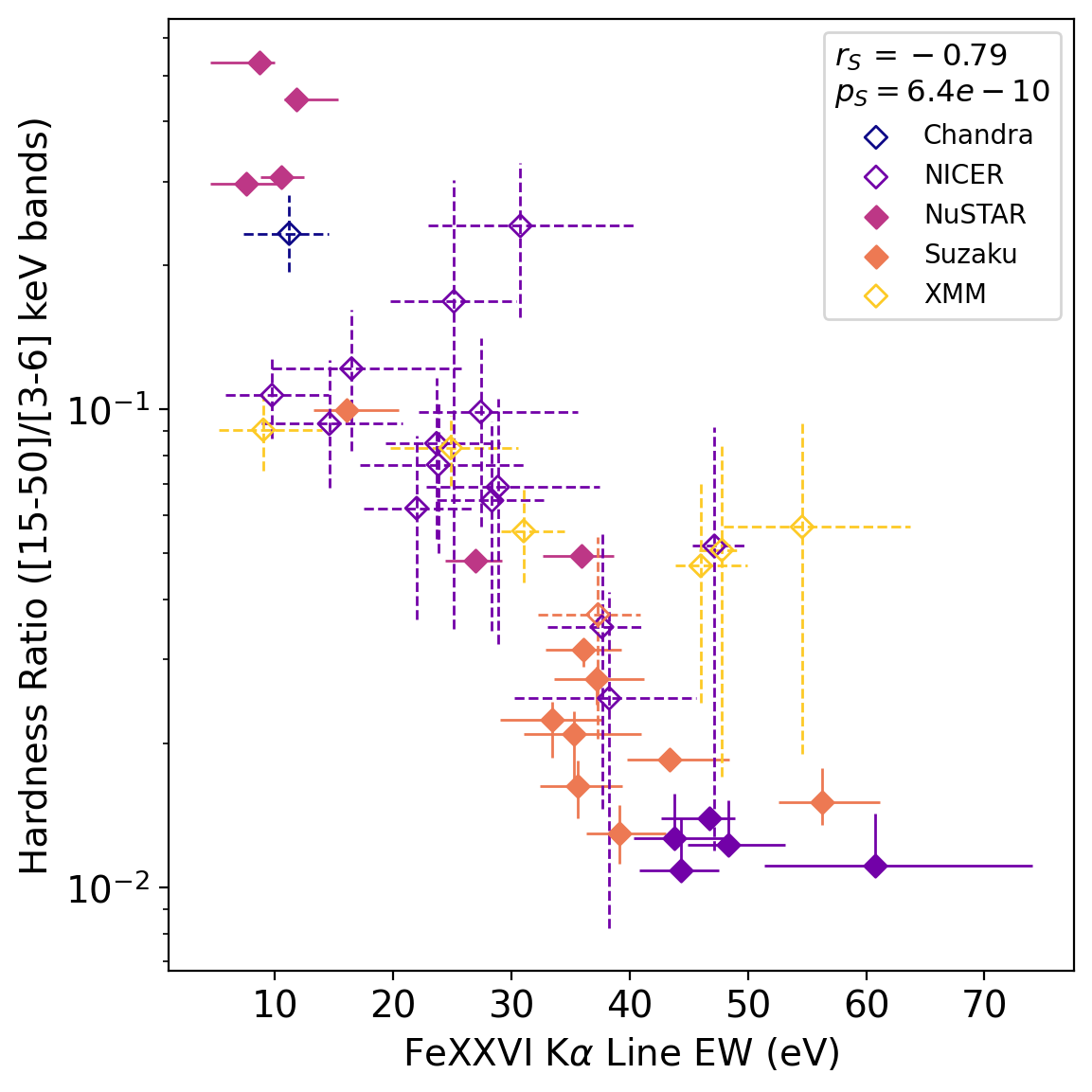}
\vspace{-2.em}
\caption{Global scatter plots of line parameters against soft X-ray luminosity (top) and the hard X-ray Hardness Ratios (bottom). In the bottom panels, the markers are dashed for epochs with a high-energy flux converted from BAT or \integral{}, and full markers  observations with simultaneous high-energy coverage from \nustar{} or \suzaku{}.
}\label{fig:4U_correl_observ}
\end{figure*}

The velocity shift distributions confirm the bias of \xmm{} towards high blueshifts reported in P24, with \nustar{} showing similar behavior, at odds with the rest of the instruments. When restricting to the observations of \chandra{}, \nicer{}, and \suzaku{}, the velocity shift distribution retains an average of $\overline{v_{out}}\sim-560\pm60$ km/s with a standard deviation of $\sim700$ km/s. However, this result does not factor in the 5 eV absolute energy accuracy of \nicer{} \citep{Markwardt2023_NICER_calibration_energy}, or $\sim220$ km/s at the K$\alpha$ lines. The absolute energy accuracy of Suzaku is slightly worse at $\sim$14eV \citep{Koyama2007_Suzaku_XIS}, or $\sim600$ km/s at the K$\alpha$ lines, but the vast majority of the Suzaku blueshifts are much higher than -600km/s, as can be seen in the bottom left panel of Fig.~\ref{fig:glob_distrib}. Thus, even adding these systematics, the average of the distribution remains very significantly distinct from 0, pointing at very low but significant blueshifts in this source, fully compatible with the results obtained in P24 with \chandra{}-HETG, for a larger sample of wind-emitting sources. 

The individual distributions of \FeKav{} and \FeKavi{}, which we show in Fig.~\ref{fig:4U_correl_bshift}, remain extremely similar, with $\overline{v_{out}}_{K\alpha,25}\sim-590\pm90$ km/s, $\overline{v_{out}}_{K\alpha,26}\sim-550\pm75$ km/s, and standard deviations of $\sim700$ km/s for both. 

Finally, we show in the bottom right panel of Fig.~\ref{fig:glob_distrib} the distribution of the EW ratio of the Fe K$\alpha$ complex, defined as the ratio of the EWs of the \FeKavi{} and \FeKav{} lines, and used as a proxy of the ionization parameter. The present sample confirms the trend previously seen in other standard wind-emitting sources: almost all common detections of the $K\alpha$ complex have an EW ratio above 1, with a single detection (NuSTAR in 2023) below as a possible outlier. 

\subsubsection{Significant correlations}\label{sub:correlations}

We do not obtain any significant correlation between the line parameters themselves and, notably, no link between the EW of each line and their velocity. However, this may be due to the very high uncertainty of the velocity measurements with all instruments except \chandra{}. When restricting to \chandra{} only, we see a hint of correlation between
the \FeKav{} velocity and EWs (higher EWs being associated to redshifts). However, this is a natural consequence of the contamination by lower-E satellite lines for lower ionization when using a single gaussian to model the line, as seen in P24 for GRS 1915+105. The number of observations with line detections (6) remains too low for any other conclusion.

As there are no notable correlations between the line parameters themselves, we focus on their behavior compared with the continuum, both at low ($HR_{soft}$,$L_{3-10}$) and high ($HR_{hard}$,$L_{15-50}$,$\Gamma_{nthcomp}$) energies. We first note a lack of correlation between the EW of the lines and the soft Hardness Ratio, similarly to what we obtained for the global wind-emitting sample. This is, however, not the case for the luminosity: as we show in the upper panels of Fig.~\ref{fig:4U_correl_observ}, the \FeKav{} EW, \FeKavi{} and $K\alpha$ EW ratio are all significantly correlated with the soft X-ray luminosity, with $p_{s}\sim2\times10^{-5}$, $r_s=0.49$ for the EW ratio correlation,  $p_s\sim3\times10^{-4}$, $r_s=-0.43$ and $p_s\sim3\times10^{-5}$, $r_s=-0.4$ for the \FeKav{} and \FeKavi{} EWs, respectively.
These trends match both what was found for the global wind-emitting sample in P24 and the individual correlations that were already present using only \xmm{} and \chandra{}, but not significant due to the low number of observations. 

\begin{figure*}[h]
\includegraphics[width=0.49\textwidth]{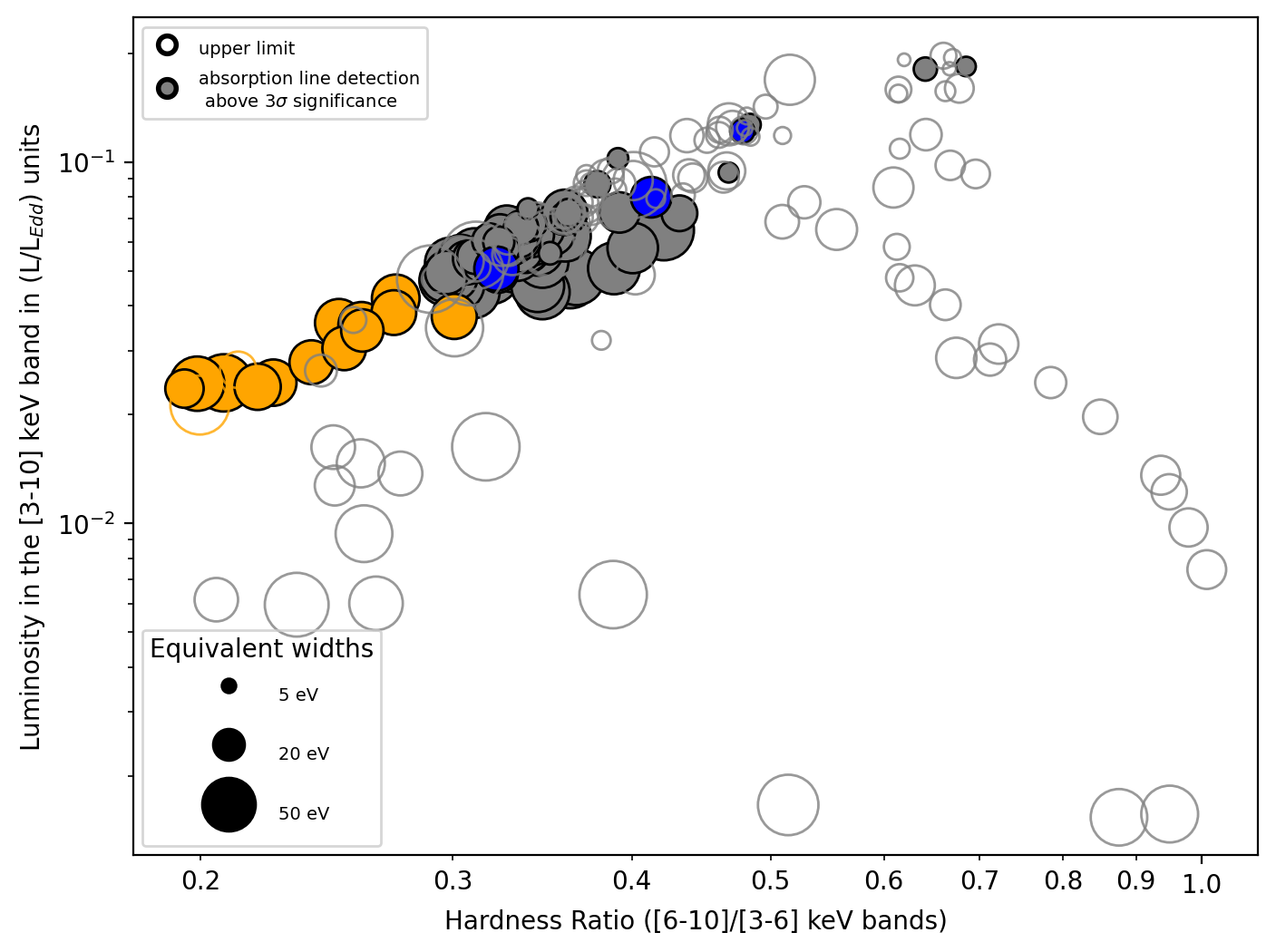}
\includegraphics[width=0.49\textwidth]{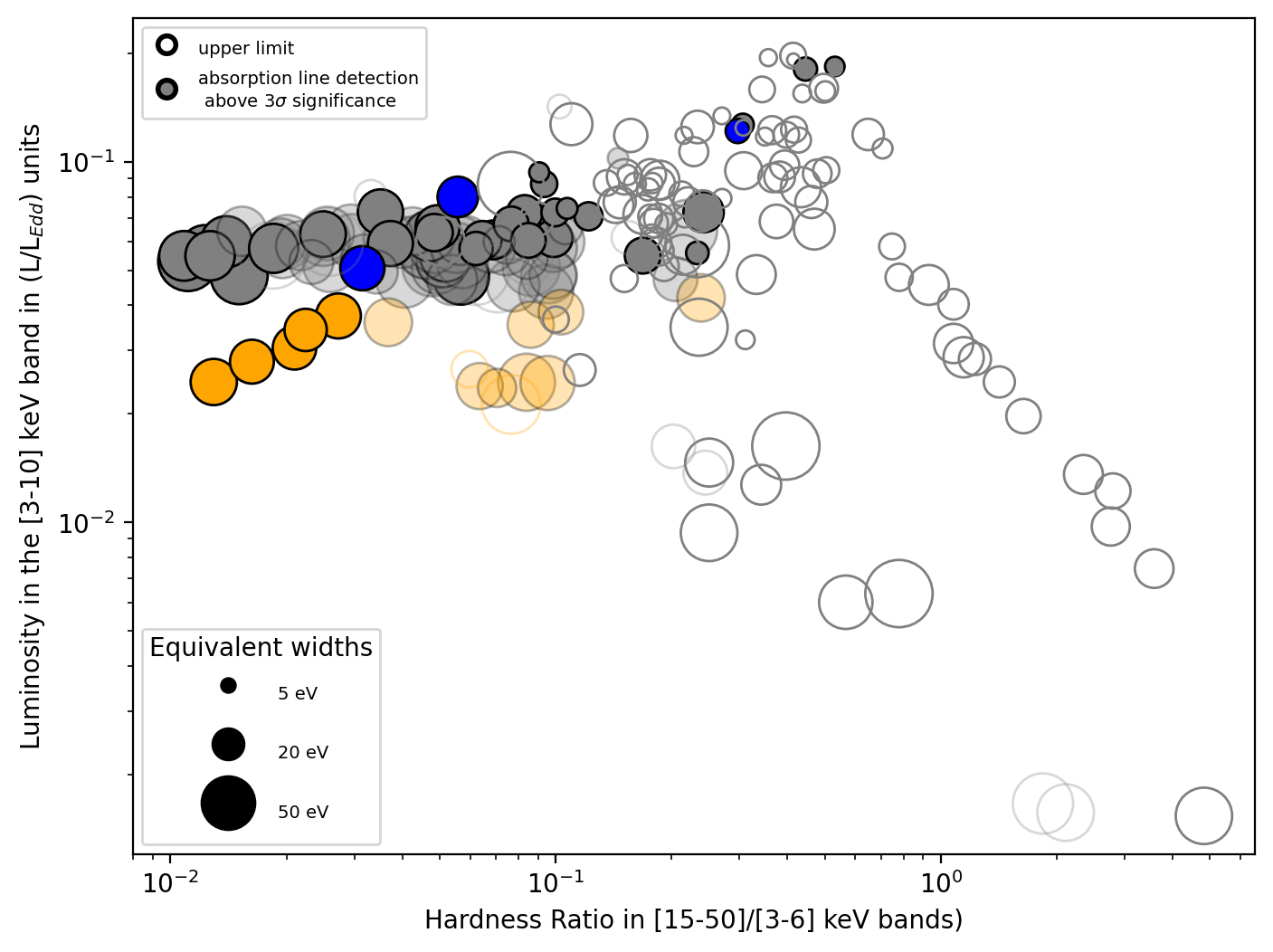}
\includegraphics[width=0.49\textwidth]{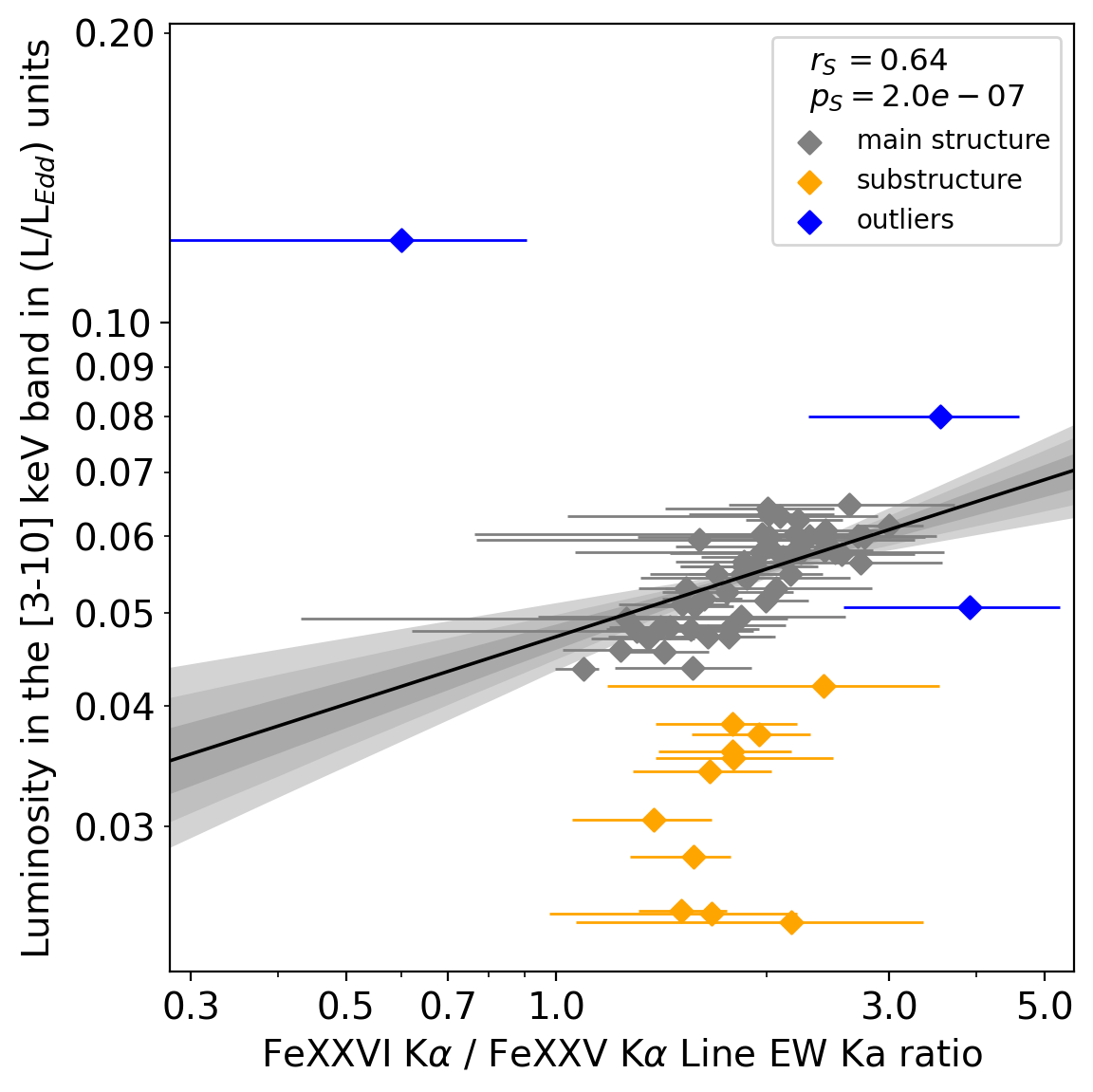}
\includegraphics[width=0.49\textwidth]{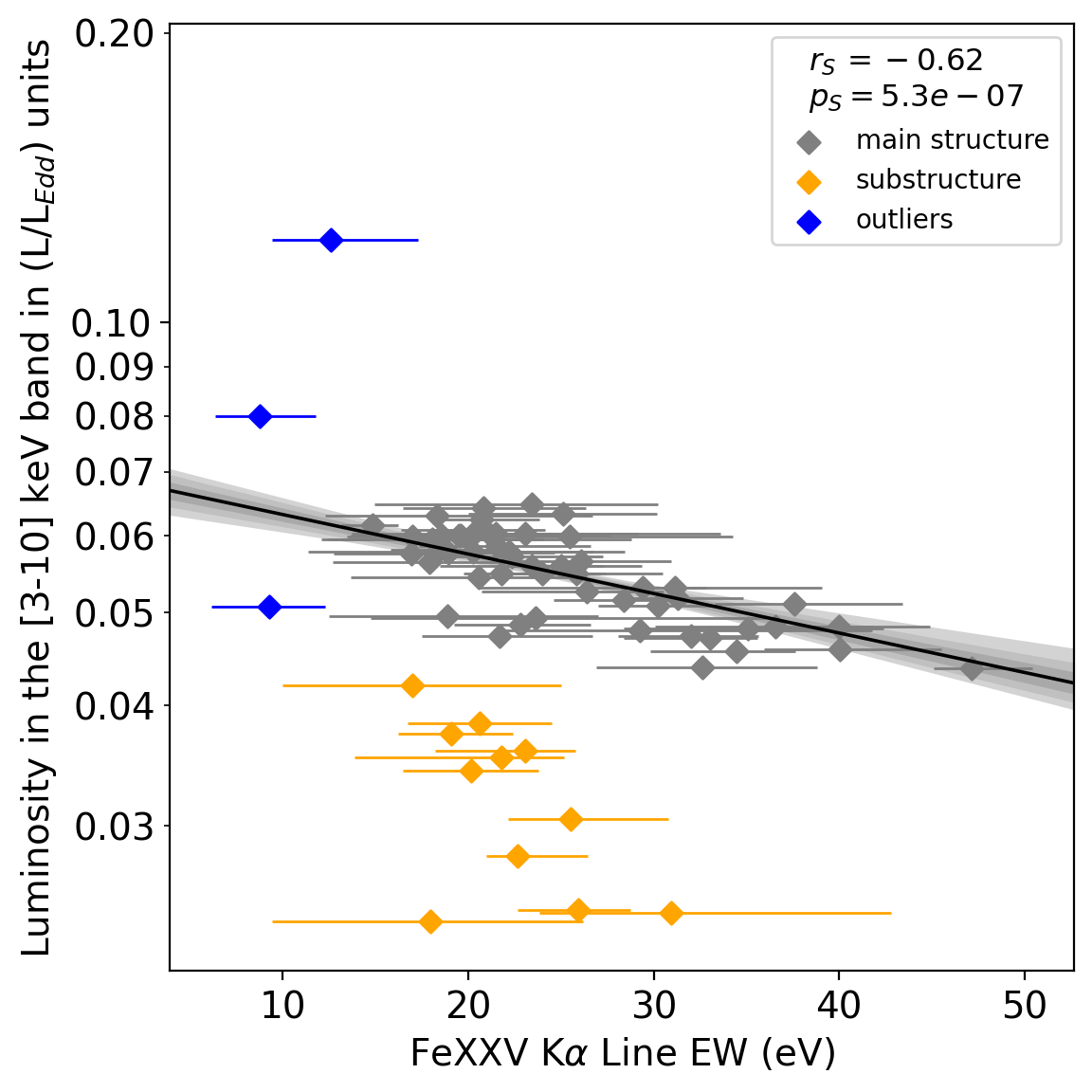}
\vspace{-1.5em}
\caption{\textbf{(Top)} Multi instrument "soft" (left) and  "hard" (right) HLDs of 4U~1630$-$47, colored according to the substructure and outliers defined in Sect.~\ref{sub:correlations}. \textbf{(Bottom)} Global scatter plots of the Fe K$\alpha$ EW ratio \textbf{(Left)} and \FeKav{} EW \textbf{(right)} against luminosity. In the left panel, EW ratio lower limits are plotted for observations with only \FeKavi{} detected, and the black line and gray region show the extent of the log-log linear correlation and its 1,2 and 3$\sigma$ confidence intervals. Both these and the spearman rank are computed only from observations of the main structure (in gray).}\label{fig:4U_plots_substructure}
\end{figure*}

The main difference from the global sample of P24 is the number of observations that clearly depart from this common structure, notably at the highest and lowest luminosities. A simple way to quantify their effect on the correlations is to compare the correlation coefficients with and without them: by removing all observations below $L/L_{Edd}\sim4\cdot10^{-2}$ and above $L/L_{Edd}\sim10^{-1}$, the spearman rank p-values drop from $p_{S}=2.3\cdot10^{-5}$ to $p_{S}=5\cdot10^{-7}$ for the K$\alpha$ EW ratio, and from $p_{S}=2.8\cdot10^{-4}$ to $p_{S}=7.5\cdot10^{-7}$ for \FeKav{}. In contrast, the p-value of \FeKavi{} only decreases by a factor $\sim2$. Due to the very high significance of this evolution, we discuss these regions in more detail in Sect.~\ref{sub:correlations}.

When focusing on correlations with characteristics of the high energy (>15 keV) continuum, the correlations of the line parameters with $L_{15-50}$ turn out mostly identical to correlations with $HR_{hard}$, albeit with more spread, and we thus focus on this parameter, whose correlation with the main EW parameters are shown in the lower panels of \ref{fig:4U_correl_observ}. We see a very strong correlation between $HR_{hard}$ and the \FeKavi{} EW, with $p_s=6.4\cdot10^{-10}$ and $r_s=-0.79$, in contrast to the lack of correlations of any EW line with $HR_{soft}$. 

The \FeKav{} EW and K$\alpha$ EW ratio still show hints of structured behavior, but the sample size remains too limited for any definitive conclusion. All of these aspects strongly contrast with the behavior of the low energy-energy part of the continuum, as in the 3-10keV band, the\FeKavi{} EW has significantly more spread than the other line parameters, which all correlate with the luminosity. 

Finally, we do not see a single significant correlation between the \texttt{nthcomp} $\Gamma$ and the line EW parameters. This may be due to the small number of observations with \nustar{} or \suzaku{}-PIN spectra, combined with the line detections largely favoring soft states, where the hard X-ray flux is too low for $\Gamma$ to be well constrained.

\begin{figure*}[h!]
\includegraphics[width=0.33\textwidth]{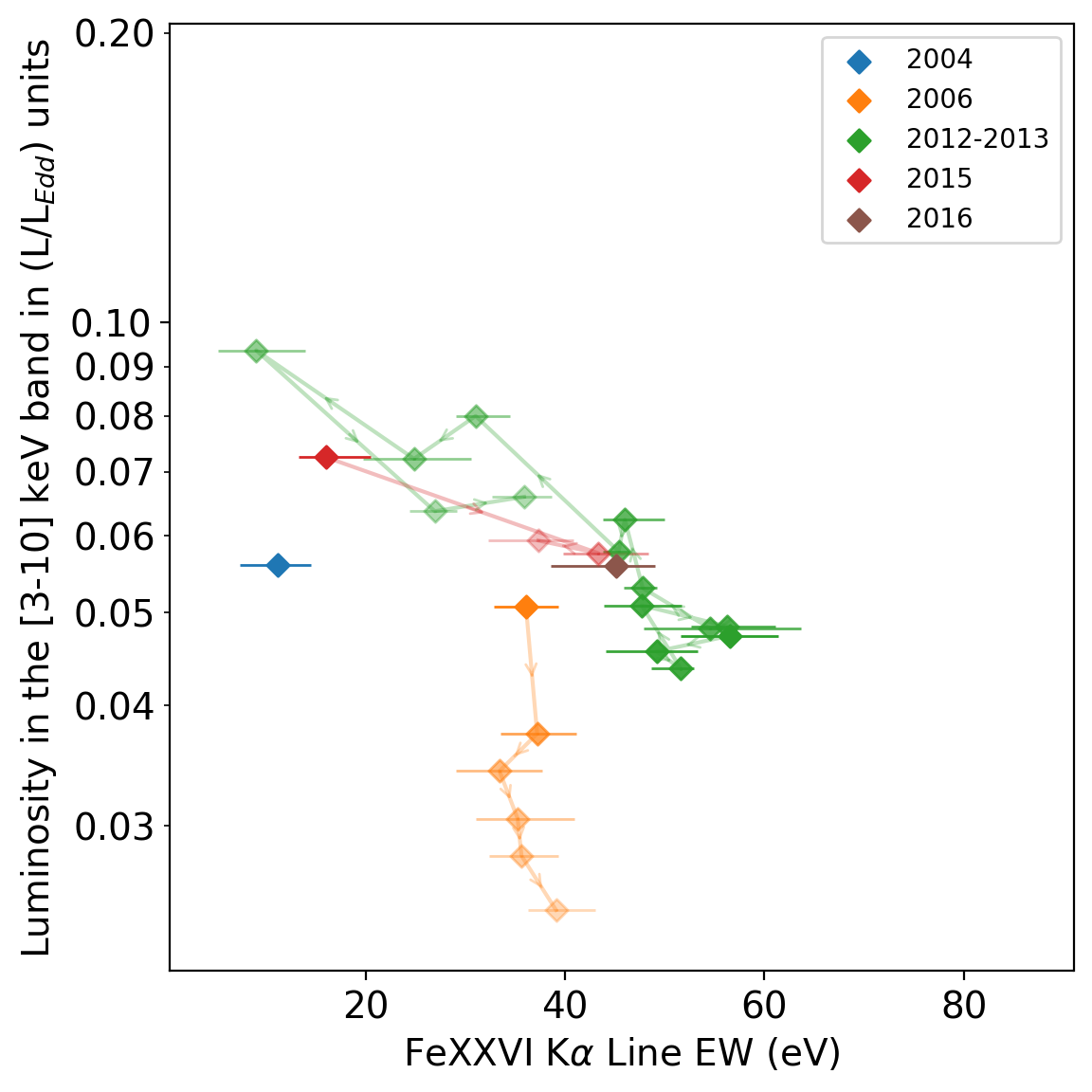}
\includegraphics[width=0.33\textwidth]{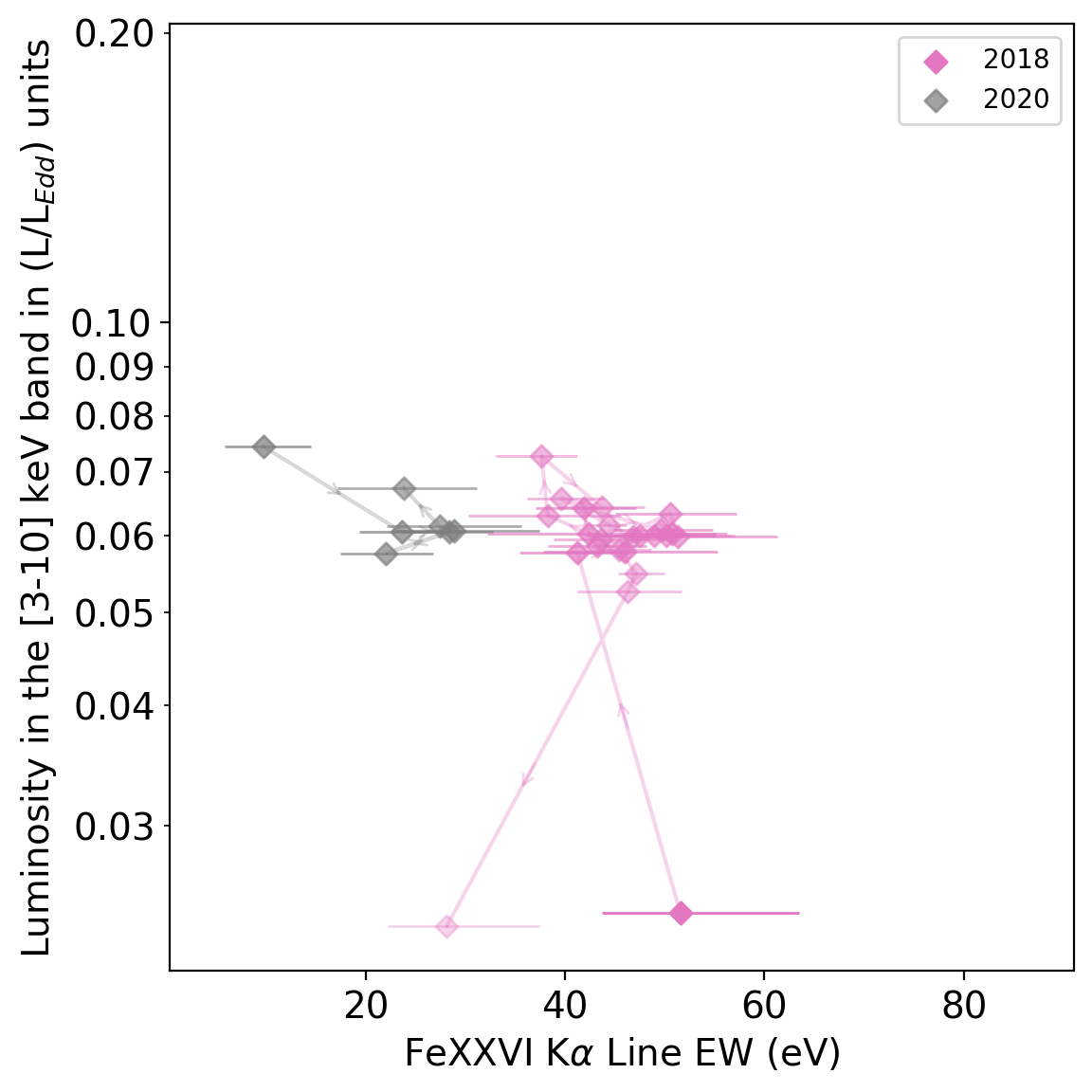}
\includegraphics[width=0.33\textwidth]{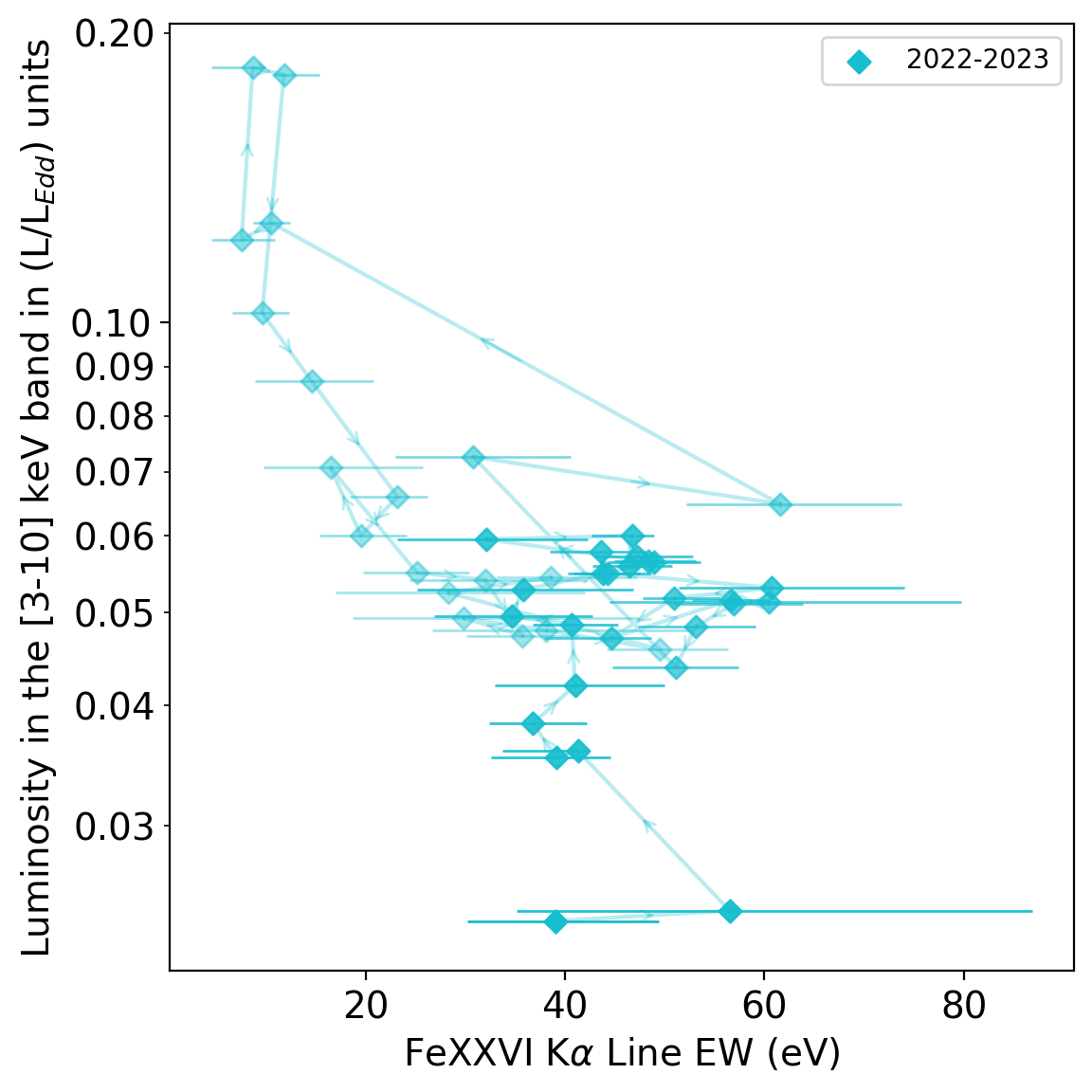}
\vspace{-2.5em}
\caption{Global scatter plots of the \FeKavi{} EW and [3-10] keV band luminosity, color-coded according to their outburst and split for legibility. Arrows and increasing transparency highlight the time evolution in each outburst.} \label{fig:4U_scatter_FeKa26_time}
\end{figure*}

We now focus on the upper left and upper middle panels of Fig.~\ref{fig:4U_correl_observ}, where a portion of the detections with low luminosity and low EWs seems to detach from the main structure, significantly lowering the Spearman rank and increasing the p-values (by more than two orders of magnitudes for the latter). In Fig.~\ref{fig:4U_plots_substructure}, we visually identify in yellow the observations of this ``substructure'', defined as below $L_[3-10]/L_{Edd}=4.2\cdot10^{-2}$. To compare the behavior of both groups, we compute log-scale linear regressions for the observations in the main structures (in gray), along with their confidence intervals. All observations in the substructure end up very distant from the linear regression and it's 3 $\sigma$ envelope.

We also highlight three distinct outliers in blue, which, substructure aside, are the most distant observations from the regressions, being the only ones further than 3$\sigma$ (all are beyond 7 $\sigma$) from the regression. The outlier at highest luminosity, above $L_{3-10}/L_{Edd}=10^{-1}$, is the \nustar{} SPL observation with a \FeKav{} detection (ObsID 80902312002), from March 2023. The detailed behavior of this observation will be addressed in a separate work. The second observation, at $L_[3-10]/L_{Edd}\sim 8\cdot10^{-2}$, is a \xmm{} observation from 2013 (ObsID 0670673001\_S003), at the beginning of a state transition \citep{DiazTrigo2014_4U1630-47_wind_2012-13XMM}. The third, at $L_[3-10]/L_{Edd}\sim 5\cdot10^{-2}$, is the first and brightest of the set of 2006 \suzaku{} observations (ObsID 400010010) during the source's declining soft state.
Interestingly, the behavior of this observation in the scatter plots (higher EW ratio and lower \FeKav{} EW than the main structure) matches the line behavior of the low-luminosity substructure.

In parallel, we also look at the disposition of these observations in the HLDs, which is plotted in with the same color coding in Fig.~\ref{fig:4U_plots_substructure}. In the soft HLD, the substructure forms the lower end of the main soft state diagonal, which is much less populated than the higher luminosity ranges above $\sim4\cdot10{-2}L_{Edd}$. In the hard HLD, the observations in the substructure with good constrains on the high-energy flux(represented by fully colored circles, c.f. Fig.\ref{fig:4U_HID_full}) are much more distinct from the main group, with both a lower $L_{3-10}$ for the same $HR_{hard}$ and a clear correlation between the two. Meanwhile, the other substructure observations (without \suzaku{}) lack good simultaneous BAT measurements, resulting in very high $HR_hard$ upper limits. However, neighboring days of monitoring strongly suggest that their $HR_{hard}$ values are much lower than presently displayed upper limits, in line with the more constrained observations.

Finally, we can look more in depth at the behavior of the upper right  panel of Fig.~\ref{fig:4U_correl_observ}, where the evolution of the \FeKavi{} EW with L$_{3-10}$ shows a much larger spread than the two other EW parameters, in the upper left and upper middle panels. We focus on the evolution of this correlation with time, splitting different outbursts in the panels of Fig.~\ref{fig:4U_scatter_FeKa26_time}. The different outbursts evolve very differently: only the 2012-2014 super outburst (left panel, green) and post 2022 part of the 2022-2024 super outburst outburst (right panel, transparent cyan) follow a clear, structured path, with both of these periods having individual Spearman p-values below $10^{-4}$. These two are the main drivers of the low p-value for the global correlation in Fig.~\ref{fig:4U_correl_observ}. 

In constrast, the normal outbursts appear much more spread out in individual clusters, but they are sampled much more scarcely, both in terms of duration and HID evolution. This is partly due to super outbursts remaining in the soft state for much longer periods than normal outbursts. Some differences between individual periods (such as 2018 and 2020 in the middle panel) remain clear, but could be imputable to changes in other parameters, such as $HR_{hard}$. A more complete sampling of normal outbursts in both soft and hard X-rays is thus necessary to disentangle potential wind evolution between outbursts from changes in SED of individual soft states.

\section{Wind evolution along the spectral states}\label{sec:wind_evol}

The evolution of the absorption lines observed can indicate intrinsic changes in the outflow properties, but could also be the consequence of changes in the SED. To distinguish the two, two main effects need to be considered: the stability of the plasma, and the evolution of its ionization. Here, we thus assume that both the SEDs and the properties of the outflow themselves do not vary significantly on the timescale of the thermal equilibrium of the plasma.

\begin{figure*}[h!]
\includegraphics[width=0.5\textwidth]{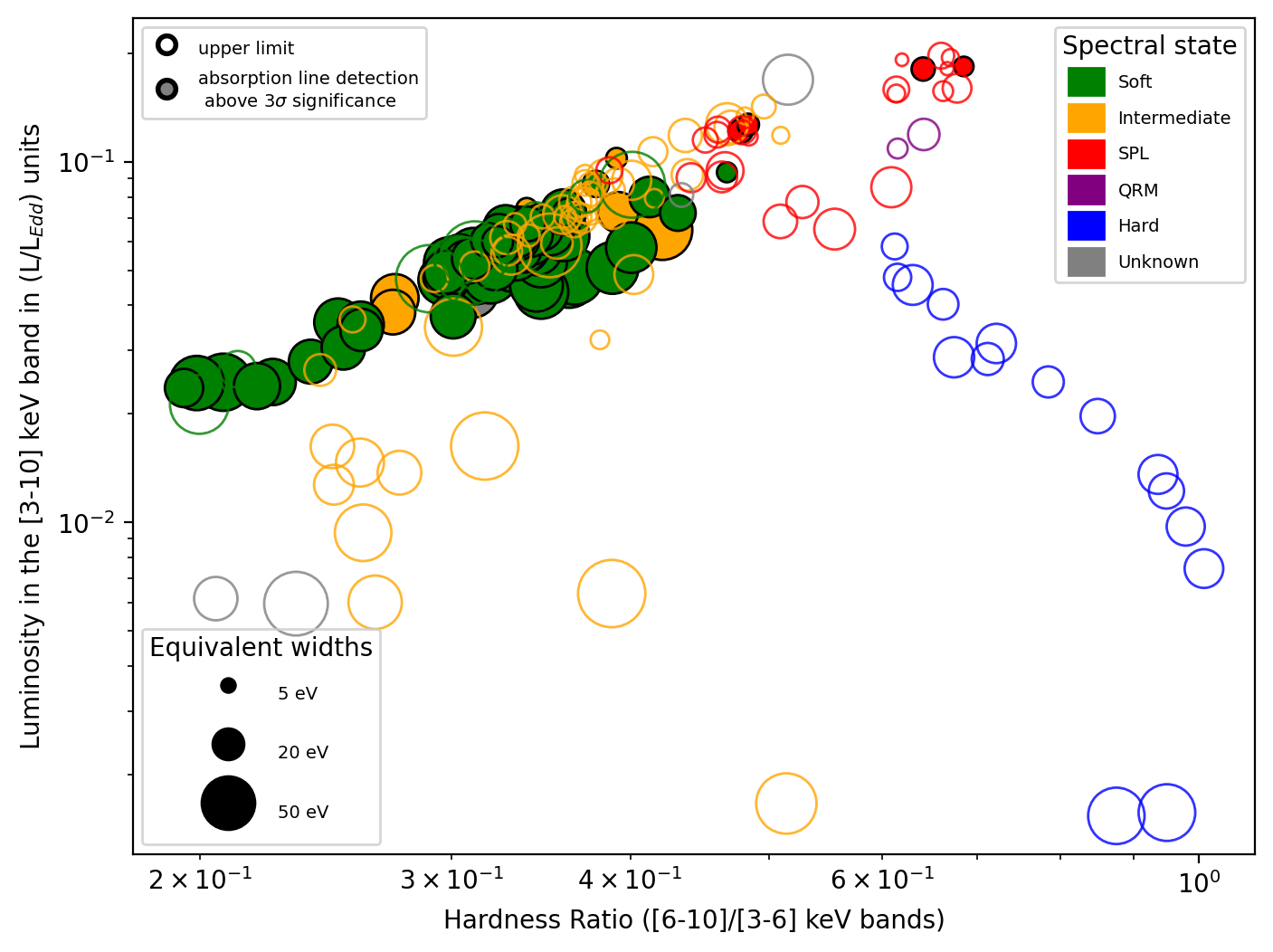}
\includegraphics[width=0.5\textwidth]{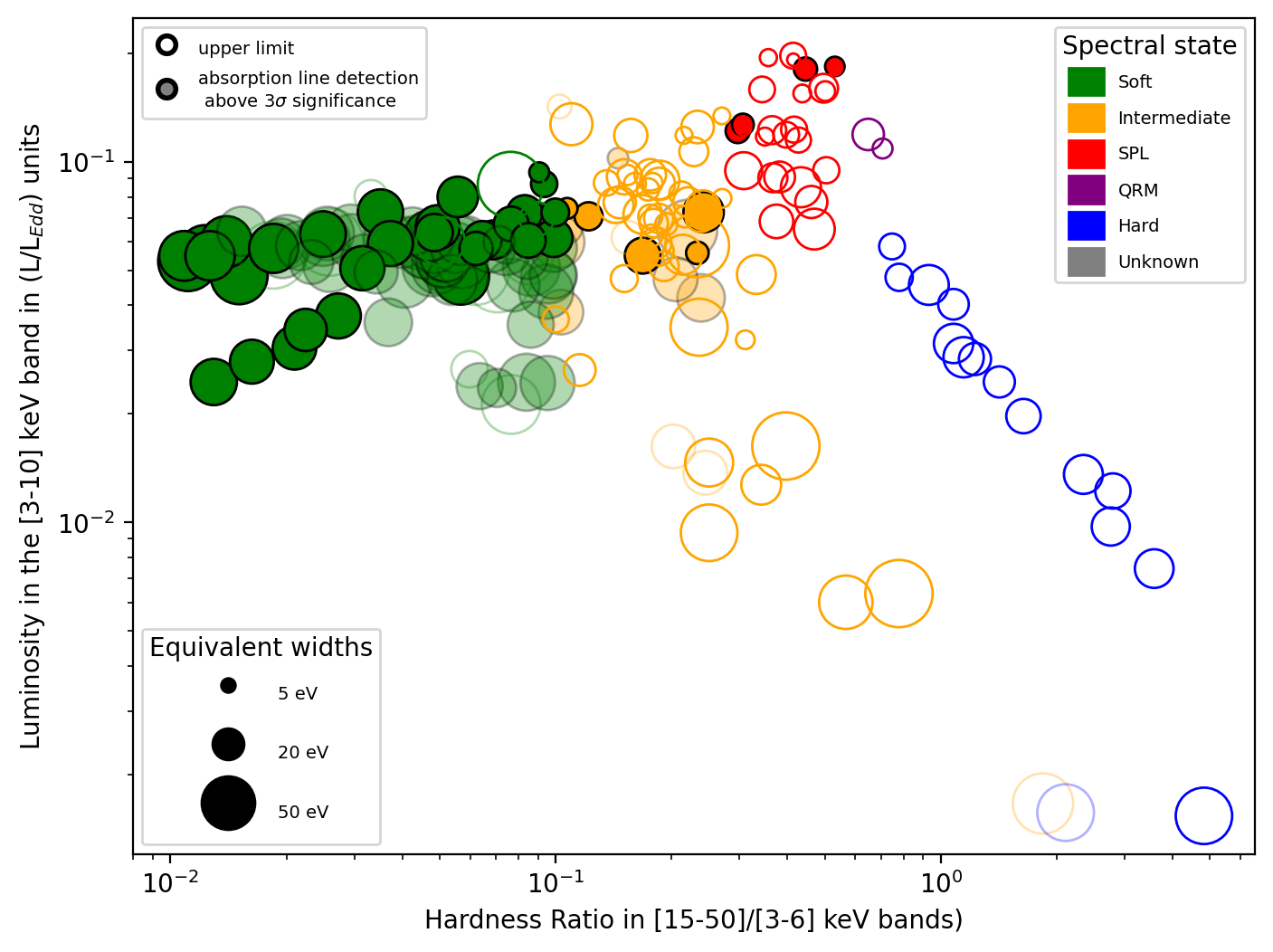}
\includegraphics[width=0.5\textwidth]{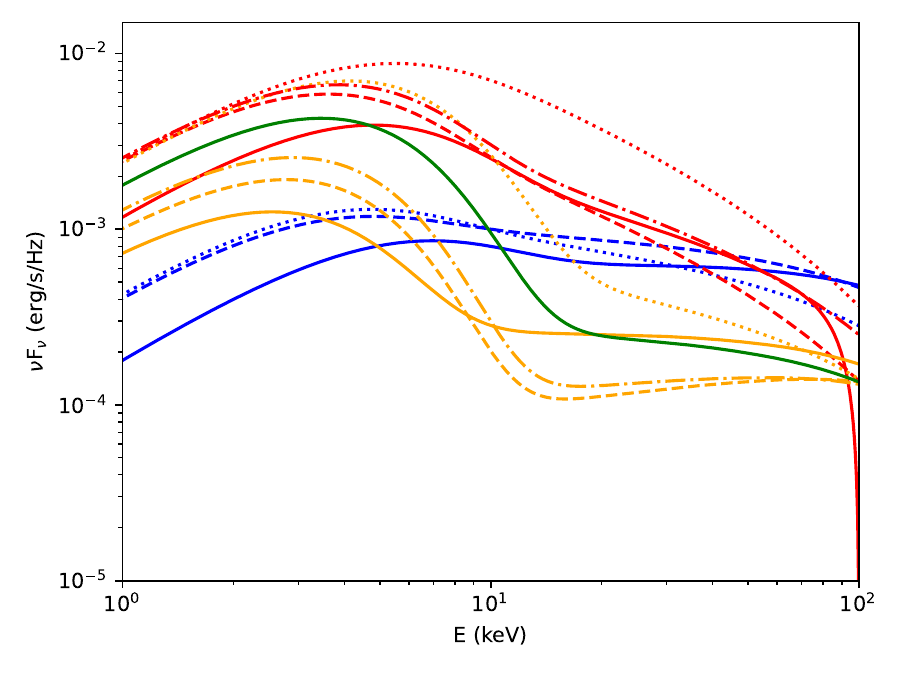}
\includegraphics[width=0.5\textwidth]{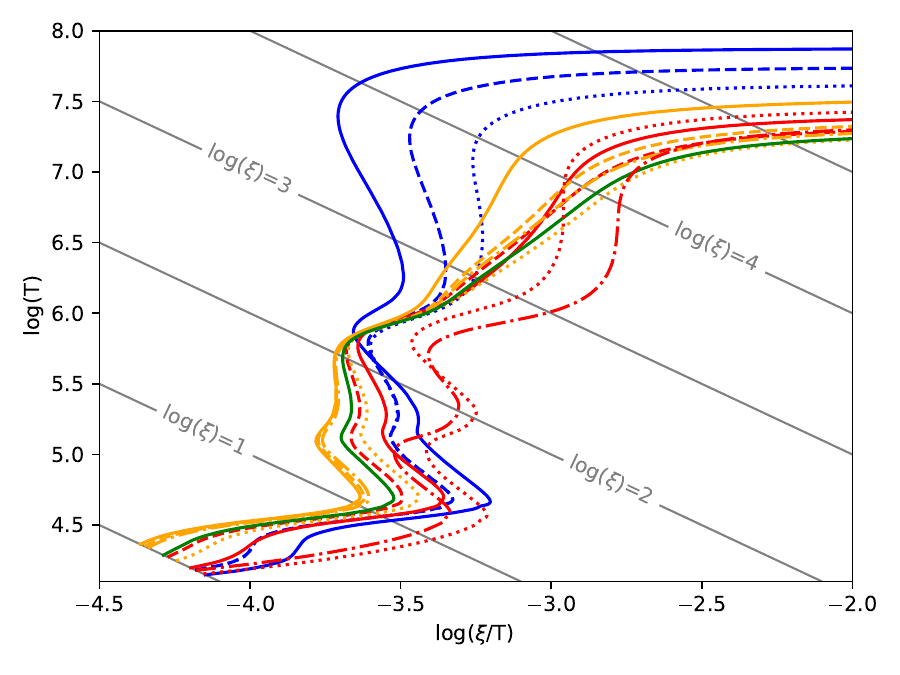}

\caption{Upper panels: Soft and Hard HLD of 4U1630$-$47, colored according to the spectral state. Lower panels: unabsorbed SEDs and corresponding stability curves for few SEDs in each states.} \label{fig:4U_HLDs_accretion_states_stability}
\end{figure*}

4U~1630$-$47 evolves between complex accretion states, with a combination of specific spectral and timing properties, which time-integrated spectra and the HLDs alone do not fully encompass. However, when computing the products of each instrument, we also computed individual lightcurves in the 1-3, 3-6 and 6-10 keV bands. This allowed us to verify that the sources did not significantly vary on the timescale of the observation, in the overwhelming majority of cases. After discarding the very few exceptions to this rule, which will be discussed in a subsequent work, and the observations in which no good hard X-ray constrain is available, we can regroup the behavior of the source in five basic states. We display the different states in both HLDs in the upper panels of Fig.~\ref{fig:4U_HLDs_accretion_states_stability}. Our distinction follows the following criteria:

-The "soft" state (green) is restricted to observations with $HR_{hard}<0.1$, which is the limit below which virtually all observations show significant absorption lines. It is characterized by a spectrum dominated by a thermal component and a very small (if at all) hard tail. 

-The "hard" state (blue) is restricted to observations with a negligible disk component, dominated by a hard component with $\Gamma\lesssim2.5$. 

-The "intermediate" (orange) and SPL (red) states correspond to observations where there is still an important disk component, but the spectrum shows a noticeable contribution at high energies. In the canonical definitions of \cite{Tomsick2005_4U1630-47_2002-2004_outburst_integral}, three different states (intermediate, flaring, and SPL) are distinguished by their disk contribution, flux, high energy $\Gamma$, and temporal properties. Here, because we lack the temporal information to perform proper distinctions, we regroup together the intermediate and flaring states (which are very similar except for their variability) and define the SPL states as the hardest ($HR_{hard}\gtrsim0.3$) and brightest ($L_{[3-10]}/L_{Edd}\gtrsim0.05$) of the intermediate states, which matches the \cite{Tomsick2005_4U1630-47_2002-2004_outburst_integral} spectral definition.

-Finally, we highlight in purple the observations in which the source exhibits Quasi-Regular Modulations (QRM, purple), very low-frequency QPOs (also called mHZ QPOs) with high RMS ($\sim 10-20\%$). These timing properties have previously been seen in a few other sources in hard states: in X-rays for BHLMXBs, GROJ1655-40 \citep{Remillard1999_GROJ1655-40_mHZ_QPO}, H1743-322 \citep{Altamirano2012_H1743-322_mHZ_QPO}, MAXI J1348-630 \citep{Wang2024_MAXIJ1348-630_mHZ_QPOs}, and previously in 4U~1630$-$47 in 1998 \citep{Trudolyubov2001_4U1630-47_1998_outburst_mHZ_QPO_QRM,Zhao2023_4U1630-47_QRM_mHZ_QPO_1998_RXTE} and 2021 \citep{Yang2022_4U1630-47_QRM_mHZ_QPO_2021_HXMT}. We distinguish these states from the so-called "heartbeat" states seen in IGRJ17091-3624 \citep{Altamirano2011_IRJ17091-3624_heartbeats,Wang2024_IGRJ17091-3624_wind_2022_exotic_V_exotic_X} and GRS1915+105 \citep{Belloni2000_GRS1915+105_states,Neilsen2011_GRS1915+105_wind_heartbeat,Zoghbi2016_GRS1915+105_rho_heartbeat_wind_Chandra-NuSTAR}, which are soft, with systematic wind detections, and even higher RMS (up to $\geq40\%$). On the other hand, QRM states are hard, powerlaw dominated, with no signs of absorption lines. In 4U~1630$-$47 , QRM states occupy a very well defined region of the HLD. We consider this as a "transition" state because the two previously reported detections in 4U~1630$-$47 signaled the transition from the canonical "hard" state into a SPL-like state. In several other sources, the QRMs are also seen just before or just after state transitions, e.g., H1743-322 \citep{Altamirano2012_H1743-322_mHZ_QPO} and MAXI J138-630 \citep{Wang2024_MAXIJ1348-630_mHZ_QPOs}. We note that in our sample, beside the 2021 QRM-state seen with \nicer{}, we discovered another observation with clear QRMs during the 2022-2024 outburst (obsid 6130010109), which this time occurred before the transition from SPL and soft states. A detailed analysis of this observation is out of scope of this paper, but we note that its spectral properties match very well that of the 2021 QRM period.

\begin{figure*}[h!]
\includegraphics[clip,trim=1cm 0.cm 6.5cm 0cm,width=0.5\textwidth]{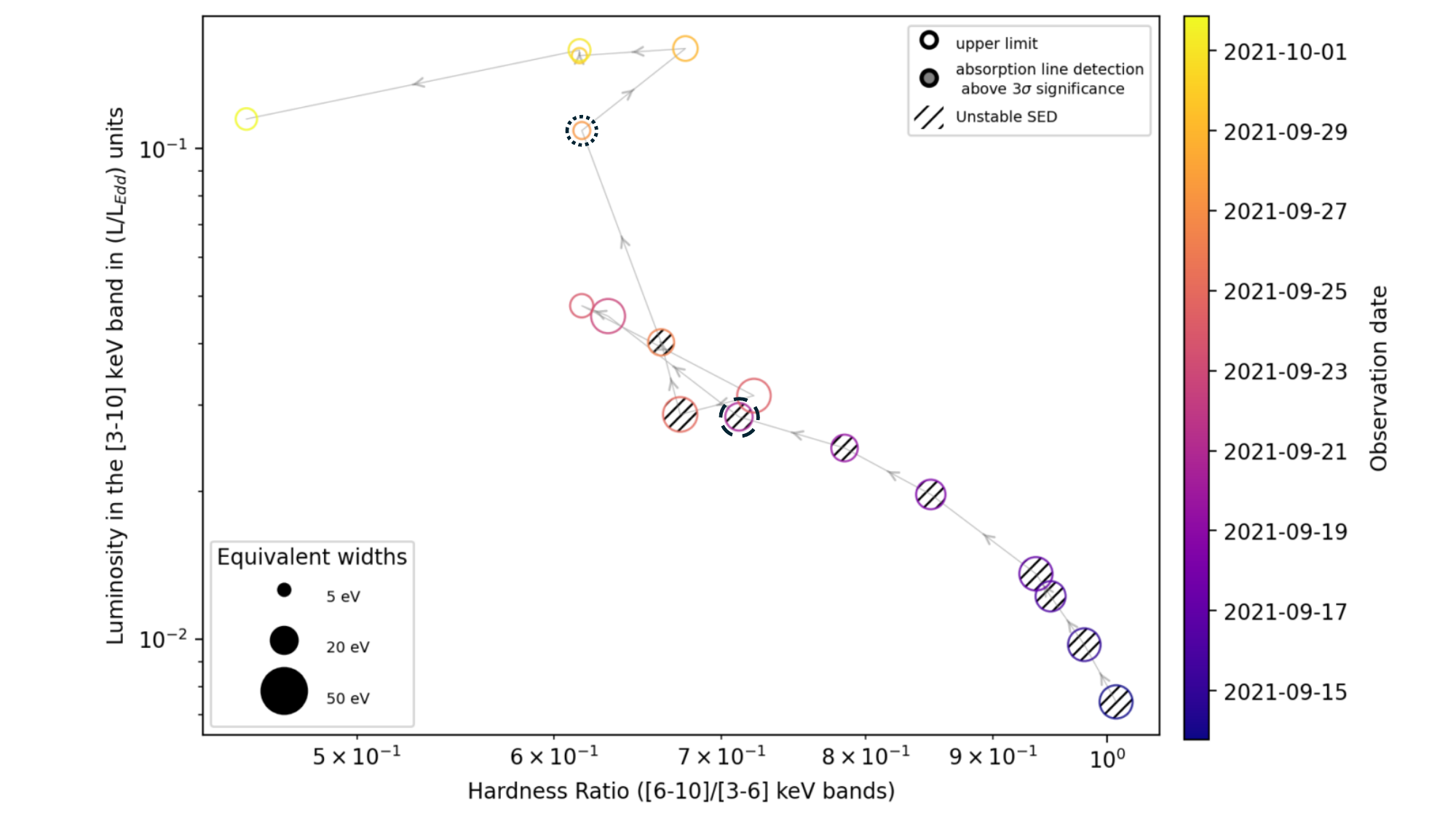}
\includegraphics[clip,trim=1cm 0.cm 6.5cm 0cm,width=0.5\textwidth]{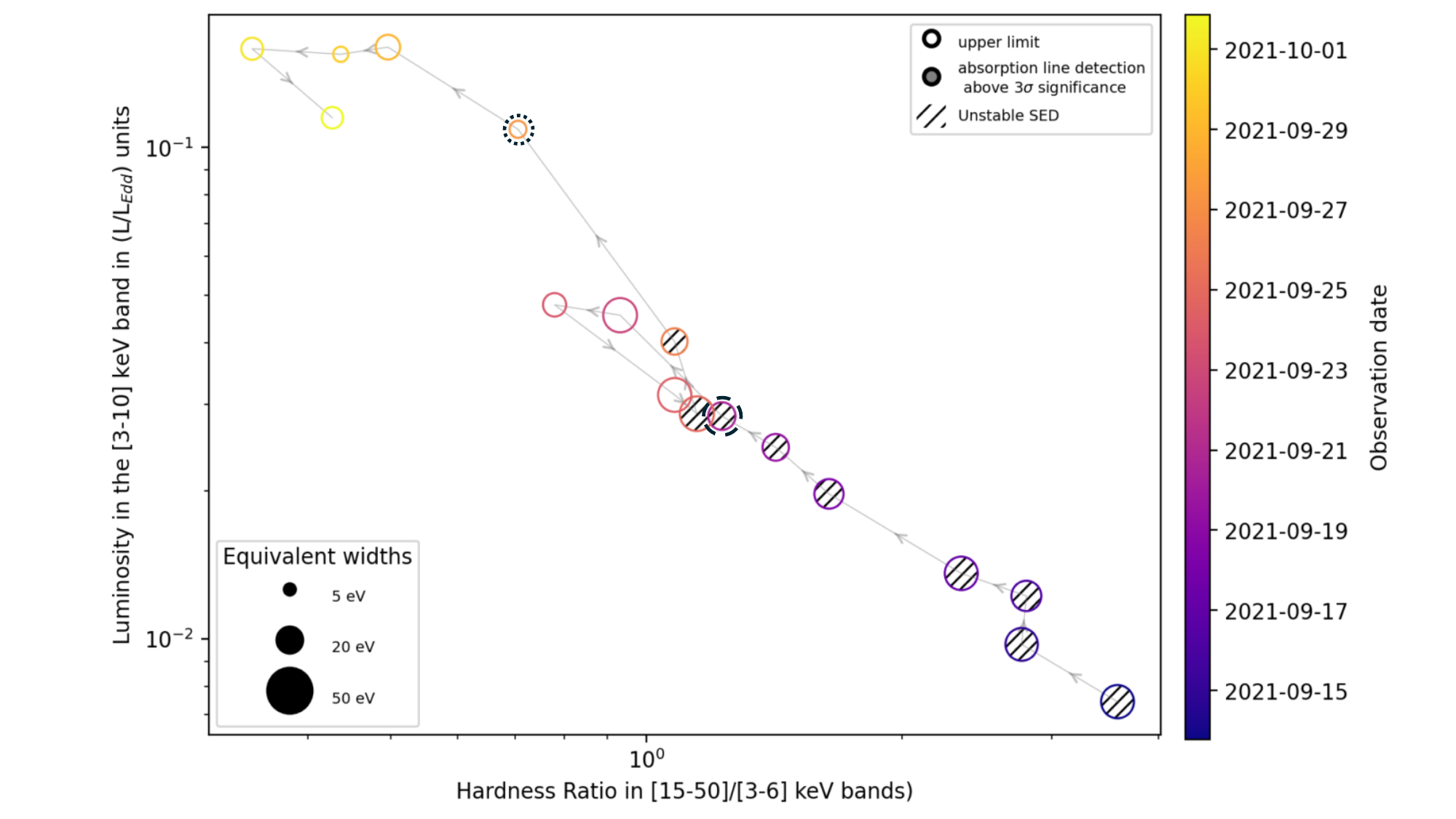}
\includegraphics[width=0.5\textwidth]{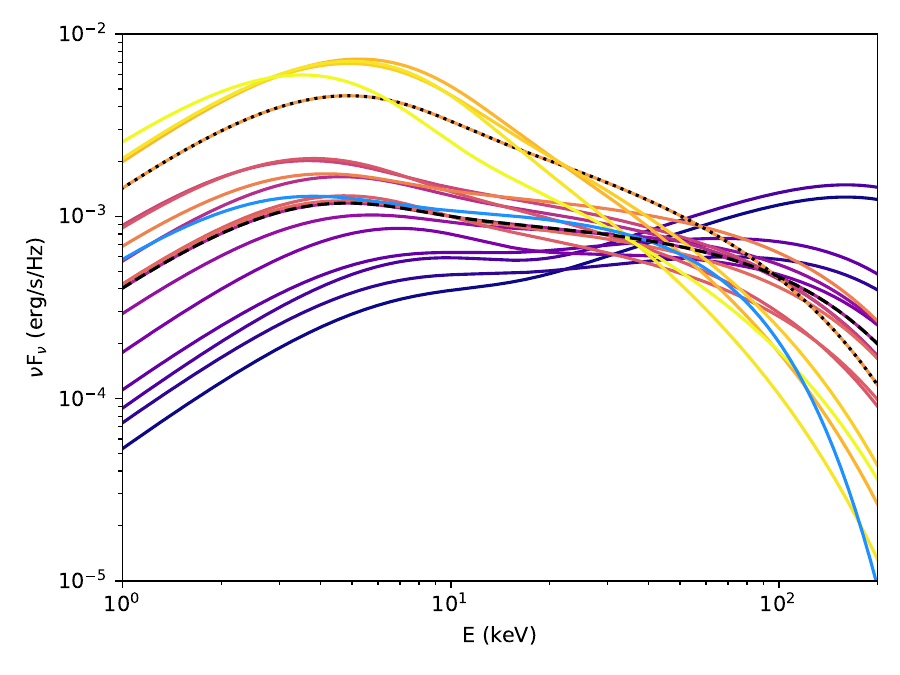}
\includegraphics[width=0.5\textwidth]{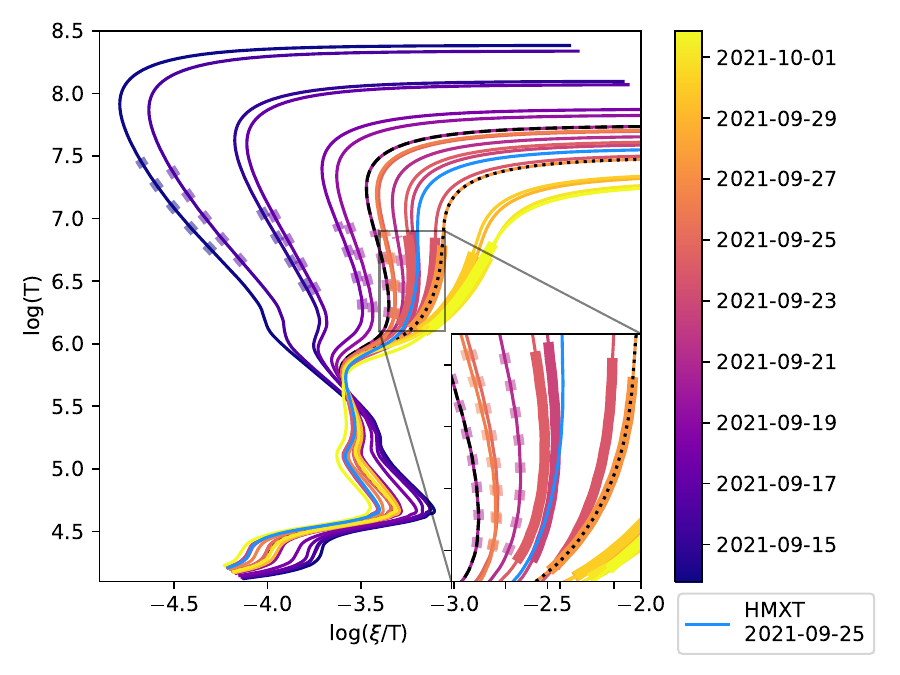}
\vspace{-2.5em}
\caption{Evolution of the beginning of the 2021 Outburst of 4U1630$-$47, seen in the Soft (upper left) and Hard (upper right) HLDs, corresponding unabsorbed SEDs (lower left), and stability curves (lower right). The regions bolded in the stability curves shows the 90\% ionization range of \FeKav{} and \FeKavi{}, and are dotted when mostly unstable. In all panels, the black overplotted lines highlight when the observation occured in the presence of a hard-state compact jet (dashed) or detached radio ejecta (dotted). The light blue SEDs and stability curves are derived from the fit parameters of \cite{Yang2022_4U1630-47_QRM_mHZ_QPO_2021_HXMT} in their last pre-transition observation.} \label{fig:4U_stability_2021}
\end{figure*}

We leave in gray in the soft HLD all observations without simultaneous hard X-ray coverage and thus no identification.  We also remove from both HLDs two hard state observations during the 2021 outburst, where the \nicer{} observation happened during very short flares. In both cases, the BAT daily average count rates are much lower than the peak seen in individual snapshots matching the \nicer{} observation periods, making the hard HLD artificially softer. We stress that this change is only important for the detailed stability analysis below, and does not affect the results of all the previous sections .\\

\subsection{Influence of plasma stability}\label{sub:stability}

\begin{figure*}[h!]
\includegraphics[width=0.49\textwidth]{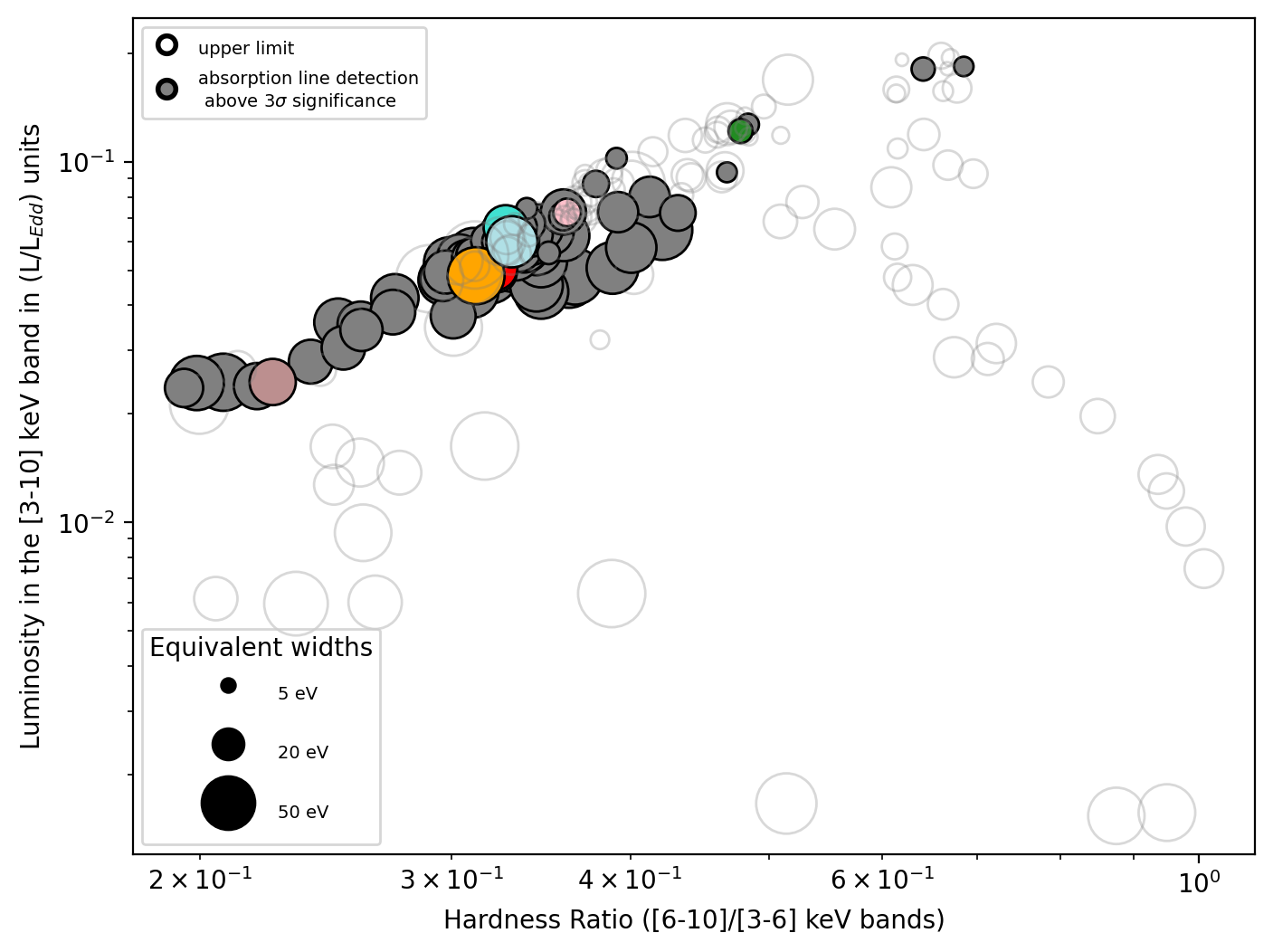}
\includegraphics[width=0.49\textwidth]{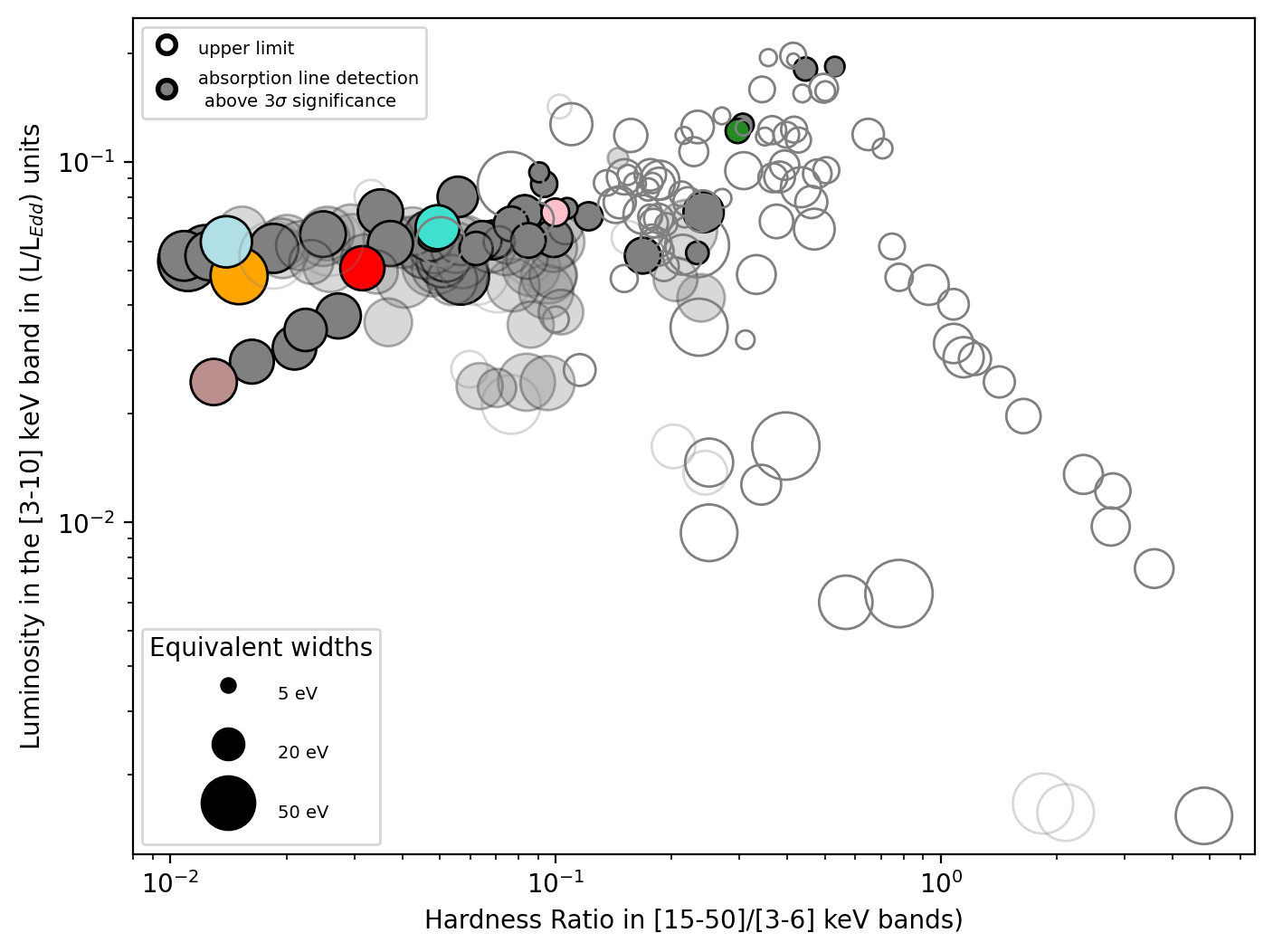}
\includegraphics[width=0.33\textwidth]{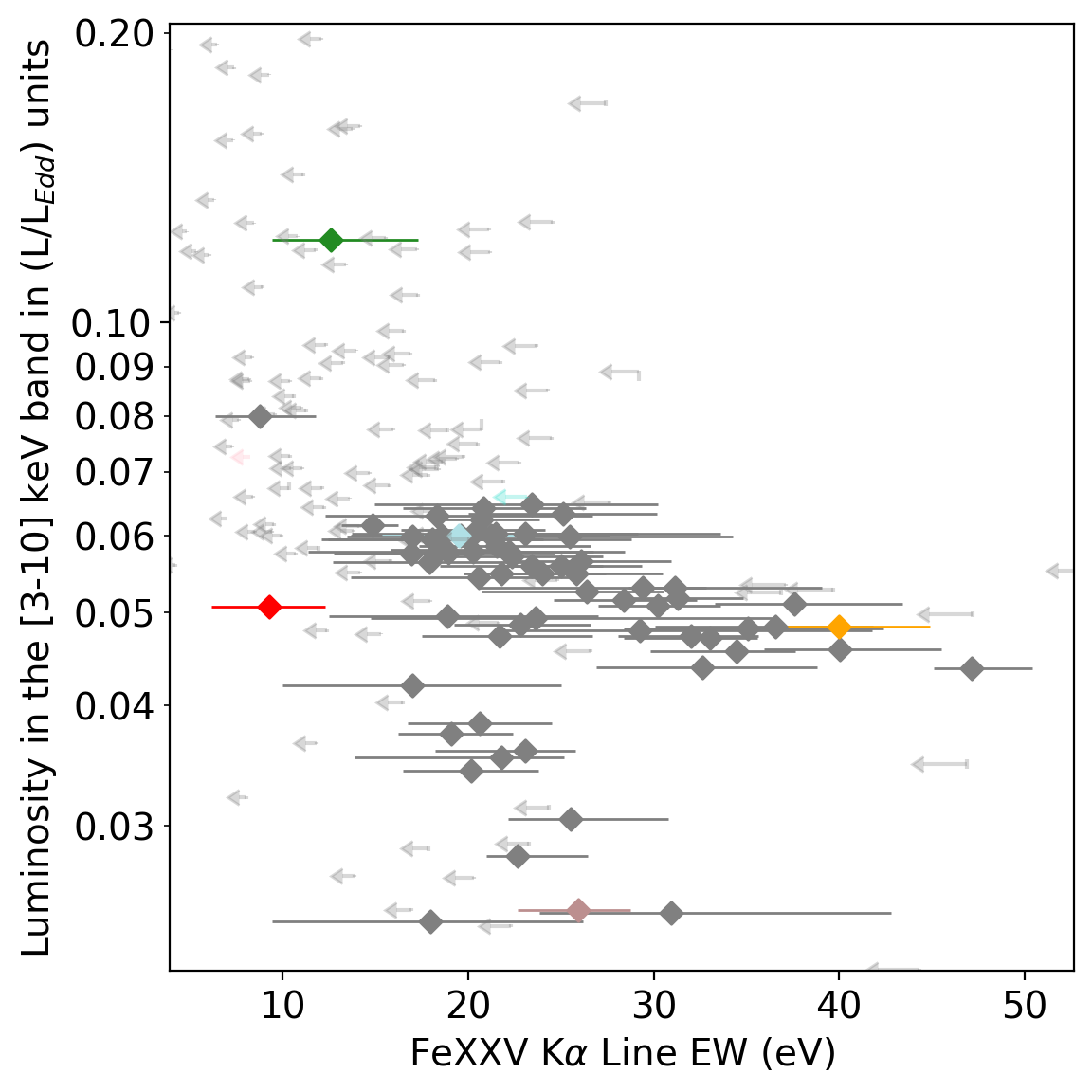}
\includegraphics[width=0.33\textwidth]{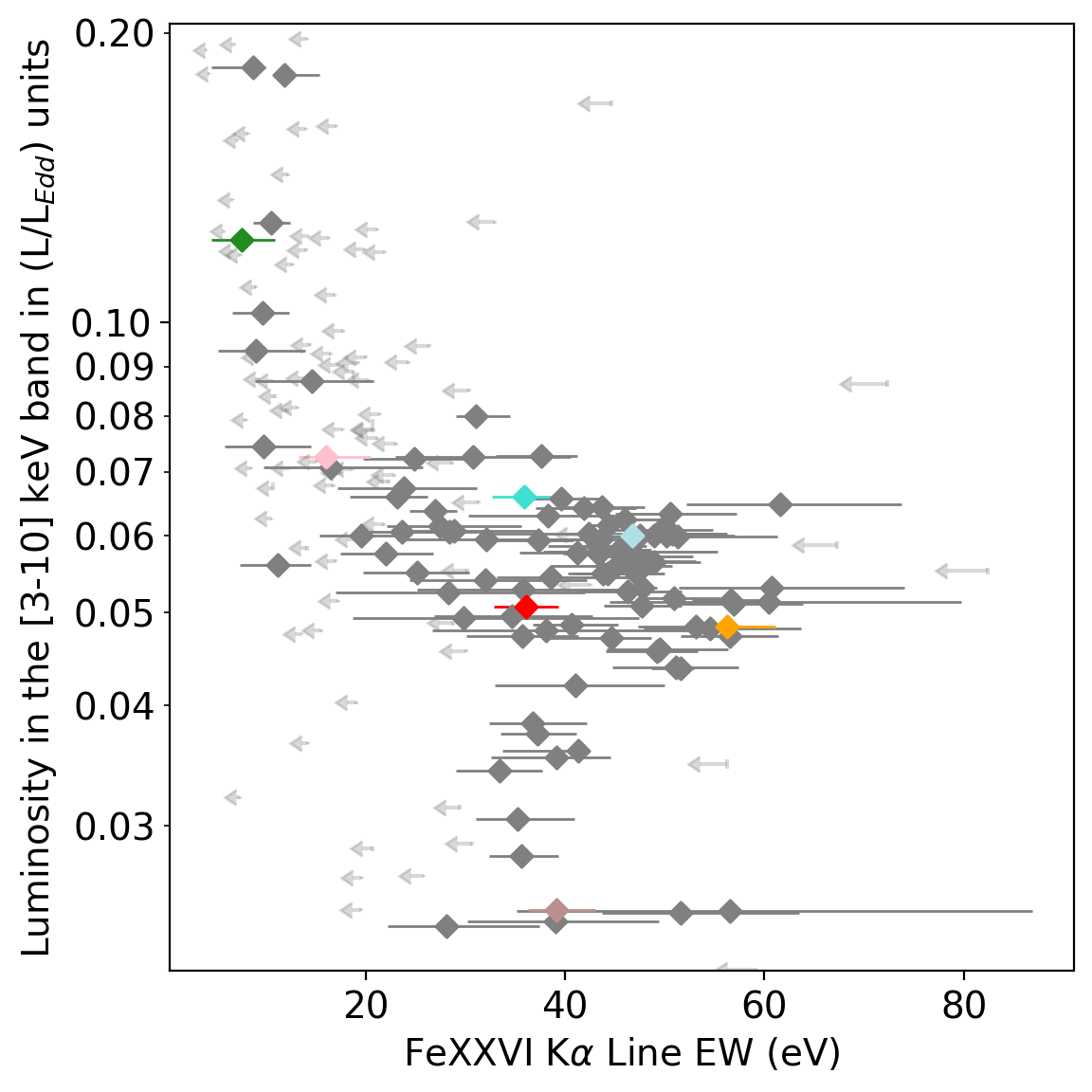}
\includegraphics[width=0.33\textwidth]{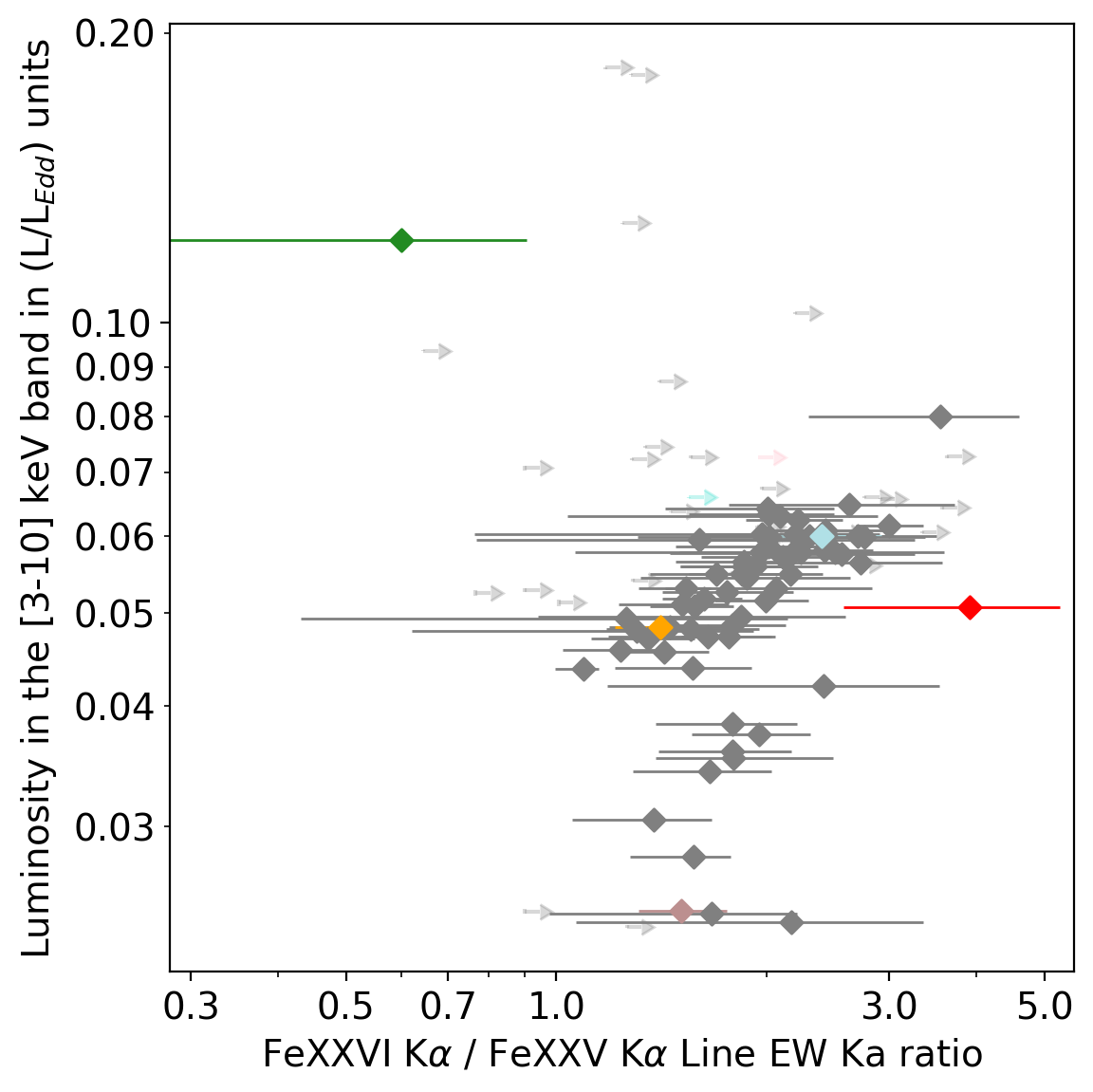}
\vspace{-2em}
\caption{\textbf{(Top)} soft \textbf{(left)} and hard \textbf{(right)} HLDs highlighting our selection of observations sampling different portions of the wind structure. \textbf{(Bottom)} Global scatter plots of the K$\alpha$ line EWs and EW ratios, including upper limits and color-coded according to the observations highlighted above.} \label{fig:4U_ionization_sampling}
\end{figure*}

The global hard X-ray coverage provided by \nustar{}, \suzaku{}, \swift{}/BAT and \integral{} allows, for the first time in an XRB, to compute the evolution of the plasma stability \citep{Krolik1981_stability} along the entire path of the source in the HID. We can then assess whether the correlation between the disappearance of the lines and the increase in HR$_{\mathrm{hard}}$ is the consequence of the favorable ionization states for Fe XXV and Fe XXVI becoming progressively unstable as we move to harder states and SEDs. 

We thus computed stability curves using CLOUDY 23.01 \citep{Chatzikos2023_CLOUDY}, from a range of observations in each of the previously defined accretion, prioritizing observations with good high energy coverage, from which we extracted broad band, unabsorbed SEDs in the 0.01-1000 keV band. We show the results in the lower panels of Fig.~\ref{fig:4U_HLDs_accretion_states_stability},  highlighting different ionization parameters. Previous studies with detailed photoionization modeling of the absorption line features seen in \chandra{} spectra have led to estimates of  log$\xi\sim3.5-4$ for the main absorption zone in soft and intermediate states \citep{Gatuzz2019_4U1630-47_wind_2012-13Chandra,Trueba2019_4U1630-47_wind_2012-13Chandra}. In this ionization range, all observations in the soft, intermediate, SPL and QRM states are completely thermally stable, in accordance with previous results for this source \citep{Gatuzz2019_4U1630-47_wind_2012-13Chandra}. Stability effects thus cannot explain the decrease in absorption line EW between soft and intermediate/SPL states.

Unexpectedly, the hard states also retain a stable region around log$\xi\sim$2.5-3., which corresponds to a non-negligible ionic fraction of (notably) Fe XXV for such SEDs (see e.g. \citealt{Chakravorty2013_thermal_stability,Petrucci2021_outburst_wind_stability}). To the best of our knowledge, this is the first time that a stable region at this log$\xi$ range is found in a BHLMXB hard state. Its existence stems from the unexpectedly steep comptonized component of 4U-1630$-$47: in most LMXBs, both NS and BHs, hard states SEDs are dominated by a $\Gamma\sim1.5-2$ high-energy component, and are completely thermally unstable (see e.g. \citep{Bianchi2017_stability_NS,Petrucci2021_outburst_wind_stability}). Here, on the other hand, the "softer" hard states reach up to $\Gamma\gtrsim2.3$ before the transitions to the QRM state. We stress that these high photon indexes are not a specificity of the recent outbursts, whose hard state was sampled by \nicer{}, as our results remain in line with other hard state measurements obtained during previous outbursts of this source \citep{Seifina2014_4U1630-47_properties}.

This thermally stable $\xi$ region has strong implications on the detectability of wind signatures via highly ionized absorption lines in hard states, which could give new constrains on the disk-wind geometry and allow for direct comparisons with the cold winds seen in OIR. We thus investigate the transition from unstable to stable SEDs in more details. For that, we take advantage of the detailed \nicer{} coverage of the hard state rise at the beginning of the 2021 outburst, and compute the stability curves of the first 18 observations, sampling from very unstable hard states to the SPL, well after the SEDs have become stable. To maximize our constrain on the broad band SEDs, here, we directly fit the \nicer{} spectra together with daily BAT survey spectra derived that we derive using the automatized pipeline of the BaTAnalysis package \citep{Parsotan2023_BatAnalysis}, for a total energy coverage of $0.3-195$ keV. In order to account for potential BAT calibration uncertainties, we allow for a variation of $30\%$ in the constant factor of the BAT datagroup during the fit, and keep the \texttt{thcomp} cutoff frozen at 100 keV since it remains unconstrained even with BAT spectra.

We plot the HLDs, SEDs and stability curves of these observations in Fig.~\ref{fig:4U_stability_2021}, highlighting stable SEDs in the HLDs for the ionization range dominated by Fe XXV and Fe XXVI. The first half of the hard state observations exhibit more standard (although relatively steep compared to other binaries) $\Gamma$ values of $\sim2$, and are all largely unstable. The latter observations before the QRM, with $\Gamma\sim2.2-2.4$, are either very close to stability or barely stable, and after the QRM state, all observations in the SPL are much softer ($\Gamma\gtrsim2.7$) and completely stable down to much lower ionization parameters. This would indicate that there is a short period at the very end of the hard states (below HR$_{\mathrm{hard}}\sim0.1$, as seen in the HLDs) where the SEDs would not prevent the apparition of wind signatures from highly ionized iron.

\begin{figure*}[h!]
\begin{center}
\includegraphics[clip,trim=2cm 2cm 6cm 5cm,width=0.36\textwidth]{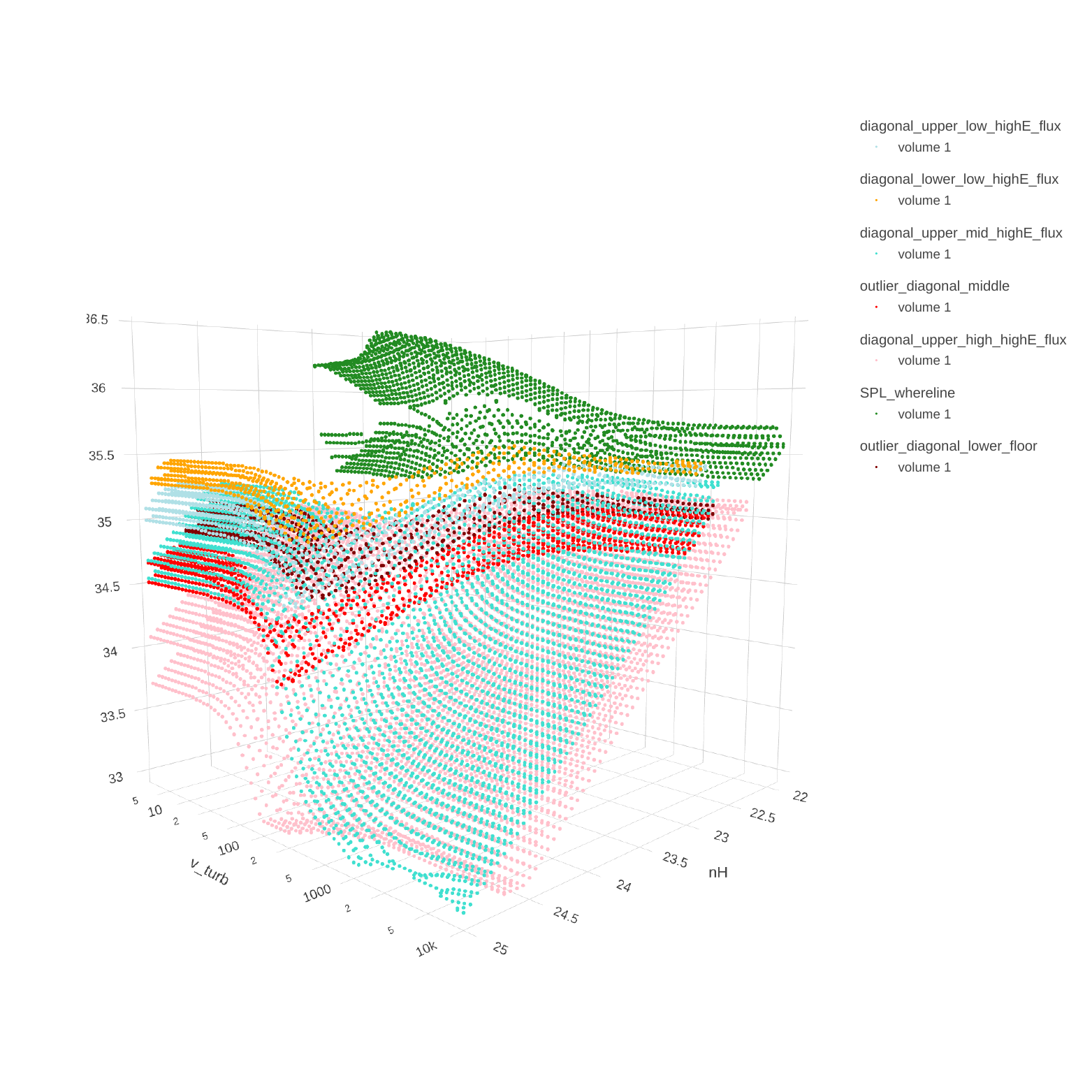}
\includegraphics[clip,trim=2cm 2cm 6cm 5cm,width=0.31\textwidth]{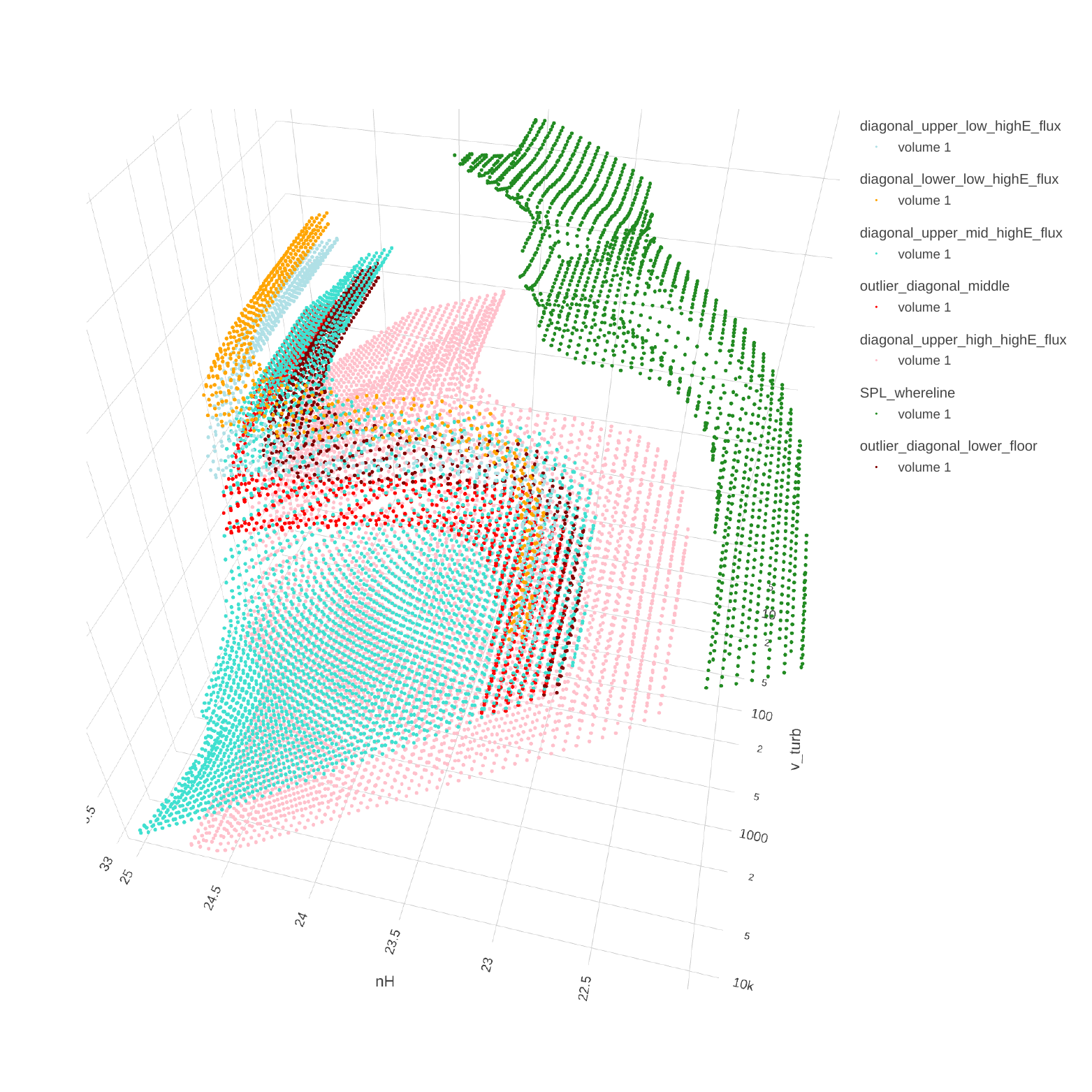}
\includegraphics[clip,trim=2cm 2.3cm 6cm 5cm,width=0.31\textwidth]{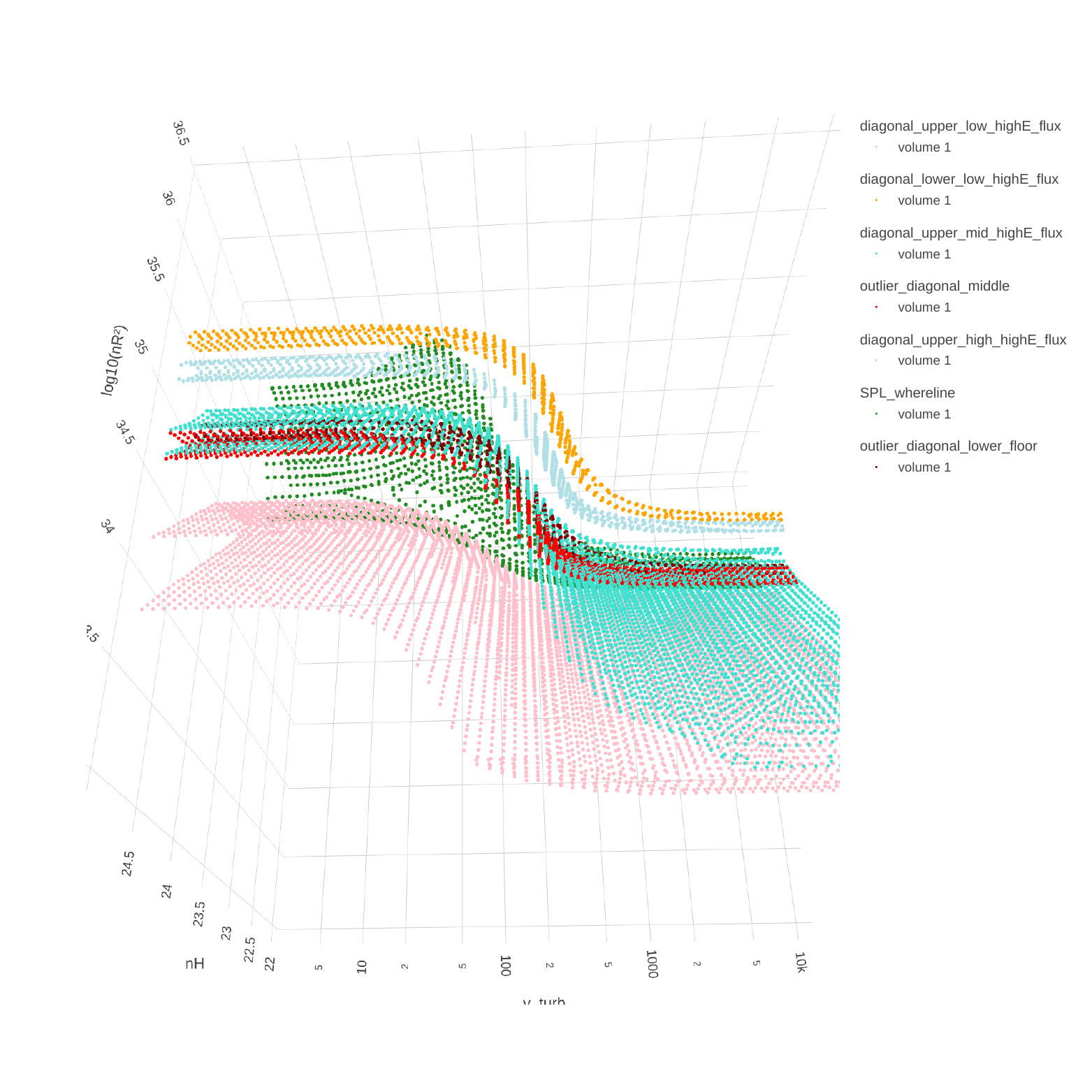}
\end{center}
\vspace{-1em}
\caption{Wind parameter space available for each SED highlighted in Fig.~\ref{fig:4U_ionization_sampling}, and colored accordingly. For the sake of clarity, we only show the outside envelope, namely the highest and lowest possible nR$^{2}$ available for each combination of NH and $v_{turb}$. For better visualization, different panels showcase different angles of the same figure.} \label{fig:4U_wind_parameter_plots_3D}
\end{figure*}

There are nevertheless some caveats in our SED derivation, due to the large uncertainties of BAT measurements at high energies. In comparison, when analyzing \hxmt{} observations during the same period (just before the transition to the QRM state), \cite{Yang2022_4U1630-47_QRM_mHZ_QPO_2021_HXMT} report high energy rollovers down to $kT_e\sim20$ keV. A lower energy cutoff would tend to reduce the amount of hard photons, and thus stabilize the plasma, but this effect may be negligible as the physical cut-off happens at E $\gtrsim2-3 kT_{e}$ (see \citealt{Petrucci2001_cutoff,Miyakawa2008_cutoff,Zdziarski2021_thcomp}). As a sanity check, we thus compute the stability curve of the average "pre-QRM" epoch of \citep{Yang2022_4U1630-47_QRM_mHZ_QPO_2021_HXMT}, using their reported model parameters to reproduce the SED. The results, displayed in cyan in the lower panels of Fig.~\ref{fig:4U_stability_2021}, are in very good agreement with our own SEDs for the last hard state periods, cutoff aside, and most notably result in almost identical (stable) stability curve in the range of ionization parameter tied to Fe xxv and Fe xxvi.\\

We note that \cite{Yang2022_4U1630-47_QRM_mHZ_QPO_2021_HXMT} classify the "pre-QRM" observations as "intermediate states", notably from their position in the HLD. This raises an important point of whether such "soft" SEDs, although dominated by a comptonized component, should be interpreted as the signature of canonically "hard" accretion states. First, the timing properties they report for this period, such as the type-C QPO quality factor and important continuum RMS ($>20\%$), are much more in line with the hard state than with the SPL state according to their definition in \citep{Tomsick2005_4U1630-47_2002-2004_outburst_integral}. Secondly, weekly radio monitoring was performed during this period (Zhang et al. in prep), including observations on 2021-09-20, which corresponds to our seventh observation, before the state transition, and on the 27-09, during the QRM state. We highlight the corresponding X-ray observations in black in Fig.~\ref{fig:4U_stability_2021}. The strong evolution in radio spectral index between the radio detection signals a change from a compact jet on the 2021-09-20  (black dashes), to a radio ejecta on the 2021-09-27 (black dots). This provides a very strong argument to consider all of the observations before 2021-09-27 (which all show SEDs and timing properties similar to the 2021-09-20 observation) as coming from a canonically "hard" accretion-ejection structure, and for the QRM state to be the consequence of a significant change in the accretion flow.

\subsection{Photoionization modeling}\label{sub:photo_mod}

The second element affecting the ionization structure is the influence of the SED on the ionization range with high ionic fractions of Fe XXV and Fe XXVI. 
Here, we adopt a qualitative approach, whose aim is to detect \textit{any} change in wind parameters.

\begin{figure*}[h!]
\begin{center}
\includegraphics[clip,trim=0cm 3cm 6cm 5cm,width=0.34\textwidth]{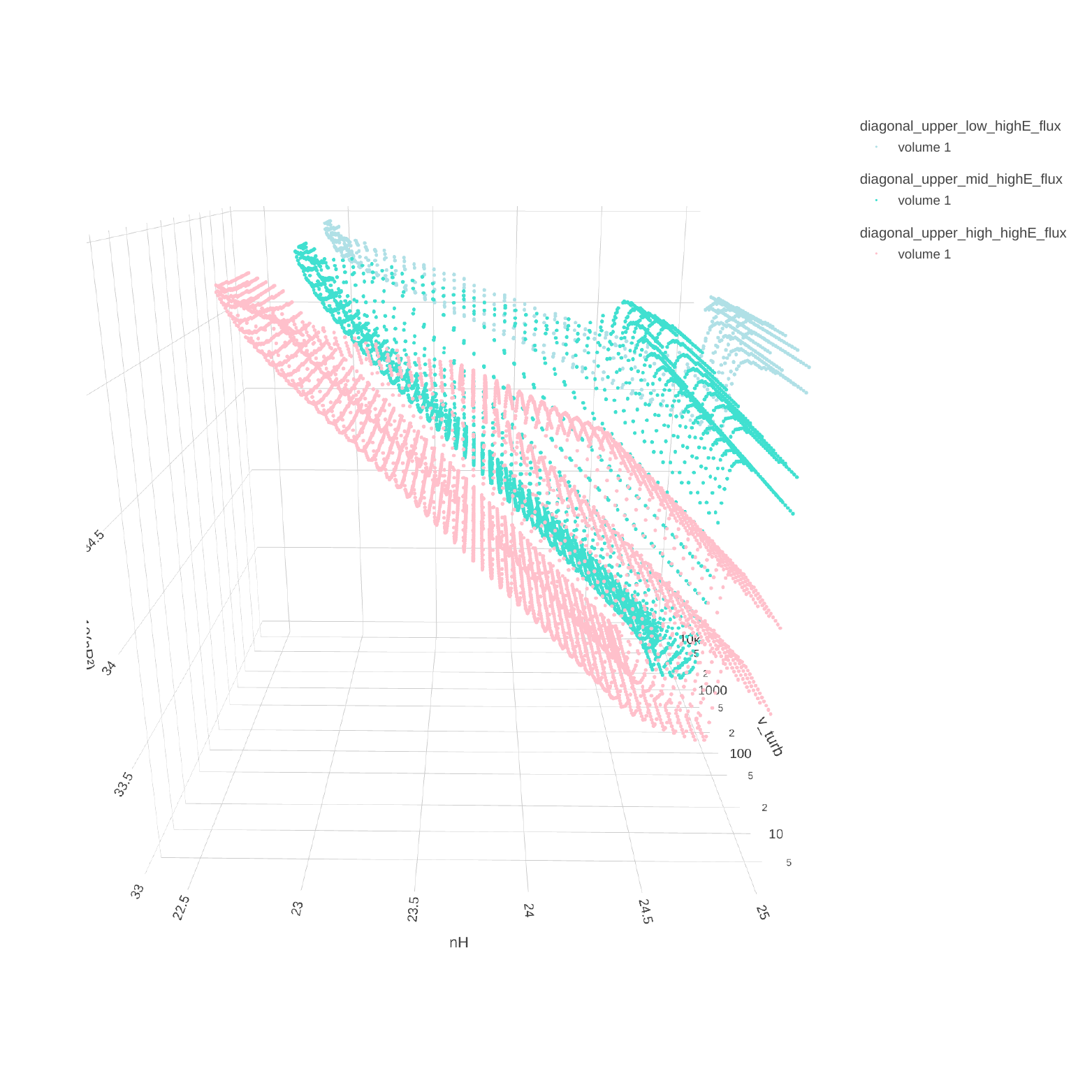}
\includegraphics[clip,trim=2cm 3cm 5.6cm 5cm,width=0.34\textwidth]{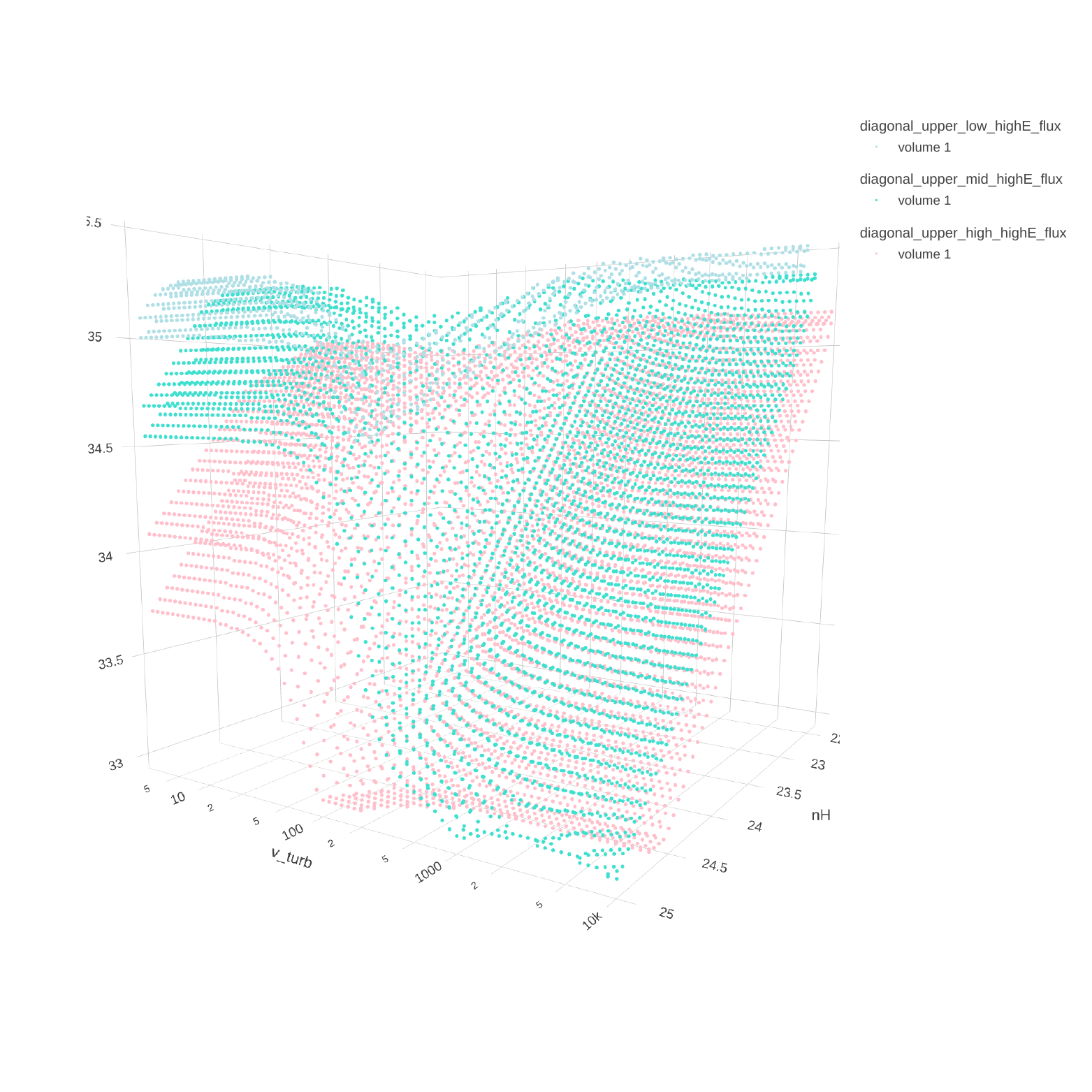}
\includegraphics[clip,trim=1cm 2cm 6cm 5cm,width=0.31\textwidth]{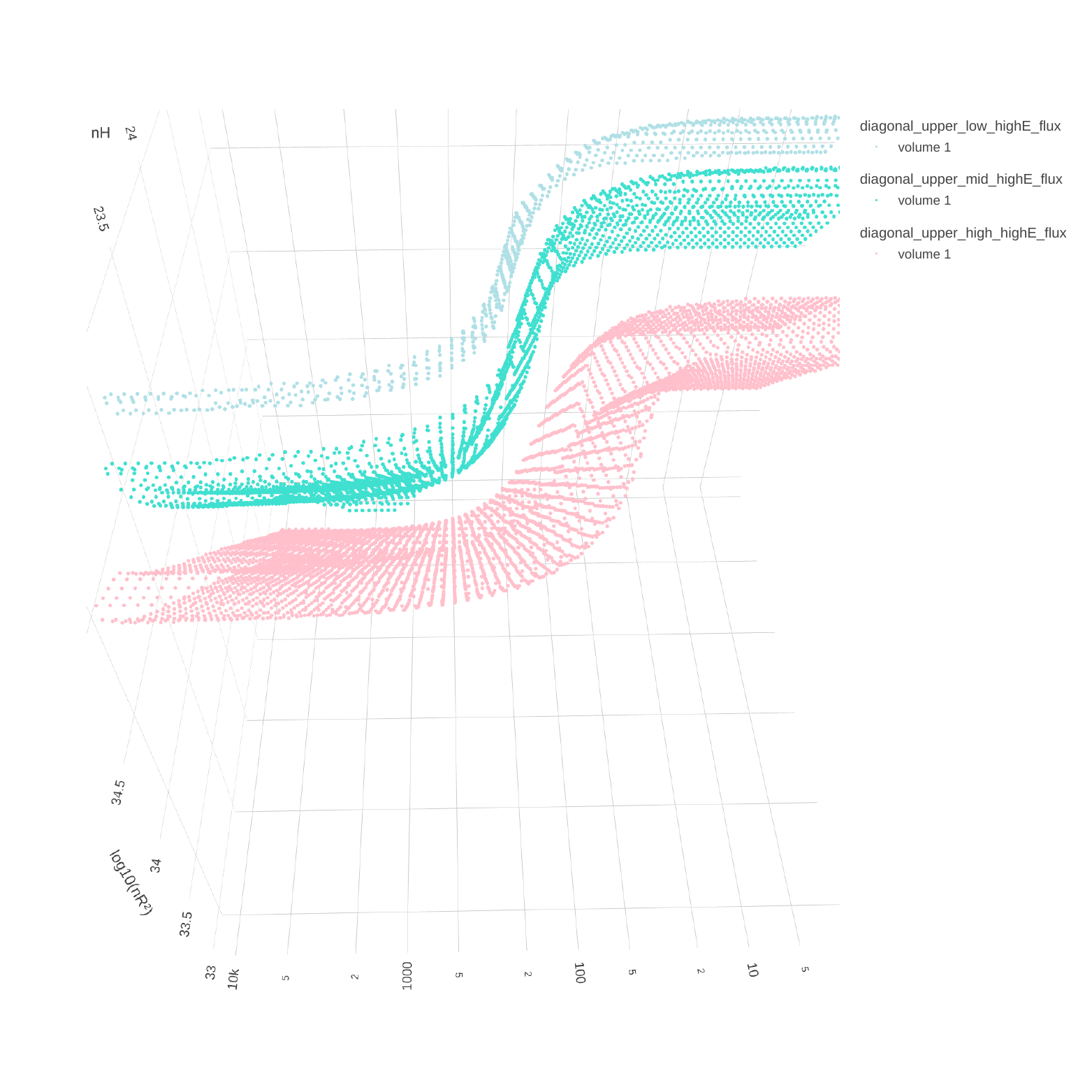}
\includegraphics[width=0.33\textwidth]{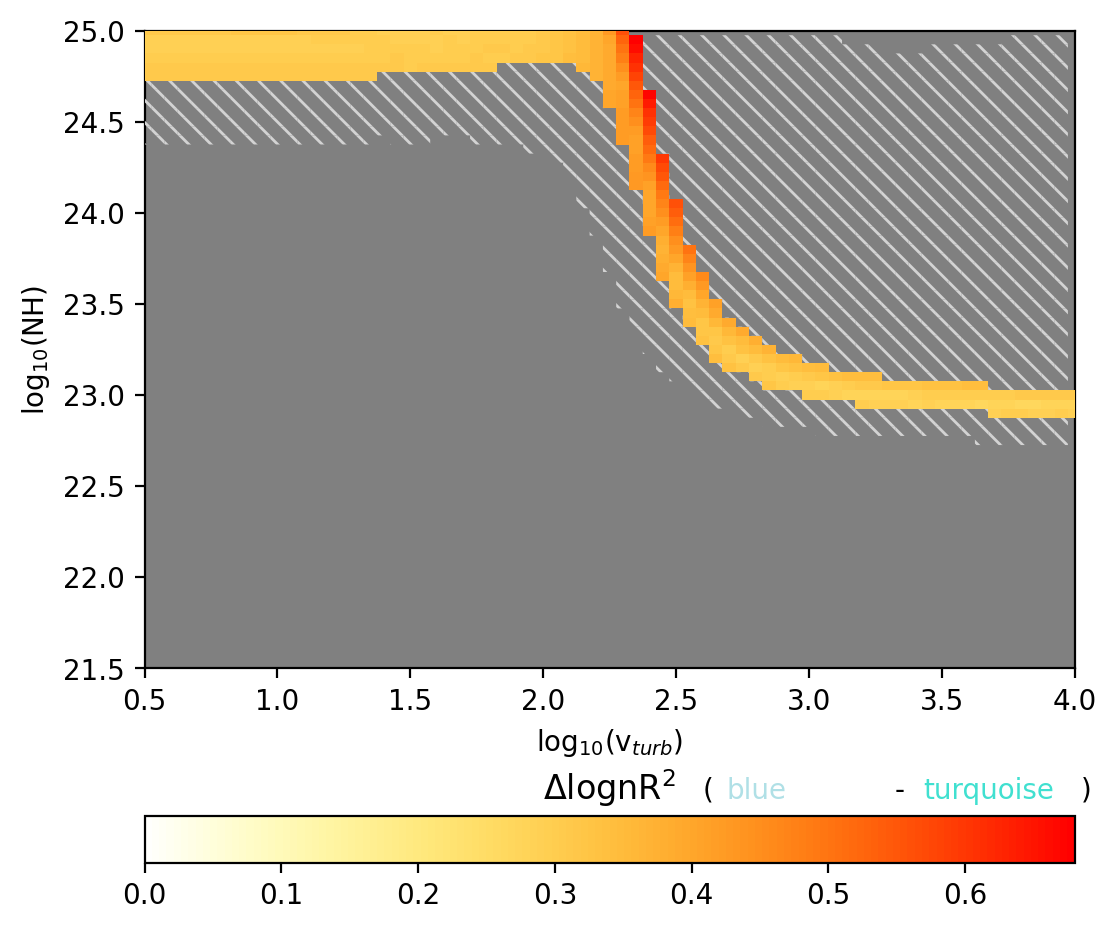}
\includegraphics[width=0.33\textwidth]{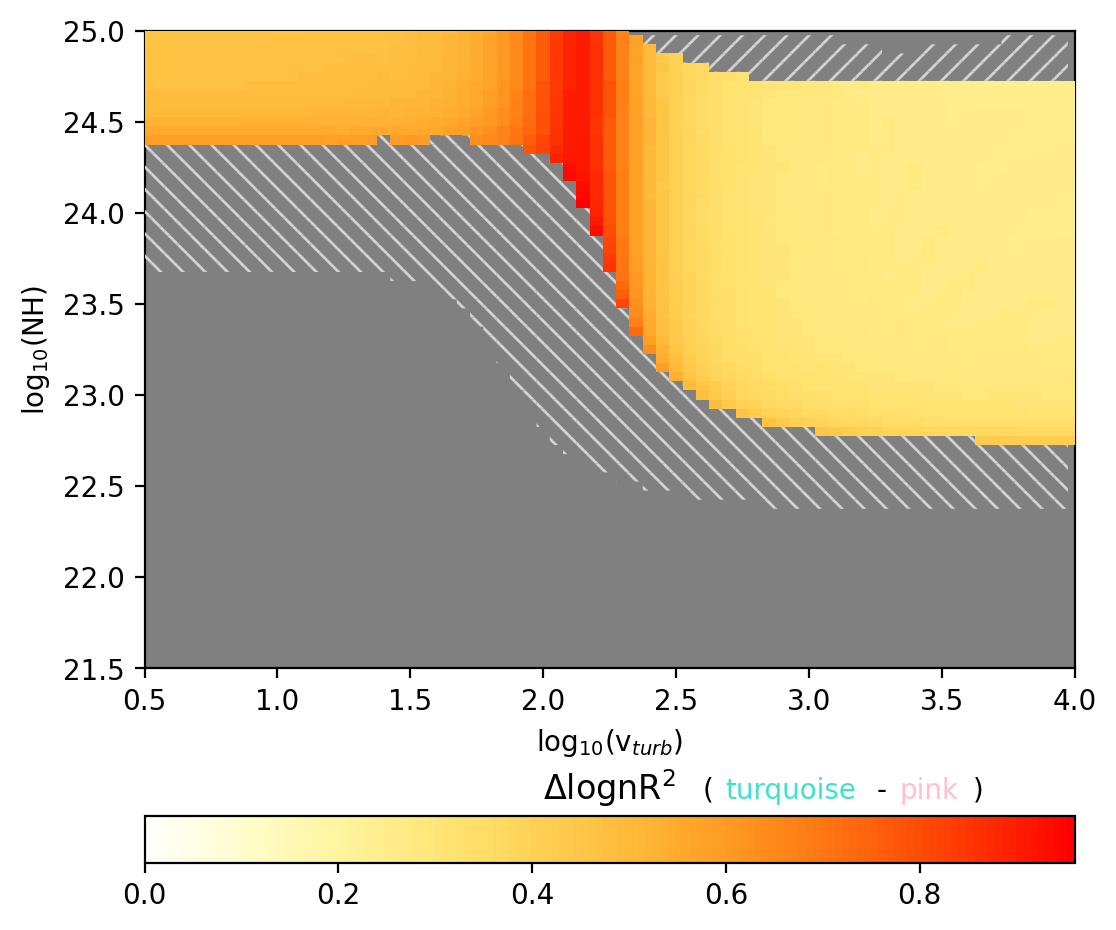}
\includegraphics[width=0.33\textwidth]{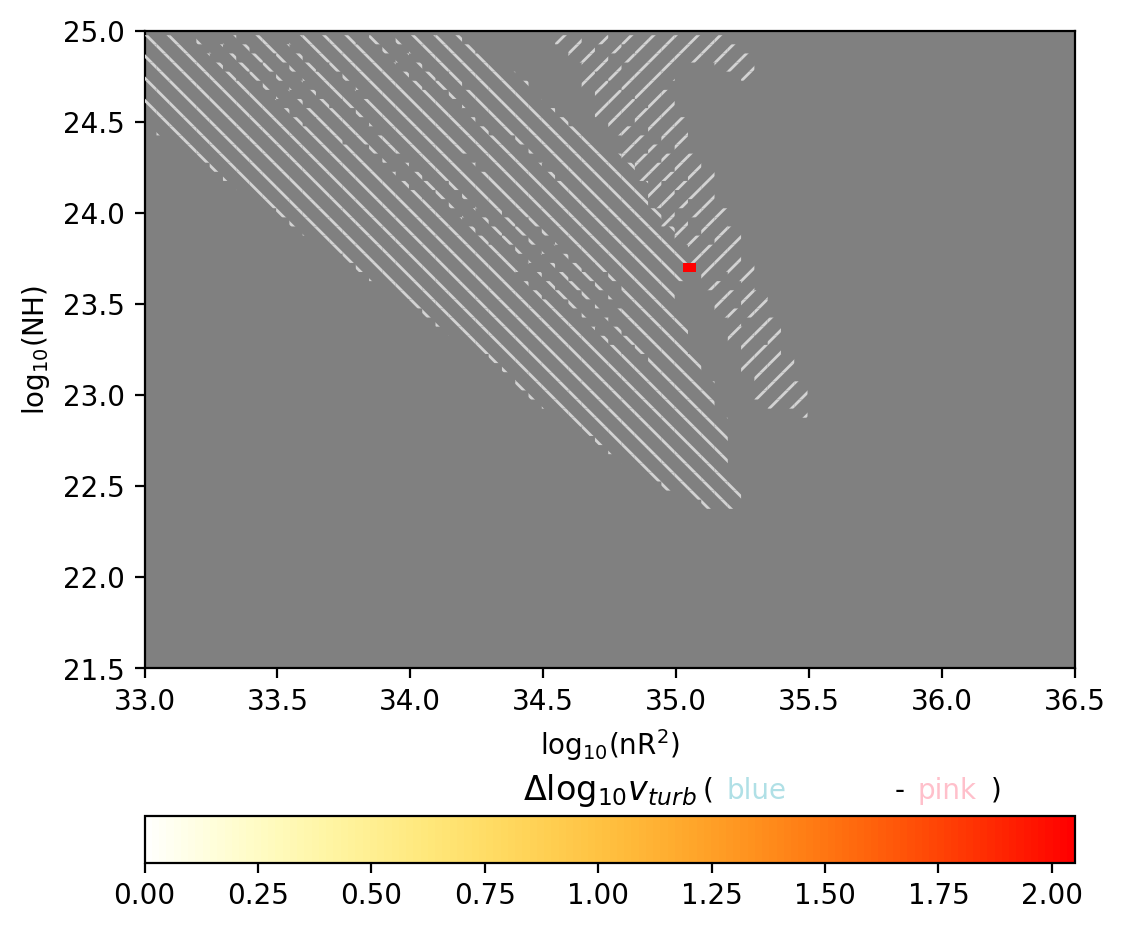}
\end{center}
\vspace{-1em}
\caption{\textbf{(Top)} Wind parameter space available for several SEDs that only differ in HR$_{hard}$. For better visualization, different panels showcase different angles of the same figure. \textbf{(Bottom)} Projections along two wind parameters, highlighting the differences between pairs of SEDs among the one displayed above. In the plane, combinations of the wind parameters only valid for the first (second) SED are shown with /// (\textbackslash\textbackslash\textbackslash) hashes. When the first two wind parameters are compatible, the difference in the third wind parameter is highlighted with a colormap in shades of orange (blue) if it is positive (negative).} \label{fig:4U_wind_parameter_plots_HRevol}
\end{figure*}

We start by selecting a range of observations with significant variations in luminosity, hardness ratio, and line properties across the HLDs. We highlight the characteristics of these observations in Figure \ref{fig:4U_ionization_sampling}. The observation highlighted in red is taken at the beginning of the decay of the 2006 outburst, and previously highlighted as an outlier (blue, at $\sim5\cdot10^{-2}$ L$_{Edd}$) in the EW plots of Fig.~\ref{fig:4U_plots_substructure}. The observation in brown is the latest observation taken during the same outburst, with much lower luminosity, and part of the potential substructure described in Sect.~\ref{sub:correlations}. 
The observations in orange and pale blue are from the main portion of the soft state diagonal above $\sim4\cdot10^{-2}$ L$_{Edd}$, both at very low HR$_{\mathrm{hard}}$, but with luminosities covering the lower and upper parts of the so-called ``main structure''. The observations in turquoise and pink are chosen for their higher HR$_{\mathrm{hard}}$ but similar luminosities to the previous observations. Finally, the green observation is from the outlier \nustar{} SPL state where both a \FeKav{} and \FeKavi{} lines were detected.

We compute the ionization fractions of \ion{Fe}{XXV} and \ion{Fe}{XXVI} for a thin slab, with each SED, using \texttt{CLOUDY} 23.01, for a wide range of ionization parameters $\xi$. We then use these ionic fractions to compute the EWs of the \FeKav{} and \FeKavi{} lines, constructing the curve of growths as a function of column density N$_\mathrm{H}$ and turbulent velocity v$_{turb}$ (see \citealt{Bianchi2005_AGN_EW_ionisation} for details). For each SED, this results in a grid of the EW of each line as a function of nR², $v_{turb}$, and $\xi$, which can be compared to the EWs measured in the corresponding observation. Finally, to obtain an SED-independent parameter space, we convert $\xi$ to nR² using the unabsorbed luminosity measured in each observation extrapolated in the 1-1000 Ry range. Thus, for each observation, the resulting space of "possible" wind parameters span a set 3D volumes in a nR²-NH-$v_{turb}$ space, of which we show the sampled envelope in Fig.~\ref{fig:4U_wind_parameter_plots_3D}. In this graph, the volumes overlap for parameters that can produce the observed features in different observations.

 The shape of each volume is broadly similar in each observation: a region at low turbulence ($v_{turb}\lesssim100$ km/s) restricted to extremely high column densities (NH$\gtrsim 10^{24.5}$ cm$^{-2}$), a region at high turbulence ($v_{turb}\gtrsim1000$km/s), restricted to lower column densities (NH$\lesssim 10^{23}$cm$^{-2}$), and a transition region between the two. We highlight that the volumes of the cyan and pink observations are significantly wider than the rest due to the lack of detections of \FeKav{} in these 2 observations. In these cases, the upper limits measured for the \FeKav{} line were used to derive the parameter constraints. The rest of the volume, aside from the outlier green observation from the SPL state, remains constrained to nR² values of $10^{34}-10^{35.5}$. Outside of the red, brown, and cyan volumes, which show significant overlap, every other volume is significantly shifted in the nR²-NH plane and remain completely distinct. This implies that at least one wind property (either nR², NH, or $v_{turb}$) must evolve between all these observations. To highlight and quantify the regions where a pure change in a single parameter is enough to explain the evolution of the wind properties in two individual observations, we complete the 3D graphs with 2D projections in each plane. The entire set of projections, for all parameters and all pairs of observations, is made available in the Appendix~\ref{app:2D_corner}. 

We can now relate the evolution of the wind parameters to the evolution of the SED along the outburst, with $HR_{hard}$ and $L_{[3-10]}$. For this, we consider two groups of three observations that were chosen for this specific purpose: the light blue, turquoise, and pink observations, which all have very similar luminosities but progressively higher $HR_{hard}$ values, and the brown, orange, and light blue observations, with almost identical $HR_{hard}$ values but progressively higher luminosities. 

In the first group, the disposition of the volumes, which we show in the upper panels of Fig.~\ref{fig:4U_wind_parameter_plots_HRevol}, reflects the evolution of $HR_{hard}$ between the SEDs: for both NH and nR², the light blue volume rests systematically higher than the (harder) turquoise, itself above the (harder) pink. This confirms that the evolution of the accretion flow that leads to the change of $HR_{hard}$ also has an effect on the structure of the wind. We further show several projections highlighting the changes of nR² and $v_{turb}$ in the bottom panels of  Fig.~\ref{fig:4U_wind_parameter_plots_HRevol}. The first two panels highlight that the minimal change in nR² between each observation mostly depends on $v_{turb}$. However, we stress that, since we do not know where the outflow parameters lie within each volume, we cannot quantify the value itself. As seen in the right panel, aside from a single combination in nR² and NH, the light blue and pink observations (with the lowest and highest $HR_{hard}$, respectively) are completely incompatible with a pure change in $v_{turb}$. Considering the drastic change in $v_{turb}$ required (2 orders of magnitude), this all but confirms that at least NH or nR² must evolve along with $HR_{hard}$.

\begin{figure*}[h!]
\begin{center}
\includegraphics[clip,trim=2cm 5cm 6cm 5cm,width=0.34\textwidth]{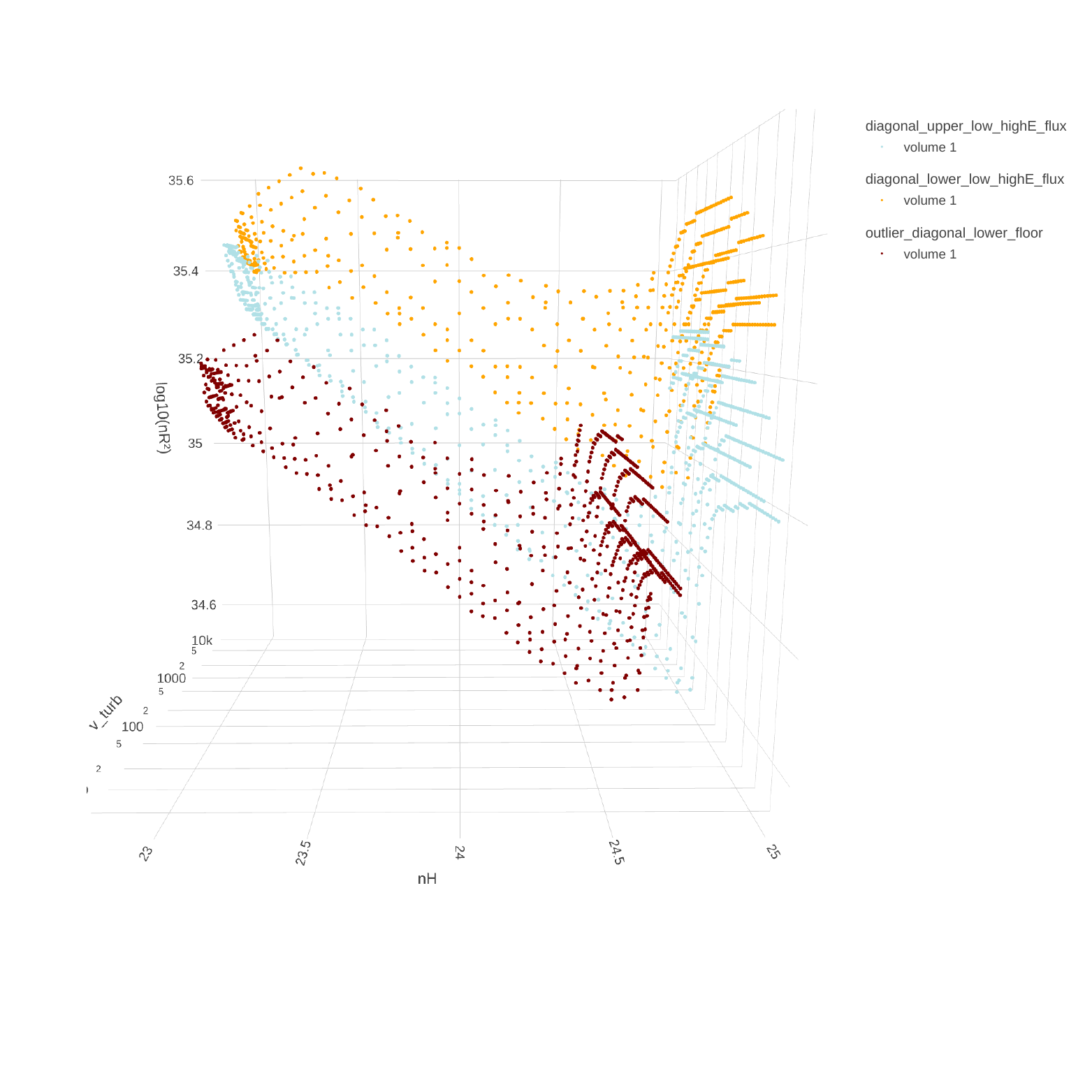}
\includegraphics[clip,trim=4cm 6cm 6cm 5cm,width=0.34\textwidth]{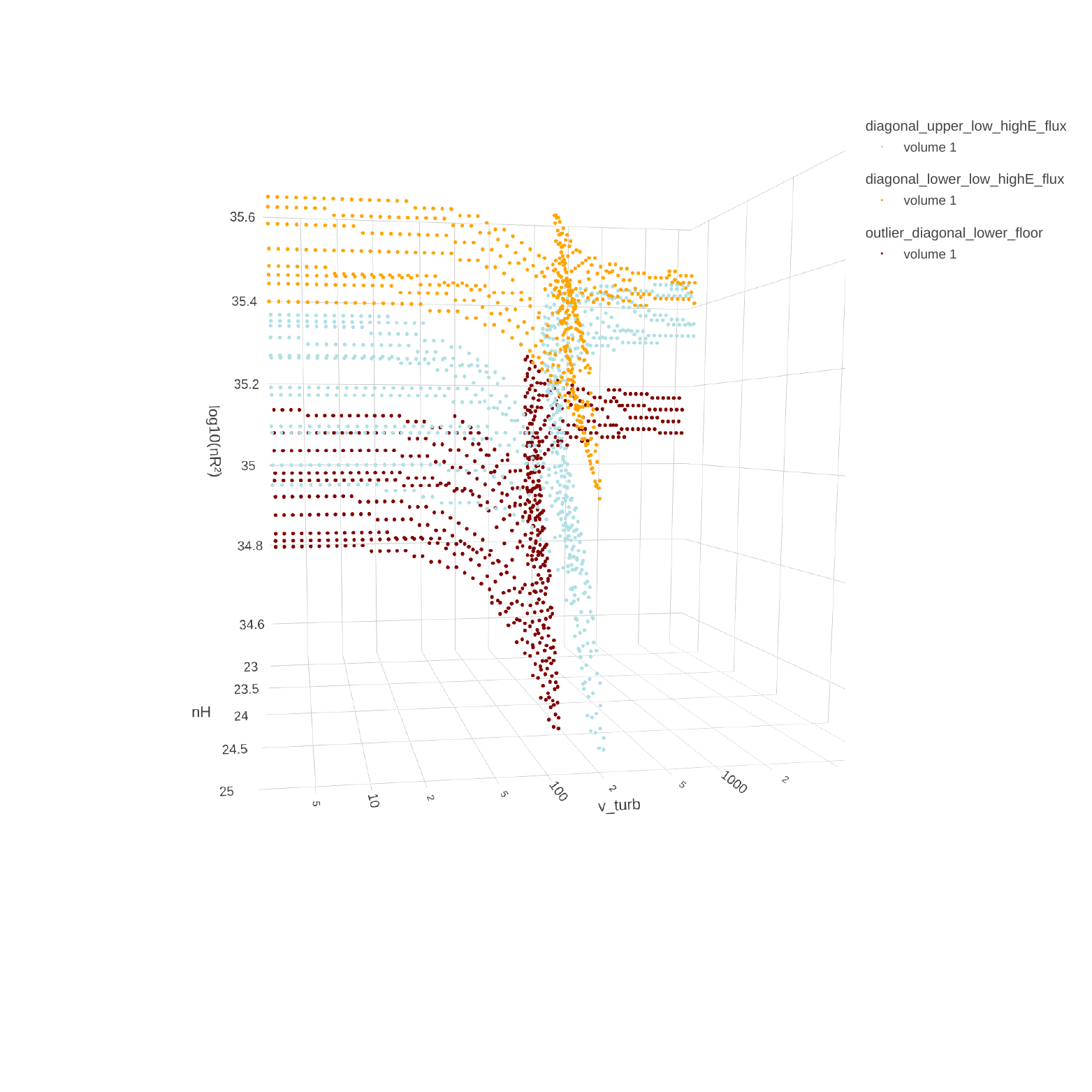}
\includegraphics[clip,trim=2cm 2cm 6cm 5cm,width=0.31\textwidth]{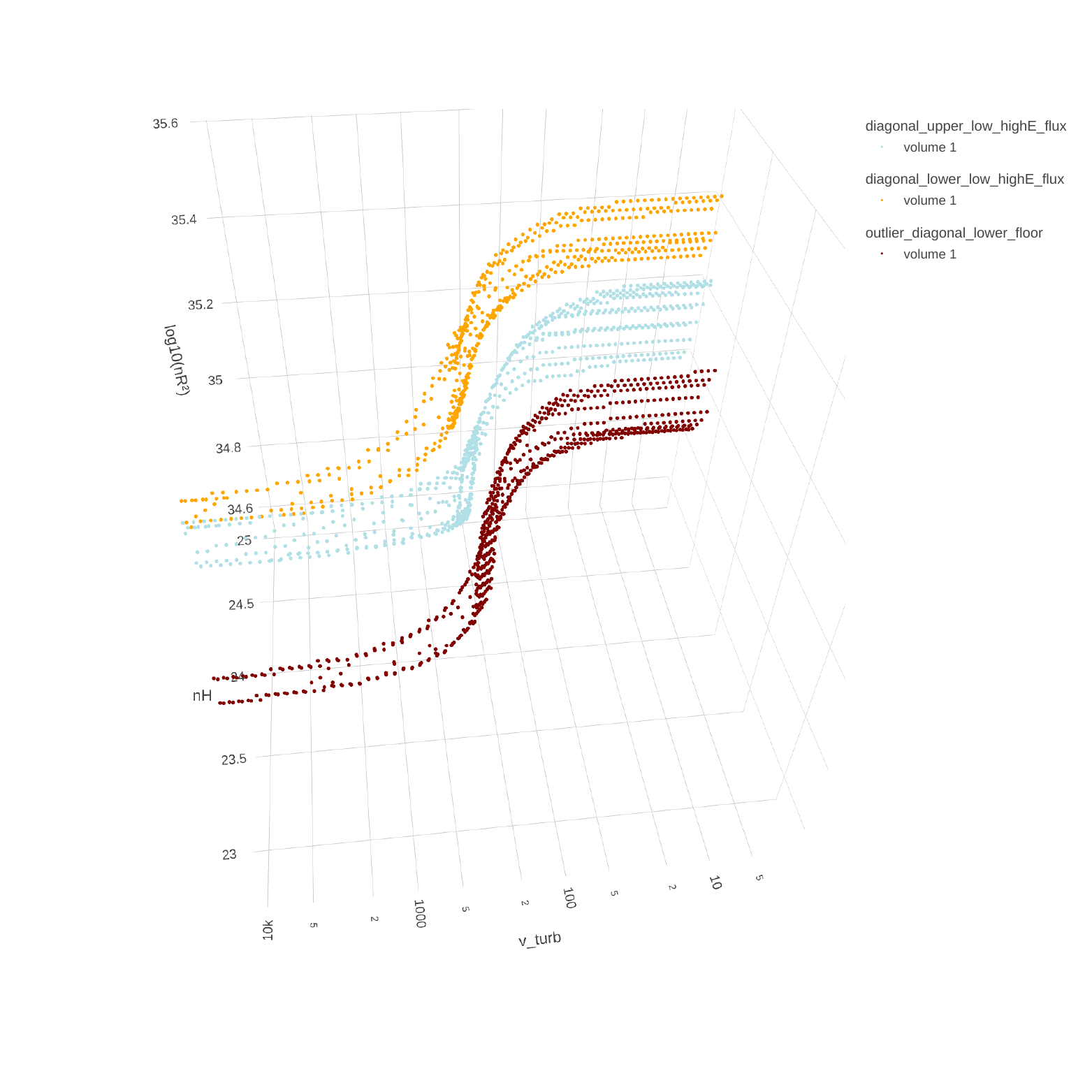}
\includegraphics[width=0.33\textwidth]{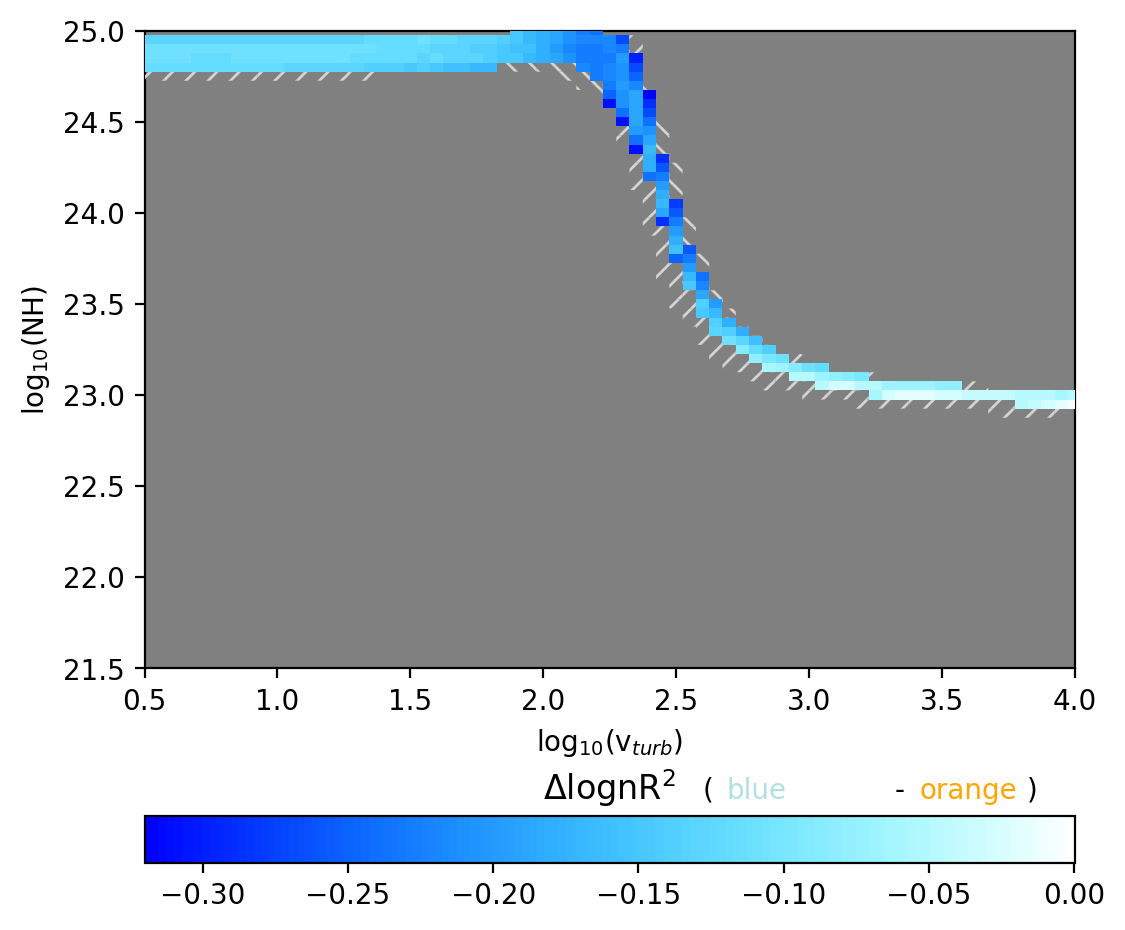}
\includegraphics[width=0.33\textwidth]{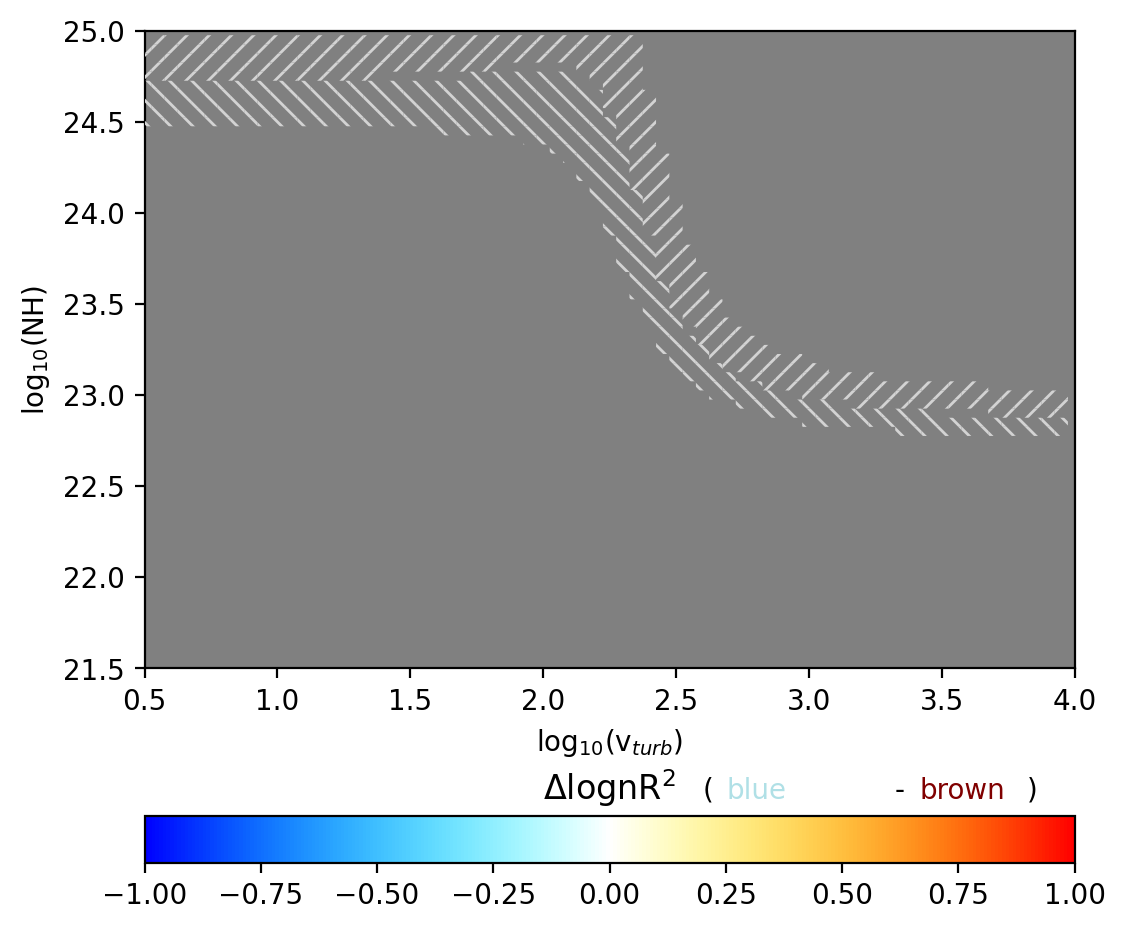}
\includegraphics[width=0.33\textwidth]{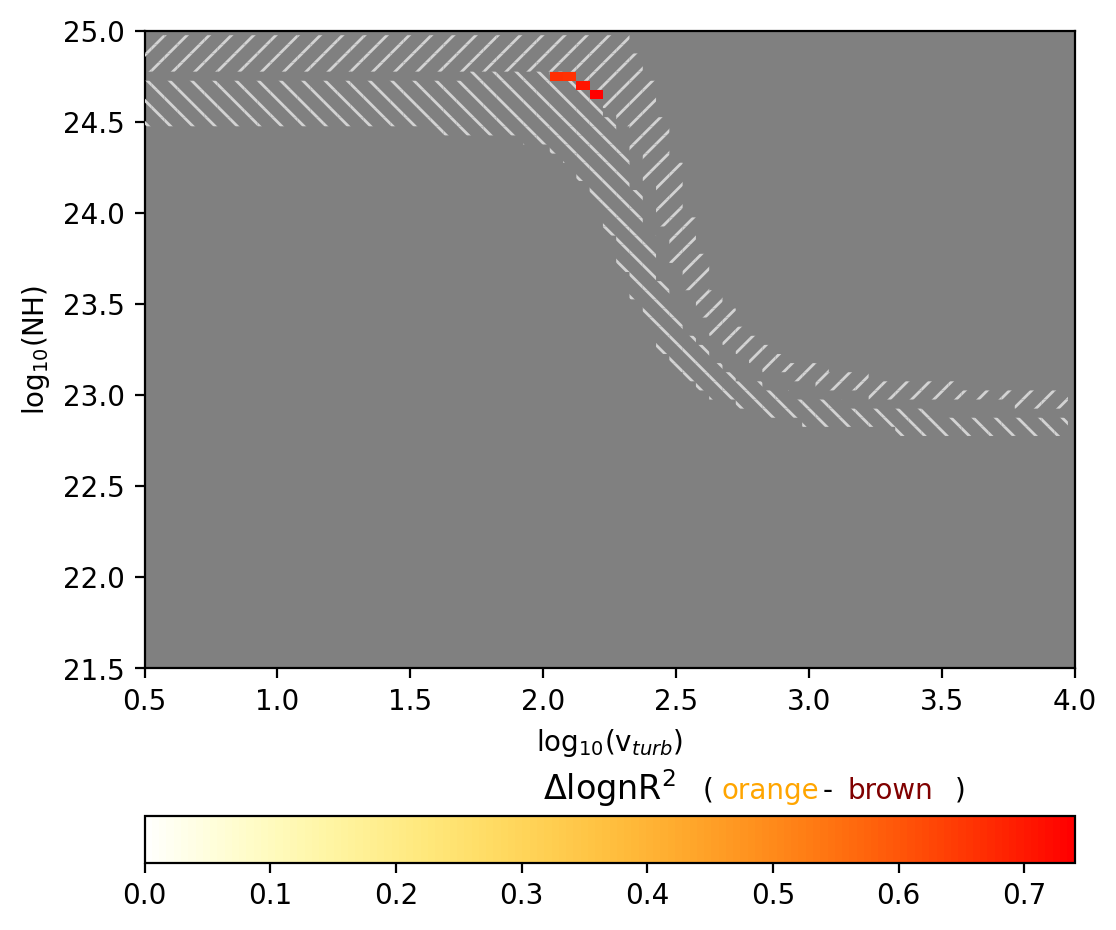}
\end{center}
\vspace{-1em}
\caption{\textbf{(Top)} Wind parameter space available for several SEDs which only differ in luminosity. For better visualization, different panels showcase different angles of the same figure. \textbf{(Bottom)} Projections along two wind parameters, highlighting the differences between pairs of SEDs among the one displayed above. In the plane, combinations of the wind parameters only valid for the first (second) SED are shown with /// (\textbackslash\textbackslash\textbackslash) hashes. When the first two wind parameters are compatible, the difference in the third wind parameter is highlighted with a colormap in shades of orange (blue) if it is positive (negative).} \label{fig:4U_wind_parameter_plots_lumevol}
\end{figure*}

In the second group, which we highlight in Fig.~\ref{fig:4U_wind_parameter_plots_lumevol}, the shift between the volumes confirms a change in wind properties between the main and low luminosity structures presented in Sec.~\ref{sub:correlations}. However, the evolution of the parameter space does not directly follow the evolution in luminosity: instead, the three observations must have opposite changes in at least one wind property. Indeed, the shape of the orange volume (intermediate luminosity) is virtually identical to that of the light blue (high luminosity), with the latter shifted to higher nR² values. In contrast, the brown (faintest) observation's parameter space has a different shape from the first two, and is shifted this time to lower nR², as well as to lower NH values. This disposition requires an inversion in the evolution of at least one wind parameter, between the brightest, intermediate, and faintest observations. Individual projections, of which we show examples in the bottom panels of Fig.\ref{fig:4U_wind_parameter_plots_lumevol} provide limited information due to the complex evolution of the 3D shapes between in each volumes. The first panel reveals a region at high turbulence with almost no difference in nR² between the light blue and orange observations. However, such high turbulence (above $\gtrsim10^3$ km/s) has yet to be observed in BHXRB wind signatures and thus, remain disfavored. On the other hand, the blue and brown observations are never compatible with a pure change in nR², showing that either NH or $v_{turb}$ must change between these two observations.

Finally, it is difficult to constrain the behavior of the two outliers seen in Fig.~\ref{fig:4U_wind_parameter_plots_3D}. The soft observation in red, taken during the decay of the 2006 outburst, likely requires a strong decrease in nR². Indeed. its volume, although similar to the observations highlighted in Fig.~\ref{fig:4U_wind_parameter_plots_lumevol}, remains at nR² values even lower than the brown SED. Meanwhile, the SPL observation in green, when compared to the rest of the dataset, heavily favors a significant increase in either nR² or $NH$, barring a shift in $v_{turb}$ of several orders of magnitude.

\section{Discussion}\label{sec:discussion}

\subsection{Physical implications}\label{sub:physical_implications}

The results of the previous sections not only have strong implications on the evolution of 4U~1630$-$47, but help contextualize the previous wind detections (and lack thereof) in the rest of the BHLMXB population. 

First, the observations chosen to test the influence of $HR_{hard}$ and the luminosity are representative of the general evolution of the SEDS and wind parameters in the source. Thus, the correlation between $HR_{hard}$ and the \FeKavi{} EW can now be interpreted physically as indicative of the continuous evolution of at least one wind parameter linked to the hardening of the spectrum. Futhermore, as $HR_{hard}$ is a proxy of the evolution of the hard tail component along the soft state and SPL, this part of the outflow evolution must be exclusively tied to the evolution of the corona. In parallel, we can now interpret the structures seen when comparing $L_{3-10}$ and \FeKav{} EW or the K$\alpha$ EW ratio (see Fig.~\ref{fig:4U_plots_substructure}) as the sign of the evolution of at least one wind parameter, linked to the luminosity of the source. Since we compared observations with identical values of $HR_{hard}$, this aspect of the outflow evolution must be linked to the thermal disk. Interestingly, we can confirm that this evolution is not monotonic: for $L\lesssim4.5\cdot10^{-2}L_{Edd}$, the link between luminosity and wind parameter reverses, compared to higher luminosities. This reflects the very strong changes in behavior of the lines between the "main structure" above $4.5\cdot10^{-2}L_{Edd}$ and the "substructure" below that value. This "substructure" may be the consequence of a change in the entirety of the accretion flow (disk, corona, and wind), hinted at by the fact that most of the low-luminosity soft state observations show a distinct and structured $HR_{hard}$ evolution, unlike what is seen at higher luminosity (see upper right panel of Fig.~\ref{fig:4U_plots_substructure}). Finally, we note that the evolutions discussed above are, at least at first order, consistent between each outburst. 

On a larger scale, the lack of knowledge of hard X-rays and the underlying influence of $HR_{hard}$ may explain two important unknowns in the detection of BHXRB wind signatures. The first is the coexistence of wind detections and non-detections, as well as the lack of structured wind evolution with the Hardness Ratio in the soft state (see e.g. P24). The part of our study focusing on HR$_{soft}$ finds similar results, with both a much stronger spread in the $HR_{soft}$ correlation and a significant overlap between detections and non-detections in the soft HLD. The second unknown is the lack of \textit{any} X-ray wind detection in many well-observed, high-inclined BHLMXBs, such as MAXI J1820+070 or Swift J1658-4242, which may be entirely caused by a naturally higher fraction of hard X-rays during the soft state. 

We note that the influence of the corona and disk on the wind can be explained differently for different wind launching mechanisms. In physically motivated magnetic wind models such as the WED (see e.g. \citealt{Jacquemin-Ide2019_wind_weak_magnetic_JEDSAD_modeling,Datta2024_WED_signatures}), while $\xi$ is independent of the luminosity, NH scales linearly with $\dot m$, itself scaling linearly the disk luminosity (assuming a constant radiative efficiency). Meanwhile, the gradual increase in $HR_{hard}$ can be linked to a gradually bigger inner flow (e.g. higher $r_J$ in the JED-SAD framework, see \citep{Marcel2018_JED-SAD_III}), pushing the inner radius of the wind emitting disk further and thus comparatively reducing $\dot m$ and $NH$ by a factor $(r_{ISCO}/r_{J})^p$ in the region where the wind would be detectable, with p the ejection index of the wind. Since the extension of the corona is constrained by fits in other systems (see, e.g. \citealt{Marino2021_JED-SAD_1820,Demarco2021_1820_Timing}), this effect is unlikely to be sufficient, but the turbulence, approximated by the non-radial velocity component $v_\perp$ in magnetic models, would decrease by a factor $(r_{ISCO}/r_{J})^{p/4}$, due to the region of a similar $\xi$ moving to higher radii, further reducing the EW.
For thermal winds, the effect of changes in luminosity and coronal geometry are harder to assess quantitatively. However, a decrease in luminosity below $L_{CRIT}$ increases the launching radius of the wind, reducing NH \citep{Done2018_thermal_winds_modeling_H1743_GROJ1655}. Meanwhile, while a hardening of the spectrum naturally decreases the Compton Radius\footnote{defined as the radius where the thermal velocity of the disk atmosphere reaches the escape velocity}, it may also increase the scale height of the inner atmosphere's scale height, shadowing the outer regions and thus progressively suppressing the thermal winds \citep{Tomaru2019_H1743-322_radiathermalwind}. These effects are  largely dependent on the Compton Temperature\footnote{the temperature reached by the irradiated disk surface}, and although it showed little to no evolution in our SEDs from the soft state to the SPL (as seen in Fig.~\ref{fig:4U_HLDs_accretion_states_stability}), it is difficult to estimate how much our limited datasets constrain this parameter.

Finally, the discovery of hard state SEDs with thermally stable regions above $\xi\sim3$ opens up new possibilities to bridge the gap between the different observations of winds. While optical and infrared signatures show that the outflows persist throughout the entire outburst \citep{Sanchez-Sierras2020_MAXIJ1820+070_wind_emission_infrared_soft_hard}, these signatures correspond to much lower ionization parameter (log$\xi\lesssim 1$) than X-ray winds, implying a different launching region, if not mechanism. Moreover, the restriction of direct absorption features to hard states made their detection incompatible with that of X-ray winds until now (see e.g.\citealt{MAXIJ1803-298_wind_x-ray_soft}). X-ray wind themselves are split between a few warm absorber-like signatures with log$\xi\sim2-3$, whose nature as outflows is still debated but with some clear signatures in hard states, and the canonical "hot" wind detections at log$\xi\sim3-5$, with clearly identified blueshifts but no unequivocal detections in hard states (see P24 for a review). Detecting hot winds in "stable" hard states would significantly help constrain the wind launching mechanisms and the accretion-ejection structure, especially if they can be simultaneous to other wind detections at different ionization ranges. However, the evolution of the wind with $HR_{hard}$ is likely to weaken the main iron lines, thus restricting potential detections to microcalorimeters.

\subsection{Comparison with the literature and caveats}\label{sub:lit_compa}

Our current results can be compared with previous detailed studies of a smaller number of observations of 4U~1630$-$47 . \cite{DiazTrigo2014_4U1630-47_wind_2012-13XMM} attribute the disappearance of wind signatures in a set of 2012 \xmm{} observations to the changes in disk properties and progressive overionization with an increase in luminosity, without requiring any change in wind properties. However, their photoionization computations are derived from \texttt{warmabs}, which takes the ionic population balance from a fiducial powerlaw spectrum with $\Gamma=2$. Moreover, their spectral analysis stems entirely from soft X-ray spectra, and thus ignore the hard tail in the spectrum until the very last observation, where the wind has already disappeared. In our study, the evolution of the EW of the lines matches perfectly the \FeKavi{}-$HR_{hard}$ correlations, and thus the wind seems to be responding to the increase in coronal component. 

Later, \cite{Gatuzz2019_4U1630-47_wind_2012-13Chandra} performed a detailed study of the evolution of the absorbers during the same outburst, adopting \chandra{} data exclusively, and found that the evolution of the absorption lines during the transition to harder states required a change in wind parameter. Since their observed state transition occurs from a relatively faint soft state to the low-hard state, they propose that, in the case of a thermal wind, a low-temperature disk may not be able to heat sufficiently an already outflowing atmosphere, when transiting to harder states. This matches what we obtain in our study, with their observations 6 and 7 (ObsIDs 56407 and 56439 respectively) having a much harder $HR_{hard}$. Meanwhile, their Obs5 (ObsID 14441), which they did not consider in their analysis, has spectral properties indicative of a bright but thermally stable canonical hard state (with $HR_{hard}<1$). The lack of wind signatures in this observation is perfectly in line with its very high $HR_{hard}$ values of this observation. 

Finally, analyzing \suzaku{} and \nustar{} data up to 2016, \citealt{Hori2018_4U1630-47_2015SuzakuNustar} note significant correlations between their wind and continuum properties, notably with NH. However, their correlations are driven by changes between observations in the substructure and few observations at higher luminosities, all of which span a wide range of $HR_{hard}$. Thus, the result is likely to be a blend of the different effects that we identified in Sec.~\ref{sub:photo_mod}. They also report a strong change in absorption line properties in the first of the three 2015 observations, which cannot be explained by the changes in illumination. We found that these changes are perfectly in line with the harder SED of this observation (performed during the rise of a hard flare) compared to the other 2015 exposures.
\begin{figure*}[h!]
\includegraphics[width=0.43\textwidth]{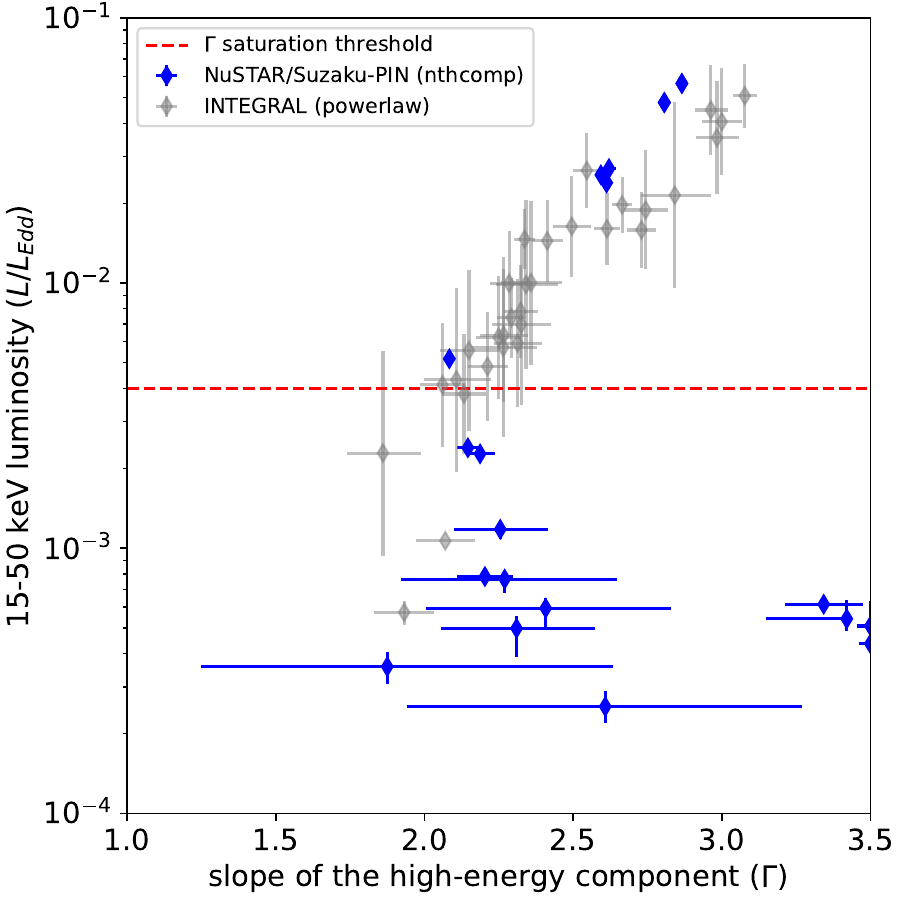}
\includegraphics[width=0.57\textwidth]{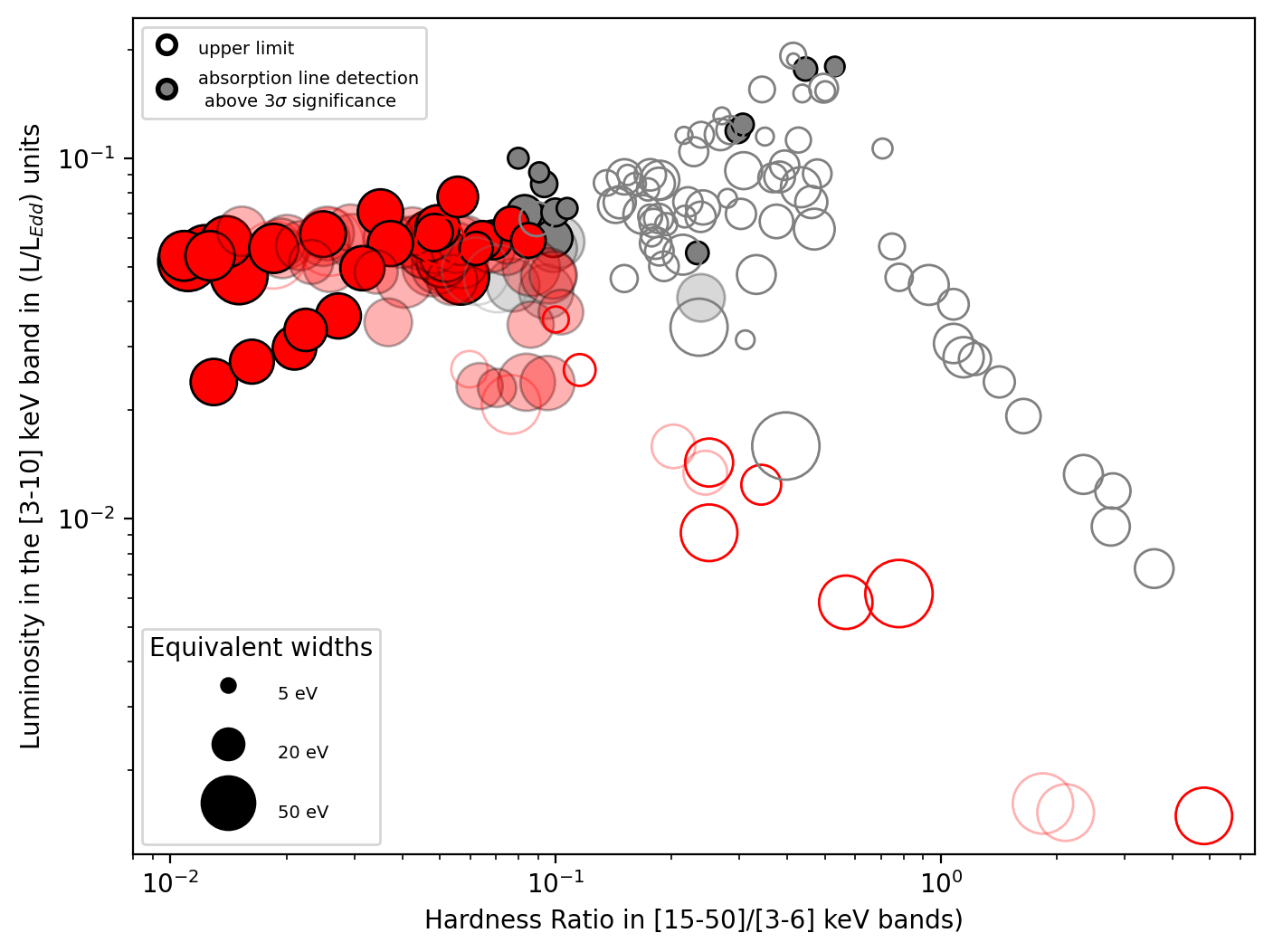}

\caption{\textbf{(Left)} Comparison of the slope of the high-energy component with the [15-50] keV luminosity, using \texttt{nthcomp} for observations with good low-energy coverage and a powerlaw for \integral{}-only epochs. The red line highlights the approximate luminosity threshold below which $\Gamma$ stops being correlated to the luminosity of the coronal component.\textbf{(Right)} Hard HLD highlighting in red the observations with $L_{15-50}$ below that threshold, or with a 1 sigma upper limit below that threshold when the BAT detection is not significant.}\label{fig:highE_intrinsic_correl}
\end{figure*}

Aside from the wind evolution itself, other studies have focused on a detailed photoionization modeling of the lines. \cite{King2014_4U1630-47_wind_2014Nustar} analyzed a \nustar{} dataset during a soft state of the 2012-2014 outburst, finding strong evidence for an absorber that may have extremely high velocities (0.043c) if it arises from the \FeKav{} line. We stress that, out of more than 200 observations, the only instance where the \FeKav{} line was preponderant occured during the anomalous SPL observation in 2023, and this absorption line was strongly variable. Since the spectral properties of this \nustar{} observation are consistent with a standard soft state, it is likely that its line properties are as well, with the line feature originating from \FeKavi{}. 

More recently, focusing on \chandra{} data, \cite{Trueba2019_4U1630-47_wind_2012-13Chandra} identified two distinct photoionization zones, with different ionization parameters, column densities, and velocities. Our analysis and results, which assume a single absorber, are largely applicable to their "zone 2", which is the most important contributor to the EW of the lines. Since the identification of a second absorber is virtually impossible without gratings or microcalorimeter-level spectral resolution, a detailed study of the wind evolution in each zone would require more observations of the source with these instruments, along a wide range of spectral states. 

We note that outside of the more detailed analysis of \ref{sec:wind_evol}, we only considered the [3-10] keV band for the luminosity, even in the hard HLD. Although this approach helps to compare with the standard HLD and avoids adding the significant uncertainties of the BAT measurements, in the future, including the hard X-rays will give a more representative proxy of the total luminosity of the source and its evolution. Later, for the photionization modeling in \ref{sub:photo_mod}, we consider wind parameters as "valid" when the resulting EWs are compatible with the 90\% uncertainty of each line and their EW ratio, or the 3$\sigma$ upper limit of the EW for states without detection. For future studies, with quantitative computation of the parameter evolution between SEDs, we will consider a wider range of confidence levels. Moreover, while the time-averaged spectrum of the SPL state (green in Fig.~\ref{fig:4U_wind_parameter_plots_3D}) was sufficient to show that the absorber properties in this observation are vastly different from the rest of the sample, the wind signature in this observation has complex time-variability.

\subsection{Evolution of the high-energy component}\label{sub:high_E_comp_evol}

Now that we have identified a strong link between the high-energy component and the main parameters of the wind  (NH, nR$^2$, $v_{turb}$), disentangling the influence of the different high-energy parameters would greatly help to understand how the corona affects the outflow. The cutoff remains too high to be unconstrained in the states where wind signatures are detected, especially with our very limited data quality, and we thus focus on the interplay of the photon index and normalization of the hard tail. To that end, we show a scatter plot of the evolution of the two parameters in Fig.~\ref{fig:highE_intrinsic_correl}, as measured with \texttt{nthcomp} in epochs with high-quality soft X-ray coverage, and with a simple powerlaw for \integral{}-only epochs. We note that all the \nustar{} and \suzaku{}-PIN observations in our sample were taken in soft to SPL states and are thus directly relevant to the wind evolution. This is also the case for the vast majority of \integral{} observations, and notably the faintest detections, which can be identified as soft states from the neighboring \maxi{} and \swift{}/BAT monitoring. In contrast, it is not always possible to distinguish bright hard states from intermediate or faint SPL with no additional coverage. In any case, the results of both instruments remain largely compatible, and at high [15-50] keV Eddington ratios, both show a very tight correlation between the two parameters, similarly to what was reported for older outbursts in \citealt{Seifina2014_4U1630-47_properties}. This may help determine the nature of the coronal emission in softer states, but also indicates that we cannot isolate the effect of either parameter on the wind. 

Below $L_{15-50}/L_{Edd}\sim4\cdot10^{-2}$, the correlation between the luminosity of the the component and $\Gamma$ disappears visibly, with the photon index saturating around a value of $\sim2$. We note that the uncertainties of the \texttt{nthcomp} component become increasingly unreliable below that limit due to insufficient statistics. This saturation could indicate that this is the rough threshold below which secondary sources of high energy flux, such as reprocessing and reflection, become significant in the [15-50] keV band (see e.g. \citealt{Connors2021_4U1630-47_reflection_wind_soft_NuSTAR_2012_2015}), or that a significant change happens in the comptonized emission itself.

This behavior is in strong contrast to the results obtained by \citealt{Seifina2014_4U1630-47_properties}, who see a monotonic evolution of the comptonized photon index in several outbursts between 1998 and 2004, down to very low accretion rates and $\Gamma\sim1$. However, this study was based on BeppoSAX, which is lacking in sensitivity and spectral resolution compared to \nustar{} and \suzaku{}. Another clear outlier in our sample is the \suzaku{} observation at the very end of the 2010 outburst. \citealt{Tomsick2014_4U1630-47_decay_2010_anomalous} studied this observation and neighboring \swift{}/XRT exposures, finding that the spectra were well reproduced by a $\gamma\sim1.5$ powerlaw, despite a luminosity more than one order of magnitude fainter than any spectrum analyzed in \citealt{Seifina2014_4U1630-47_properties}. More importantly, this study was performed without using the PIN camera and thus without coverage above 12 keV, despite it being still much higher than the background. When analyzing the XIS and PIN data together for this study, we found that the 12-40 keV spectrum very strongly deviated from a simple comptonized component, even when adding a cutoff.

To test whether this change in behavior could be related to any of the soft continuum of wind parameters, we identified all observations with a $L_{15-50}$ value below the threshold in the right panel of Fig.~\ref{fig:highE_intrinsic_correl}. A large portion of the observations align with this selection, in intermediate states without wind detections, in the main substructure, the substructure, and at lower luminosity. Thus, there is no indication of any link with the wind parameters. 

\section{Conclusion}\label{sec:conclusion}

We performed an exhaustive study of iron absorption lines in all publicly available \chandra{}, \nicer{}, \nustar{}, \suzaku{} and \xmm{} observations of the BHLMXB 4U~1630$-$47 as of 2024. The combination of more than 200 individual days of observation provides unparalleled coverage of the HLD evolution of the source above $\sim10^{-2} L_{Edd}$, as shown in the left panel of Fig.~\ref{fig:4U_HID_full}. In Sect.~\ref{sub:HLD_evol}, for the first time, we clearly identify the limitations of the canonical "soft" HR dichotomy for line detections, as tens of constraining non-detections are observed across several outbursts, with the same HLD position as significant detections. To include the influence of hard X-rays, we replace the "soft" [6-10]/[3-6] keV Hardness Ratio by a "hard" [15-50]/[3-6] keV HR, built using \swift{}/BAT, \integral{}, \nustar{}, and \suzaku{}-PIN. This new HLD is shown in the right panel of Fig.~\ref{fig:4U_HID_full}. The new HR not only provides a clear limit between observations with and without line detections ($HR_{hard}=0.1$), but reveals a continuous link with the wind signatures. This is shown by the very strong anti-correlation ($p_S=6.4\cdot10^{-10}$, $r_S=-0.79$) between $HR_{hard}$ and the \FeKavi{} line EW, which we show in the bottom panel of Fig.~\ref{fig:4U_correl_observ}. 

We focus on the line properties of the source in Sect.~\ref{sub:distrib_correl}. The distributions of line parameters, which we show in Fig.~\ref{fig:glob_distrib}, are similar to those found for other sources, with EWs in a range of $\sim 5-60$ eV and predominantly detections of  \FeKavi{} and \FeKav{}. Small but significant blueshifts were already reported for most \chandra{} observations in P24. Here, the much larger \nicer{} and \suzaku{} samples are largely similar, albeit with more spread, with the cumulative distribution averaging at $\overline{v_{out}}\sim-560\pm60$ km/s. \xmm{} and \nustar{} remain completely unable to access these low velocities due to instrumental limits. The correlations with the soft X-rays remain in line with previous studies restrained to \xmm{} and \chandra{} (e.g. P24), with no correlation between $HR_{soft}$ and the line EWs, a significant anti-correlation between \FeKav{} EW and the 3-10 keV luminosity, and a significant correlation between the \FeKavi{}/\FeKav{} EW ratio and the 3-10 keV luminosity, as listed in the upper panels of Fig.~\ref{fig:4U_correl_observ}. However, as highlighted in Fig.~\ref{fig:4U_plots_substructure}, we also detect significant departures from the standard luminosity correlations in low luminosity soft-state. 

To assess whether changes in illumination can explain the evolution in line properties without an intrinsic change in the outflow, we test two different scenarios. First, in Sect.~\ref{sub:stability}, we consider the presence of thermal instabilities for harder SEDs, which could explain the global disappearance of the lines for high $HR_{hard}$. We thus split the source in four main spectral states and compute the stability curves using the best broad band SEDs available in each states. The results, shown in Fig.~\ref{fig:4U_HLDs_accretion_states_stability}, clearly show that the \FeKav{} and \FeKavi{} ionization range remains stable all the way through the hard states, and thus that instabilities cannot explain the disappearance of the lines. We also investigated in more detail the possibility of stable stability curves for \FeKav{} and \FeKavi{} in the hard state, using a detailed monitoring of the rise in the 2021 outburst. From the results, presented in Fig.~\ref{fig:4U_stability_2021}, we conclude that, at the very end of the rise in the canonical hard state, the illuminating SED becomes soft enough to be thermally stable in this ionization range, due to the unusually steep ($\sim \Gamma=2.3$) comptonization component in this source. This result, unprecedented for an XRB, opens the door to observations of highly ionized wind signatures in the hard state and simultaneous comparisons of X-ray and OIR absorption features. 

Then, in Sect.~\ref{sub:photo_mod}, we directly disentangle the SED and luminosity's influence on the EW of the lines by using photoionization modeling. For this, we compute the curve of growths of selected observations with strongly different HLD positions or line behaviors, for a range of $NH$, $v_{turb}$, and nR², and determine the parameter space whose EWs match the observed values of \FeKav{} and \FeKavi{}, as shown in Fig.~\ref{fig:4U_wind_parameter_plots_3D}. We then compare the parameter space of different groups of observations, isolating differences in $HR_{hard}$ (Fig.~\ref{fig:4U_wind_parameter_plots_HRevol}) and luminosity (Fig.~\ref{fig:4U_wind_parameter_plots_lumevol}). We identify a continuous evolution of at least one wind parameter with $HR_{hard}$, proving that changes in the "corona" responsible for the hard tail in soft to intermediate states of BHLMXBs directly affect the structure of the outflow. We also identify opposite evolutions of at least one wind parameter between the different groups of structured wind evolution at higher and lower luminosities, proving that the evolution of the thermal disk component affects the structure of the outflow, and that a clear change in behavior occurs at a specific Eddington fraction, estimated at $L_{3-10}/L_{Edd}\sim4.5\cdot10^{-2}$ for our fiducial values of mass and distance for this source. 

We verified in Sect.~\ref{sub:high_E_comp_evol} that this change is unrelated to the change in behavior that we detect for the photon index of the hard tail component, below a certain hard X-ray luminosity threshold. Outside of the very few observations with outlier line properties, which require abnormal changes in wind parameters, the evolution of the wind signatures in the sample is overwhelmingly dominated by these two effects, regardless of time and individual outbursts. Moreover, as discussed in Sect.~\ref{sub:physical_implications}, if the influence of the hard X-rays on the wind signature is a property shared by all BHLMXBs, it could explain the lack of structure and consistent detections in soft spectral states until now, as well as the lack of X-ray wind signature in many high-inclined BHLMXBs. However, we stress that the nature of both the accretor and the companion star in 4U~1630$-$47 are still unknown, and could impact the wind signatures in a non-standard way, such as if the system is a Be X-ray binary, as proposed in \cite{Kalemci2018_4U1630-47_d}.

Our study exemplifies the potential of comprehensive, large-scale studies of archival data for wind properties, and how this rich dataset can be used for several additional wind studies. Among them, our orbit-resolved analysis revealed several epochs with abnormally fast variability. A detailed analysis of these epochs is ongoing and will be presented in a future work. In parallel, the recent launch of XRISM offers the opportunity to quantify patterns of wind evolution in an unprecedented manner. An observation of 4U~1630$-$47 has already been performed, and a comparison with a representative set of different spectral states is a natural follow-up to this study. This next work will include the latest data of the 2022-2024 outburst, which we did not include in the present analysis.

\begin{acknowledgements}
MP acknowledges support from the JSPS Postdoctoral Fellowship for Research in Japan, grant number P24712, as well as the  JSPS Grants-in-Aid for
Scientific Research-KAKENHI, grant number J24KF0244. MP acknowledges support from the 2021 IDEX UGA - IRGA project WIND.
SB and MP acknowledge support from PRIN MUR 2017 ``Black hole winds and the baryon life cycle of galaxies: the stone-guest at the galaxy evolution supper''. SB acknowledges support from the European Union Horizon 2020 Research and Innovation Framework Programme under grant agreement AHEAD2020 n. 871158. MS acknowledges support from the JSPS Grants-in-Aid for Scientific Research 19K14762, 23K03459, and 24H01812 from the Ministry of Education, Culture, Sports, Science and Technology (MEXT) of Japan. VEG acknowledges funding under NASA contract 80NSSC24K1403. FC acknowledge support from Istituto Nazionale di Astrofisica (INAF) grant 1.05.23.05.06: “Spin and Geometry in accreting X-ray binaries: The first multi frequency spectro-polarimetric campaign”.
Part of this work was made possible thanks to the financial supports from CNES and the French PNHE.
This works uses data obtained from the \chandra{} Data Archive and software (CIAO and TGCat) provided by the \chandra{} X-ray Center (CXC), as well as data obtained through the HEASARC Online Service, provided by the NASA/GSFC, in support of NASA High Energy Astrophysics Programs, and the UK swift Science Data Center. We especially thank the \nicer{} support for their availability, and Tyler Parsotan for his support on BatAnalysis. 
\end{acknowledgements}

\bibliographystyle{aa}
\bibliography{biblio_4U_wind.bib}

\begin{onecolumn}

\appendix

\begin{figure}[h!]
\centering
\includegraphics[width=0.99\textwidth]{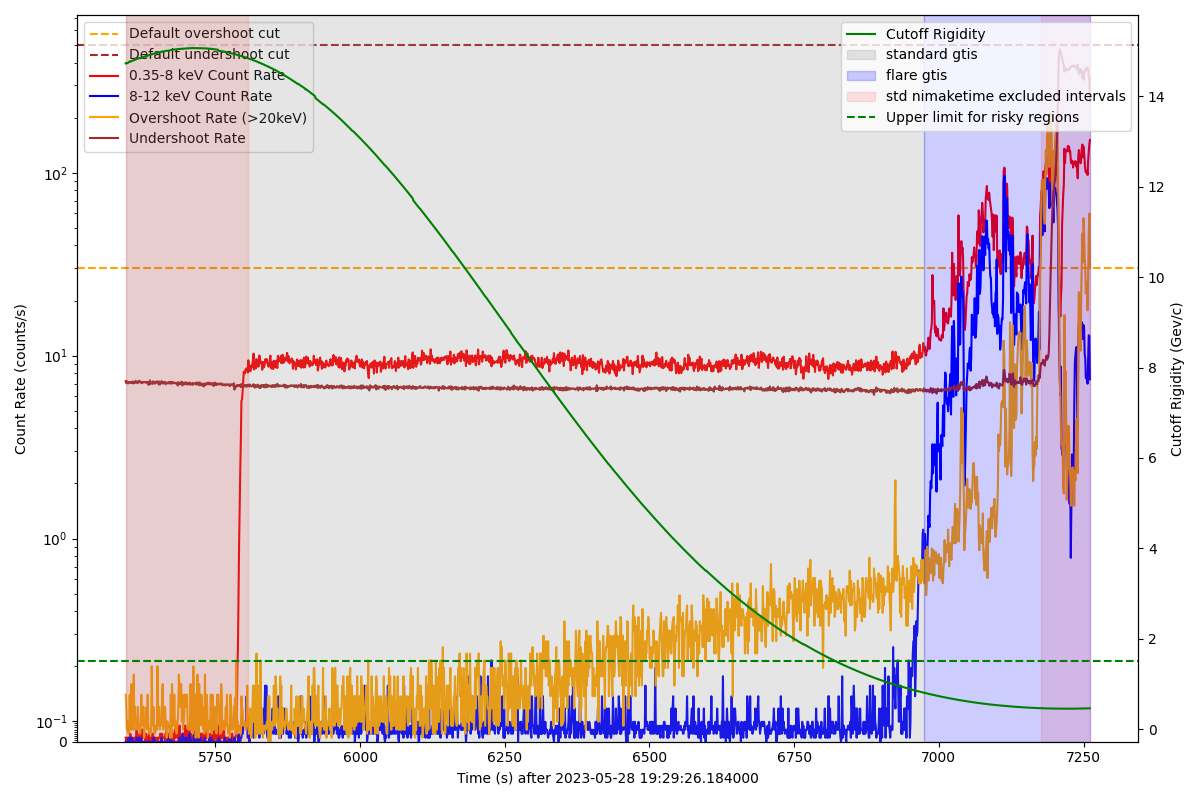}
\vspace{-1.5em}
\caption{Diagnostic plot of several quantities stored in the \nicer{} filter file for a single continuous GTI of ObsID 6130010118. A bright X-ray flare is easily visible at the end of the observation in the soft X-ray (bright red) and high-energy(blue) lightcurves. The automatic filtering (red regions) only removes a part of this flare, whereas the additional screening (blue region) removes it entirely}\label{fig:NICER_flare_example}
\end{figure}

\section{\nicer{} filtering procedure}\label{app:NICER_filtering}

NICER is sensitive to several types of non X-ray flares, which can contaminate the data and significantly change the shape of the continuum. As the standard screening criteria of \texttt{nicerl2} are not always sufficient to exclude entirely (if at all) these periods, we implement an additional screening, following the methodology proposed in the \nicer{} threads\footnote{\href{https://heasarc.gsfc.nasa.gov/docs/nicer/analysis\_threads/flares/}{https://heasarc.gsfc.nasa.gov/docs/nicer/analysis\_threads/flares/}} in order to recognize and exclude precipitating electron flares. We use the information of the filter file of each observation, split independently for each continuous gti, basing on two complementary criteria.  First, we exclude statistically distinct events, which we define as >100 times the median count rate of the 2 sigma clipped high energy (8-12keV) events. Secondly, we search and exclude  significant peaks present in both the 8-12keV and 0.35-8keV event list, using the python library findpeaks\footnote{\href{https://github.com/erdogant/findpeaks}{https://github.com/erdogant/findpeaks}}.\\

The first of these two methods allows us to exclude long flare periods that do not register as peaks through a topological analysis, and the second to extend the exclusion to the ``wings'' of each flare's peak, in order to limit residual contamination. We show an example of the screening in figure \ref{fig:NICER_flare_example}, with a plot of several parameters stored in the filter file of a continuous GTI. The manual screening (blue region) is necessary to complement the automatic \nicer{} screening (red regions), in order to remove the flare visible in all 3 lightcurves towards the end of the observation. These diagnostic plots, as well as the lightcurve products we create for the remaining "cleaned" GTIs, were then used to verify that the flares were adequately removed in all observations.

\newpage
\begin{figure}[h!]
\includegraphics[width=0.49\textwidth]{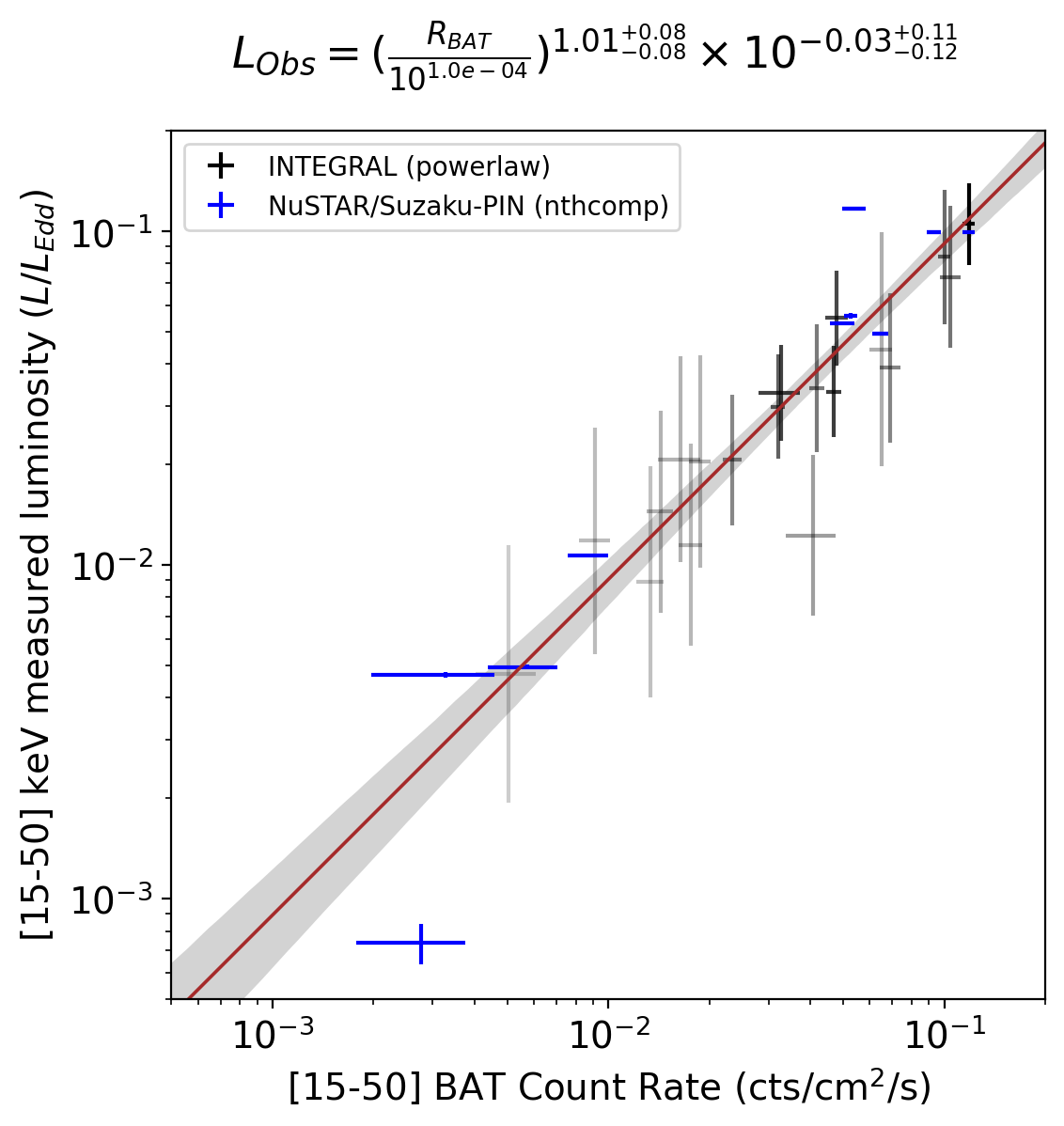}
\includegraphics[width=0.49\textwidth]{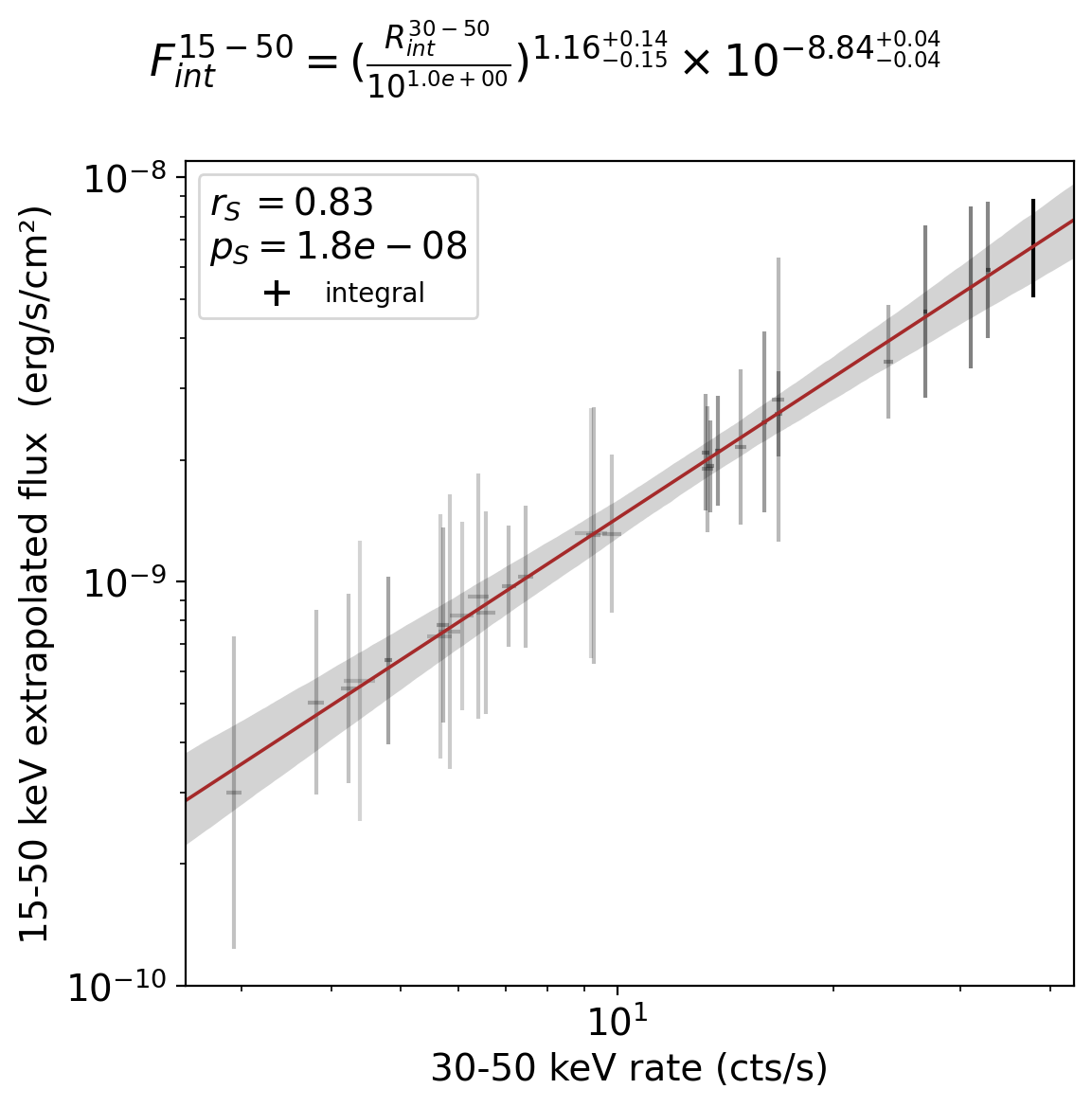}
\vspace{-1.5em}
\caption{\textbf{(Left)} Luminosity of the source in the [15-50] keV band in Eddington units, using BAT flux values estimated via WEBPIMMS, compared to actual measurements from \integral{}, \suzaku{} and \nustar{} datasets.\\
\textbf{(Right)} \integral{} 30-50 count rates versus 15-50 keV projected fluxes in observations of sufficiently high SNR to perform a fit.\\
Uncertainties on both linear regression regions and coefficients are quoted at a 1 $\sigma$ confidence level. 
}\label{fig:BAT_projected_calibration}
\end{figure}

\section{extending the high-energy coverage}\label{app:highE_coverage}

Direct fitting of observations with good SNR at high-energy shows that above $\sim15$ keV, the source is at first order well described by a simple powerlaw or \texttt{nthcomp} component, with a photon index varying between $\sim$ 2-3. Thus, we can draw estimates of the daily source flux in a given band by estimating the SED of a given day, and computing the corresponding conversion factors from the BAT count rate. 

To get the most accurate conversion conversion, we not only compare the BAT count rate with all the measurements from existing high-energy spectra of this source (from the exposures listed in Table\ref{tab:outburst_list}), but also use all of the \integral{} revolutions outside of this epochs, with good enough SNR to get a constrain on the spectral index. For these, we compute flux measurements in the BAT band (15-50 keV) from the measurements in the individual \integral{} fits (obtained with a powerlaw). The result, which we show in the left panel of Fig.~\ref{fig:BAT_projected_calibration},
allows us to obtain good flux estimates down to the edge of the BAT sensitivity, which lies around few $10^{-3} L_{Edd}$ for this source, and coincides with states where the source [15-50] keV flux departs from the standard powerlaw approximation (exemplified by the point with the lowest flux value). 

INTEGRAL can also benefit from the same type of conversion, since only 29 of its 93 revolutions have a good enough SNR to create a spectrum and compute a flux directly. 
We thus directly test whether the ISGRI count rate is sufficiently well correlated with the flux measurements in the "high SNR" observations, once again extrapolating a flux measurement to the BAT band. The result, which we show in the right panel of Fig.~\ref{fig:BAT_projected_calibration}, shows once a gain a very significant and linear correlation, and we thus directly compute a linear regression to convert the \integral{} rates in fluxes in all the observations where the flux cannot be constrained from the fit. As seen in Tab.~\ref{tab:outburst_list}, this allows to add high-energy coverage to 2 observations without BAT coverage, notably in the 2004 outburst with a very high high-energy flux measurements.

\newpage

\section{
Comparison of the wind parameter space in each dimension }\label{app:2D_corner}

\begin{figure*}[h!]
\includegraphics[clip,trim=3cm 0cm 7.5cm 0cm,width=0.99\textwidth]{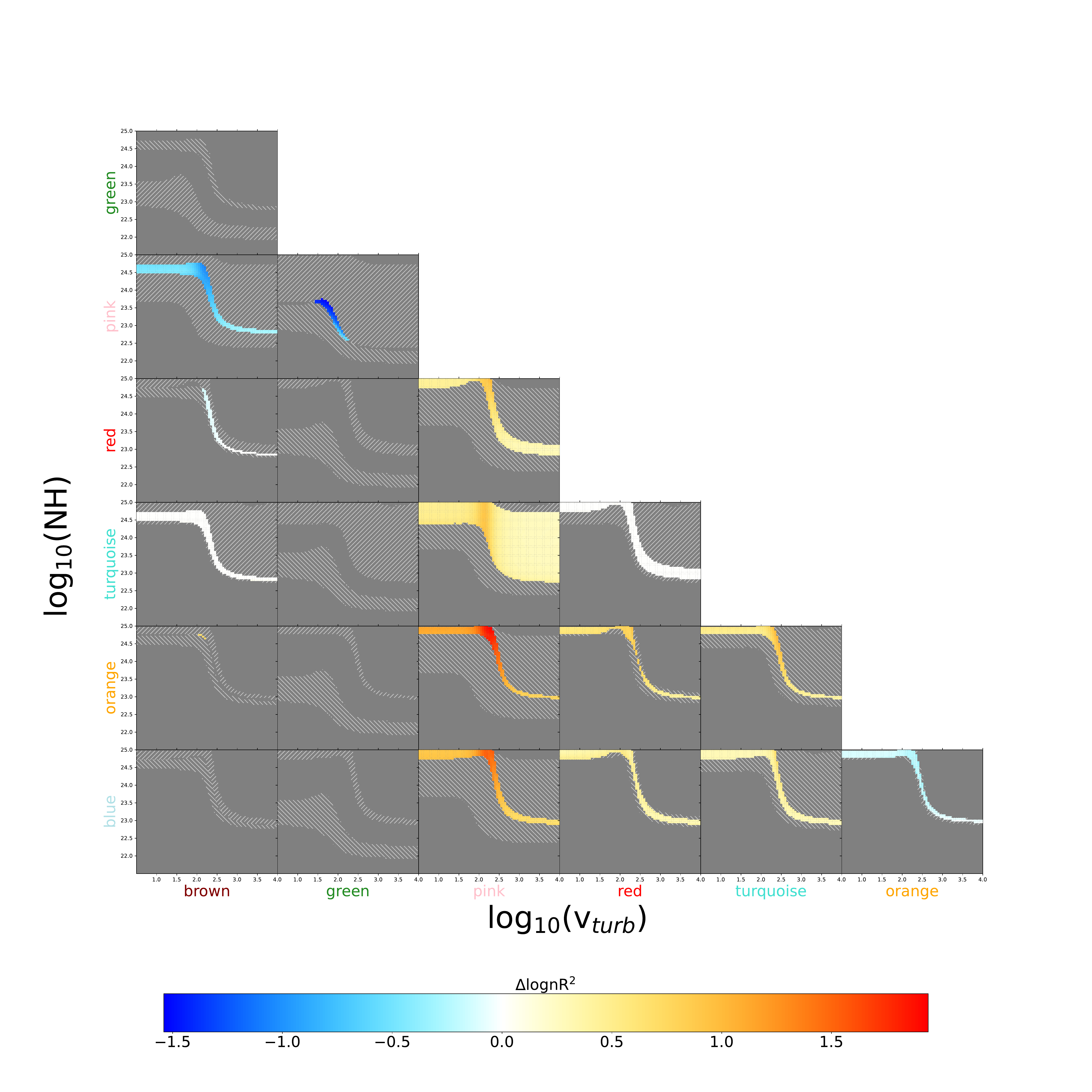}

\caption{Parameter space and potential evolution of the wind in nR² for each couple of observations.
Bottom-left to top-right (top-left to bottom-right) hashes indicate a combination of $v_{turb}$ and NH valid for only for the first (labeled on the y axis) and second (labeled on the x axis) SED, whose nR² gives the positive (negative) contribution in the colormap.
}\label{fig:corner_nR²}
\end{figure*}

\begin{figure*}[h!]
\includegraphics[clip,trim=3cm 0cm 7.5cm 0cm,width=0.99\textwidth]{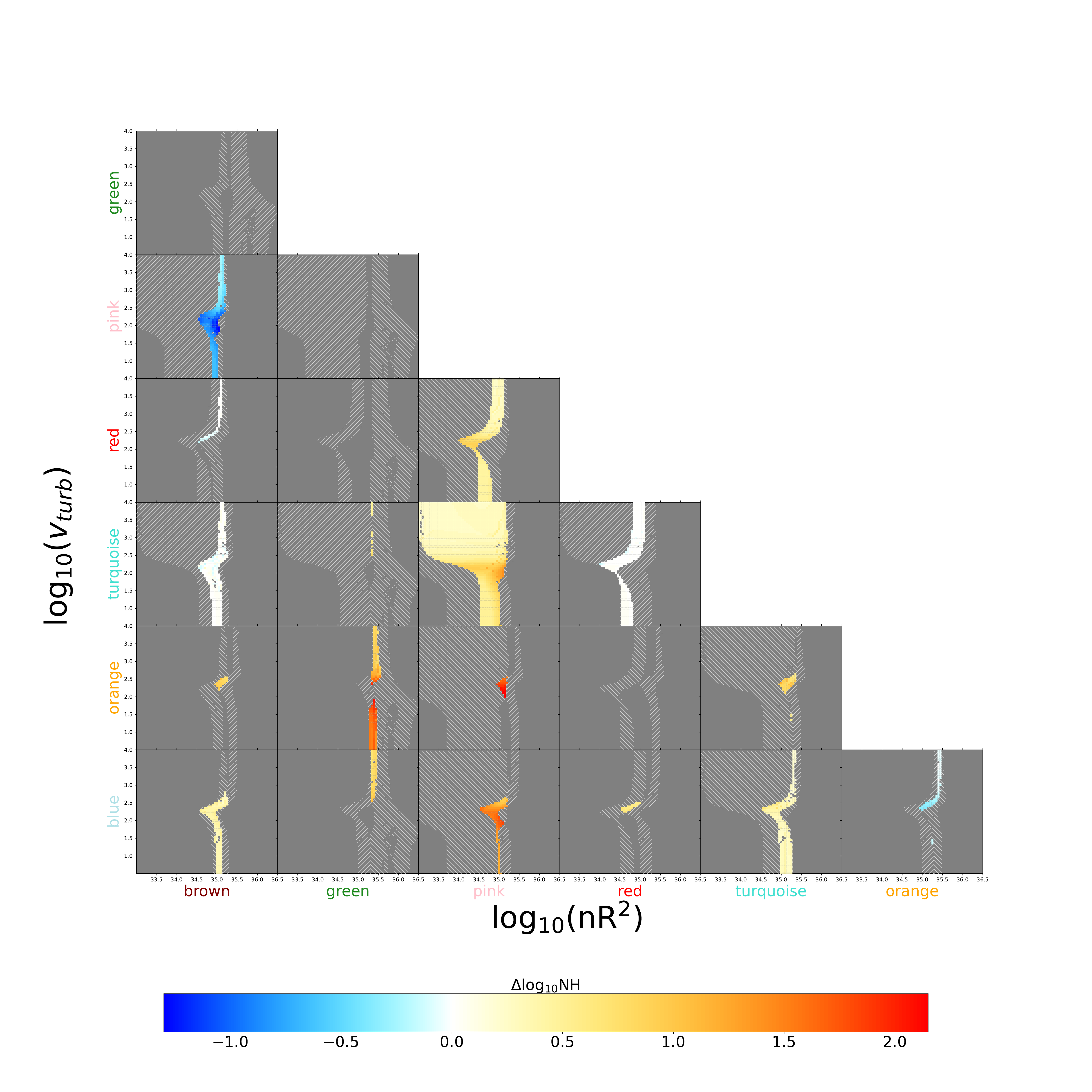}

\caption{Parameter space and potential evolution of the wind in NH for each couple of observations.
Bottom-left to top-right (top-left to bottom-right) hashes indicate a combination of $v_{turb}$ and nR² valid for only for the first (labeled on the y axis) and second (labeled on the x axis) SED, whose NH gives the positive (negative) contribution in the colormap. Since the log(nR²) values are initially continuous, we interpolate them on a grid with steps of 0.05 to give a better idea of the envelope of each shape.
}\label{fig:corner_NH}
\end{figure*}

\begin{figure*}[h!]
\includegraphics[clip,trim=3cm 0cm 7.5cm 0cm,width=0.99\textwidth]{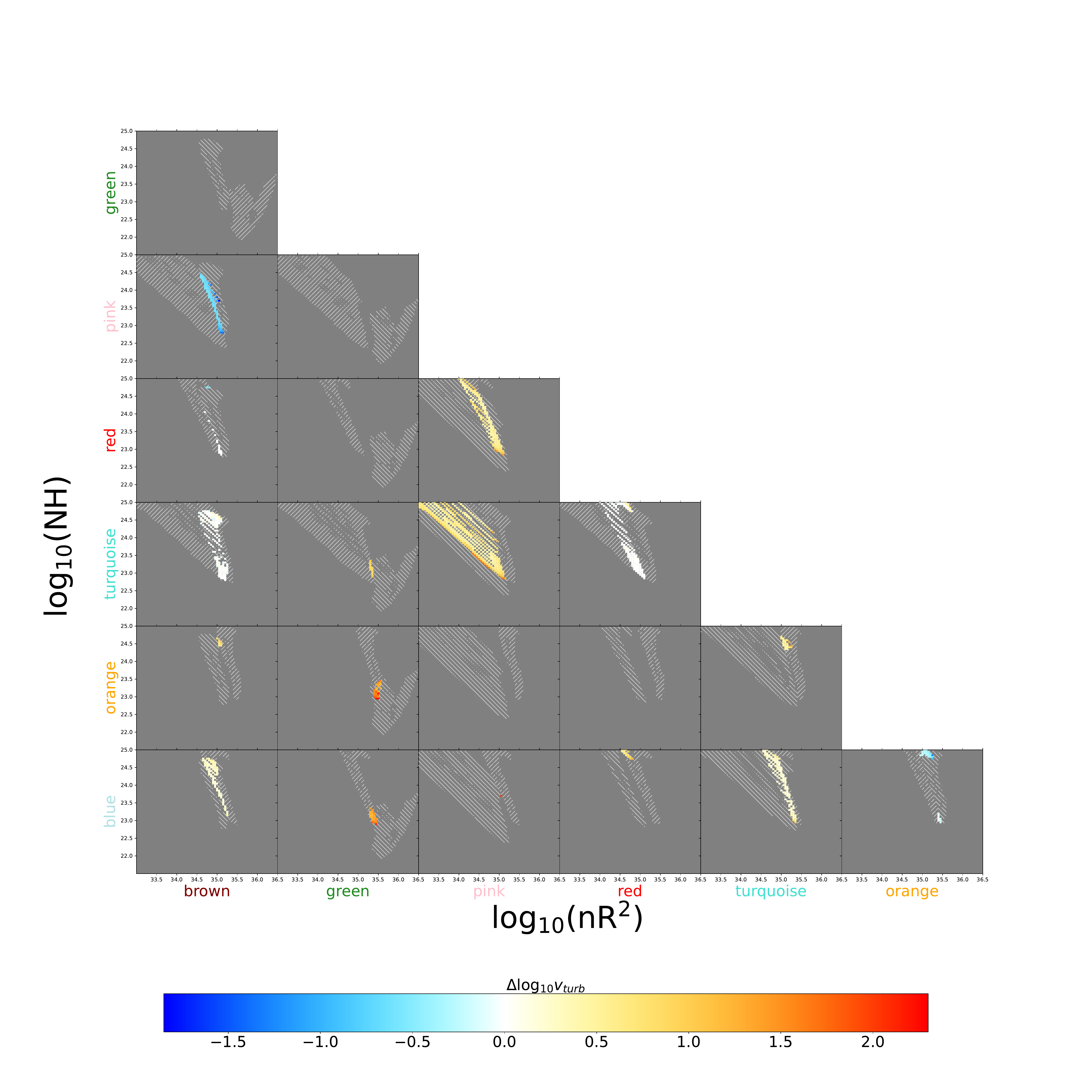}
\caption{Parameter space and potential evolution of the wind in $v_{turb}$ for each couple of observations.
Bottom-left to top-right (top-left to bottom-right) hashes indicate a combination of nR² and NH valid for only for the first (labeled on the y axis) and second (labeled on the x axis) SED, whose $v_{turb}$ gives the positive (negative) contribution in the colormap. Since the log(nR²) values are initially continuous, we interpolate them on a grid with steps of 0.05 to give a better idea of the envelope of each shape.
}\label{fig:corner_v_turb}
\end{figure*}

\section{Results of the line detection procedure for exposures analyzed in this work}\label{app:tables}

Table B.1: EW values for each line and each exposure analyzed in the sample. The sources are ordered alphabetically and with observations listed chronologically. We report EW results for detections or EW upper limits for non-detections of the main lines. Line EWs are only provided for detections above 3 $\sigma$ significance along with 90\% uncertainties. Upper limits above 100 eV are not reported. For NICER, the exposure time refers to the total exposure time of all the individual orbits selected in the epoch, and the ObsID marks the main ObsID of that epoch (see Sect.~\ref{sec:spectral_analysis}). NICER epochs marked with a dagger symbol ($\dagger$) are result of manual selection due to too high variability in the individual orbits of the day (see Sect.~\ref{sec:spectral_analysis})
{
\vspace*{-0.5em}

\tiny

\centering
\begin{longtable}[h]{c c || c c c || c c c c c}

\hline 
\hline
\multirow{2}{*}{Outburst}
& \multirow{2}{*}{Date}
& \multirow{2}{*}{Instrument}
& \multirow{2}{*}{ObsID + identifier }
& \multirow{2}{*}{exp. time (ks)}
& \multicolumn{5}{c}{Fe line Equivalent Width / 3 $\sigma$ upper limit (eV)}
\T \B \\

& 
& 
& 
& 
& xxvK$\alpha$
& xxviK$\alpha$
& xxvK$\beta$
& xxviK$\beta$
& xxviK$\gamma$
\T \B \\
\hline
\hline
\endfirsthead
\caption{continued}\\
\hline
\hline
\multirow{2}{*}{Source}
& \multirow{2}{*}{Date}
& \multirow{2}{*}{Instrument}
& \multirow{2}{*}{ObsID + identifier }
& \multirow{2}{*}{exp. time (ks)}
& \multicolumn{5}{c}{Fe line Equivalent Width / 3 $\sigma$ upper limit (eV)}
\T \B \\

& 
& 
& 
& 
& xxvK$\alpha$
& xxviK$\alpha$
& xxvK$\beta$
& xxviK$\beta$
& xxviK$\gamma$
\T \B \\
\hline
\hline
\endhead

\hline\multirow{1}{*}{2002/2004}&2004-08-04&Chandra&4568&49.99&$\leq4$&\textbf{11}$_{-4}^{+3}$&$\leq19$&$\leq76$&/\T \B \\ 
\hline\multirow{6}{*}{2005/2006}&2006-02-08&Suzaku&400010010&11.06&$\textbf{9}\pm3$&$\textbf{36}\pm3$&$\leq17$&\textbf{17}$_{-5}^{+7}$&$\leq22$\T \B \\ 
&2006-02-15&Suzaku&400010020&10.7&\textbf{19}$_{-4}^{+3}$&\textbf{37}$_{-3}^{+4}$&$\leq25$&\textbf{22}$_{-7}^{+8}$&$\leq27$\T \B \\ 
&2006-02-28&Suzaku&400010030&10.72&$\textbf{20}\pm4$&$\textbf{34}\pm4$&$\leq24$&$\leq36$&$\leq27$\T \B \\ 
&2006-03-08&Suzaku&400010040&10.66&\textbf{25}$_{-5}^{+4}$&\textbf{36}$_{-4}^{+5}$&\textbf{32}$_{-6}^{+9}$&$\leq41$&$\leq39$\T \B \\ 
&2006-03-15&Suzaku&400010050&23.18&$\textbf{22}\pm3$&$\textbf{36}\pm3$&$\leq17$&$\leq20$&$\leq19$\T \B \\ 
&2006-03-23&Suzaku&400010060&21.65&\textbf{25}$_{-3}^{+4}$&$\textbf{38}\pm4$&$\leq23$&$\leq31$&$\leq22$\T \B \\ 
\hline Out of outburst&2010-08-24&Suzaku&405051010&99.92&$\leq66$&$\leq77$&$\leq79$&/&/\T \B \\ 
\hline\multirow{19}{*}{2011/2013}&2012-01-17&Chandra&13714&28.92&$\textbf{32}\pm4$&$\textbf{57}\pm5$&/&/&/\T \B \\ 
&2012-01-20&Chandra&13715&29.28&\textbf{34}$_{-5}^{+3}$&\textbf{49}$_{-5}^{+4}$&\textbf{23}$_{-5}^{+7}$&/&/\T \B \\ 
&2012-01-26&Chandra&13716&29.28&\textbf{47}$_{-2}^{+3}$&\textbf{52}$_{-3}^{+1}$&\textbf{32}$_{-2}^{+11}$&\textbf{38}$_{-6}^{+9}$&/\T \B \\ 
&2012-01-30&Chandra&13717&29.44&$\textbf{30}\pm3$&$\textbf{48}\pm4$&\textbf{32}$_{-7}^{+14}$&\textbf{35}$_{-11}^{+10}$&\textbf{36}$_{-8}^{+13}$\T \B \\ 
&2012-02-13&Suzaku&906008010&7.71&\textbf{39}$_{-4}^{+3}$&$\textbf{56}\pm4$&\textbf{50}$_{-6}^{+9}$&\textbf{57}$_{-9}^{+7}$&\textbf{33}$_{-7}^{+8}$\T \B \\ 
&2012-03-04&XMM&0670671501\_S003&2.54&$\textbf{35}\pm7$&\textbf{55}$_{-7}^{+9}$&$\leq34$&/&/\T \B \\ 
&2012-03-04&XMM&0670671501\_U014&69.86&$\textbf{31}\pm2$&\textbf{48}$_{-2}^{+1}$&\textbf{21}$_{-1}^{+3}$&\textbf{22}$_{-0}^{+3}$&/\T \B \\ 
&2012-03-20&XMM&0670671301\_S003&22.26&$\textbf{21}\pm3$&\textbf{46}$_{-2}^{+4}$&$\leq16$&\textbf{18}$_{-5}^{+6}$&/\T \B \\ 
&2012-03-25&XMM&0670672901\_S003&62.81&$\textbf{20}\pm1$&\textbf{45}$_{-2}^{+1}$&$\textbf{9}\pm2$&$\textbf{19}\pm2$&/\T \B \\ 
&2012-06-03&Chandra&14441&19.0&$\leq12$&$\leq14$&$\leq34$&$\leq42$&$\leq62$\T \B \\ 
&2012-09-09&XMM&0670673001\_S003&22.48&\textbf{9}$_{-2}^{+3}$&\textbf{31}$_{-2}^{+3}$&$\leq10$&$\leq10$&$\leq10$\T \B \\ 
&2012-09-10&XMM&0670673001\_U002&0.8&$\leq19$&\textbf{25}$_{-5}^{+6}$&$\leq17$&$\leq22$&$\leq23$\T \B \\ 
&2012-09-11&XMM&0670673101\_S003&0.93&$\leq14$&\textbf{9}$_{-4}^{+5}$&$\leq19$&$\leq16$&$\leq19$\T \B \\ 
&2012-09-28&XMM&0670673201\_S003&1.56&$\leq5$&$\leq7$&$\leq9$&$\leq11$&$\leq19$\T \B \\ 
&2012-10-02&Suzaku&907003010&2.88&$\leq8$&$\leq9$&$\leq15$&$\leq14$&$\leq48$\T \B \\ 
&2013-02-20&NuSTAR&40014008002&16.57&$\leq21$&\textbf{30}$_{-2}^{+3}$&\textbf{13}$_{-3}^{+4}$&$\leq18$&$\leq8$\T \B \\ 
&2013-02-21&NuSTAR&40014009001&14.65&$\leq24$&\textbf{37}$_{-1}^{+0}$&$\leq6$&$\leq8$&$\leq7$\T \B \\ 
&2013-04-25&Chandra&15511&49.39&$\leq8$&$\leq7$&$\leq26$&$\leq32$&$\leq35$\T \B \\ 
&2013-05-27&Chandra&15524&48.91&$\leq44$&$\leq62$&/&/&/\T \B \\ 
\hline\multirow{3}{*}{2015}&2015-02-20&Suzaku&409007010&6.04&$\leq9$&\textbf{17}$_{-3}^{+4}$&$\leq12$&$\leq18$&$\leq18$\T \B \\ 
&2015-02-24&Suzaku&409007020&5.57&$\textbf{18}\pm3$&$\textbf{44}\pm5$&$\leq27$&\textbf{21}$_{-6}^{+9}$&$\leq19$\T \B \\ 
&2015-02-27&Suzaku&409007030&5.15&$\leq19$&\textbf{38}$_{-3}^{+4}$&$\leq23$&$\leq24$&$\leq19$\T \B \\ 
\hline\multirow{1}{*}{2016/2017}&2016-10-21&Chandra&19904&30.93&\textbf{23}$_{-5}^{+4}$&\textbf{45}$_{-7}^{+4}$&/&/&/\T \B \\ 
\hline\multirow{33}{*}{2018/2019}&2018-06-11&NICER&1130010104&1.96&\textbf{31}$_{-7}^{+12}$&\textbf{52}$_{-8}^{+12}$&/&/&/\T \B \\ 
&2018-07-02&NICER&1130010105&1.48&\textbf{19}$_{-5}^{+6}$&\textbf{41}$_{-6}^{+5}$&$\leq35$&/&/\T \B \\ 
&2018-07-04&NICER&1130010106&0.64&\textbf{18}$_{-7}^{+10}$&\textbf{46}$_{-8}^{+9}$&/&\textbf{37}$_{-12}^{+16}$&/\T \B \\ 
&2018-07-05&NICER&1130010107&0.09&$\leq69$&/&/&/&/\T \B \\ 
&2018-07-07&NICER&1130010109&0.41&\textbf{21}$_{-8}^{+12}$&\textbf{42}$_{-10}^{+13}$&$\leq54$&/&/\T \B \\ 
&2018-07-10&NICER&1130010111&1.79&\textbf{21}$_{-4}^{+6}$&$\textbf{42}\pm5$&/&/&/\T \B \\ 
&2018-07-13&NICER&1130010113&1.36&\textbf{23}$_{-4}^{+6}$&\textbf{51}$_{-5}^{+6}$&\textbf{24}$_{-9}^{+11}$&/&/\T \B \\ 
&2018-07-14&NICER&1130010114&2.07&$\textbf{20}\pm4$&$\textbf{50}\pm5$&\textbf{19}$_{-6}^{+9}$&$\textbf{37}\pm10$&/\T \B \\ 
&2018-07-16&NICER&1130010116&0.94&\textbf{25}$_{-6}^{+9}$&\textbf{51}$_{-3}^{+10}$&/&/&/\T \B \\ 
&2018-07-17&NICER&1130010117&1.52&$\textbf{25}\pm5$&\textbf{51}$_{-6}^{+7}$&/&/&/\T \B \\ 
&2018-07-21&NICER&1130010119&1.63&\textbf{22}$_{-4}^{+5}$&$\textbf{43}\pm5$&/&\textbf{28}$_{-8}^{+14}$&/\T \B \\ 
&2018-07-22&NICER&1130010120&1.01&\textbf{19}$_{-5}^{+7}$&\textbf{50}$_{-6}^{+7}$&/&/&/\T \B \\ 
&2018-07-23&NICER&1130010121&2.84&\textbf{20}$_{-5}^{+4}$&\textbf{48}$_{-5}^{+4}$&/&\textbf{28}$_{-8}^{+10}$&/\T \B \\ 
&2018-07-25&NICER&1130010123&1.19&\textbf{21}$_{-5}^{+6}$&$\textbf{49}\pm6$&/&/&/\T \B \\ 
&2018-07-26&NICER&1130010124&1.14&$\textbf{18}\pm6$&\textbf{44}$_{-5}^{+7}$&/&/&/\T \B \\ 
&2018-07-28&NICER&1130010125&0.79&\textbf{18}$_{-6}^{+8}$&\textbf{38}$_{-8}^{+7}$&$\leq41$&/&/\T \B \\ 
&2018-08-02&NICER&1130010126&2.02&$\leq10$&\textbf{38}$_{-5}^{+4}$&$\leq17$&\textbf{18}$_{-7}^{+9}$&/\T \B \\ 
&2018-08-04&NICER&1130010128&2.34&$\leq12$&\textbf{44}$_{-5}^{+4}$&$\leq25$&\textbf{23}$_{-8}^{+10}$&/\T \B \\ 
&2018-08-06&NICER&1130010130&2.04&$\leq14$&\textbf{40}$_{-3}^{+4}$&$\leq17$&$\leq25$&/\T \B \\ 
&2018-08-08&NICER&1130010132&12.06&\textbf{15}$_{-2}^{+1}$&$\textbf{44}\pm2$&/&$\textbf{18}\pm3$&/\T \B \\ 
&2018-08-10&NICER&1130010134&3.36&$\textbf{17}\pm4$&\textbf{47}$_{-3}^{+4}$&$\textbf{19}\pm5$&\textbf{20}$_{-7}^{+10}$&/\T \B \\ 
&2018-08-11&NICER&1130010135&4.82&$\textbf{19}\pm3$&\textbf{45}$_{-2}^{+3}$&/&\textbf{22}$_{-7}^{+8}$&/\T \B \\ 
&2018-08-17&NICER&1130010137&0.12&$\leq55$&$\leq82$&/&/&/\T \B \\ 
&2018-08-20&NICER&1130010139&7.75&$\textbf{22}\pm2$&\textbf{47}$_{-2}^{+3}$&/&\textbf{22}$_{-6}^{+5}$&/\T \B \\ 
&2018-08-22&NICER&1130010141&2.32&\textbf{26}$_{-6}^{+4}$&$\textbf{46}\pm5$&/&$\textbf{26}\pm10$&/\T \B \\ 
&2018-09-24&NICER&1130010142&3.64&$\leq7$&$\leq11$&$\leq15$&$\leq22$&$\leq29$\T \B \\ 
&2018-09-25&NICER&1130010143&0.94&$\leq16$&$\leq17$&$\leq32$&$\leq41$&$\leq57$\T \B \\ 
&2018-09-28&NICER&1130010144&2.23&$\leq18$&$\leq17$&$\leq32$&$\leq33$&$\leq52$\T \B \\ 
&2018-09-29&NICER&1130010145&1.99&$\leq15$&$\leq14$&$\leq23$&$\leq38$&$\leq53$\T \B \\ 
&2018-10-02&NICER&1130010146&2.65&$\leq12$&$\leq14$&$\leq32$&$\leq34$&$\leq56$\T \B \\ 
&2018-10-04&NICER&1130010147&4.16&$\leq14$&$\leq26$&$\leq29$&$\leq34$&$\leq39$\T \B \\ 
&2018-10-05&NICER&1130010148&3.06&$\leq22$&\textbf{28}$_{-6}^{+9}$&$\leq47$&$\leq44$&/\T \B \\ 
&2018-10-07&NICER&1130010150&0.9&$\leq44$&$\leq59$&$\leq74$&$\leq70$&/\T \B \\ 
\hline\multirow{42}{*}{2020}&2020-03-19&NICER&3130010101&0.57&$\leq22$&$\leq29$&$\leq36$&$\leq59$&$\leq65$\T \B \\ 
&2020-03-26&NICER&3130010102&0.88&$\leq16$&$\leq17$&$\leq25$&/&$\leq43$\T \B \\ 
&2020-03-27&NICER&3130010103&2.18&$\leq16$&$\leq18$&$\leq26$&$\leq33$&$\leq43$\T \B \\ 
&2020-03-30&NICER&3130010105&0.43&$\leq23$&$\leq29$&$\leq46$&$\leq45$&$\leq74$\T \B \\ 
&2020-04-01&NICER&3130010106&2.51&$\leq7$&\textbf{10}$_{-4}^{+5}$&$\leq14$&$\leq16$&$\leq25$\T \B \\ 
&2020-04-03&NICER&3130010108&2.47&$\leq9$&\textbf{24}$_{-4}^{+5}$&$\leq15$&$\leq19$&$\leq32$\T \B \\ 
&2020-04-06&NICER&3130010110&0.76&$\leq14$&\textbf{29}$_{-6}^{+9}$&$\leq25$&$\leq43$&$\leq42$\T \B \\ 
&2020-04-07&NICER&3130010111&2.56&$\leq11$&\textbf{22}$_{-4}^{+5}$&$\leq15$&$\leq25$&$\leq27$\T \B \\ 
&2020-04-09&NICER&3130010112&1.98&$\leq8$&\textbf{28}$_{-5}^{+4}$&$\leq16$&$\leq23$&$\leq26$\T \B \\ 
&2020-04-11&NICER&3130010114&0.39&$\leq18$&$\leq42$&$\leq54$&$\leq71$&$\leq75$\T \B \\ 
&2020-04-13&NICER&3130010116&1.06&$\leq14$&\textbf{27}$_{-5}^{+8}$&$\leq22$&$\leq40$&$\leq49$\T \B \\ 
&2020-04-14&NICER&3130010117&0.83&$\leq18$&$\leq23$&$\leq36$&$\leq42$&$\leq52$\T \B \\ 
&2020-04-15&NICER&3130010118&0.97&$\leq12$&$\textbf{24}\pm7$&$\leq32$&$\leq34$&$\leq45$\T \B \\ 
&2020-04-17&NICER&3130010119&0.4&$\leq24$&$\leq21$&$\leq39$&$\leq66$&$\leq72$\T \B \\ 
&2020-04-18&NICER&3130010120&1.47&$\leq19$&$\leq21$&$\leq37$&$\leq43$&$\leq56$\T \B \\ 
&2020-04-20&NICER&3130010122&1.05&$\leq12$&$\leq14$&$\leq22$&$\leq33$&$\leq54$\T \B \\ 
&2020-04-22&NICER&3130010124&0.62&$\leq22$&$\leq24$&$\leq26$&$\leq38$&$\leq48$\T \B \\ 
&2020-04-23&NICER&3130010125&0.95&$\leq18$&$\leq20$&$\leq22$&$\leq35$&$\leq41$\T \B \\ 
&2020-04-24&NICER&3130010126&0.6&$\leq29$&$\leq19$&$\leq46$&$\leq58$&$\leq55$\T \B \\ 
&2020-04-26&NICER&3130010127&1.49&$\leq16$&$\leq20$&$\leq30$&$\leq31$&$\leq43$\T \B \\ 
&2020-04-27&NICER&3130010128&2.64&$\leq8$&$\leq11$&$\leq14$&$\leq15$&$\leq29$\T \B \\ 
&2020-04-28&NICER&3130010129&2.04&$\leq11$&$\leq11$&$\leq15$&$\leq26$&$\leq38$\T \B \\ 
&2020-04-30&NICER&3130010131&3.15&$\leq8$&$\leq9$&$\leq13$&$\leq16$&$\leq23$\T \B \\ 
&2020-05-01&NICER&3130010132&1.4&$\leq21$&$\leq23$&$\leq35$&$\leq35$&$\leq57$\T \B \\ 
&2020-05-02&NICER&3130010133&2.61&$\leq15$&$\leq18$&$\leq26$&$\leq33$&$\leq42$\T \B \\ 
&2020-05-03&NICER&3130010134&2.04&$\leq18$&$\leq19$&$\leq30$&$\leq38$&$\leq43$\T \B \\ 
&2020-05-05&NICER&3130010136&2.6&$\leq10$&$\leq9$&$\leq16$&$\leq18$&$\leq32$\T \B \\ 
&2020-05-06&NICER&3130010137&2.25&$\leq11$&$\leq12$&$\leq26$&$\leq22$&$\leq35$\T \B \\ 
&2020-05-07&NICER&3130010138&1.87&$\leq20$&$\leq19$&$\leq34$&$\leq43$&$\leq51$\T \B \\ 
&2020-05-09&NICER&3130010140&1.99&$\leq10$&$\leq11$&$\leq18$&$\leq30$&$\leq33$\T \B \\ 
&2020-05-11&NICER&3130010141&1.04&$\leq26$&$\leq30$&$\leq51$&$\leq60$&$\leq79$\T \B \\ 
&2020-05-15&NICER&3130010143&0.17&$\leq47$&$\leq56$&/&/&/\T \B \\ 
&2020-05-22&NICER&3130010144&2.92&$\leq20$&$\leq20$&$\leq42$&$\leq45$&$\leq52$\T \B \\ 
&2020-05-28&Chandra&22376&24.5&$\leq35$&$\leq35$&$\leq75$&/&/\T \B \\ 
&2020-05-28&NICER&3130010145&1.23&$\leq35$&$\leq30$&$\leq54$&$\leq70$&$\leq78$\T \B \\ 
&2020-05-31&NICER&3130010146&0.76&$\leq39$&$\leq42$&/&/&/\T \B \\ 
&2020-06-01&NICER&3130010147&1.63&$\leq26$&$\leq30$&$\leq63$&$\leq97$&/\T \B \\ 
&2020-06-04&NICER&3130010148&1.51&$\leq34$&$\leq56$&$\leq73$&$\leq97$&/\T \B \\ 
&2020-06-06&Chandra&22377&24.5&$\leq41$&$\leq50$&/&/&/\T \B \\ 
&2020-06-08&NICER&3130010149&0.61&/&/&/&/&/\T \B \\ 
&2020-06-13&Chandra&22378&23.54&$\leq39$&$\leq55$&/&/&/\T \B \\ 
&2020-06-14&NICER&3130010151&2.05&$\leq48$&$\leq56$&/&/&/\T \B \\ 
\hline\multirow{29}{*}{2021/2022}&2021-09-13&NICER&4130010101&5.88&$\leq23$&$\leq29$&$\leq31$&$\leq42$&$\leq58$\T \B \\ 
&2021-09-14&NICER&4130010102&5.01&$\leq25$&$\leq28$&$\leq38$&$\leq36$&$\leq53$\T \B \\ 
&2021-09-16&NICER&4130010104&3.96&$\leq15$&$\leq24$&$\leq20$&$\leq25$&$\leq29$\T \B \\ 
&2021-09-17&NICER&4130010105&3.27&$\leq28$&$\leq29$&$\leq42$&$\leq48$&$\leq45$\T \B \\ 
&2021-09-18&NICER&4130010106&4.42&$\leq19$&$\leq23$&$\leq32$&$\leq35$&$\leq44$\T \B \\ 
&2021-09-19&NICER&4130010107&5.5&$\leq17$&$\leq20$&$\leq25$&$\leq29$&$\leq36$\T \B \\ 
&2021-09-20&NICER&4130010108&3.21&$\leq18$&$\leq21$&$\leq29$&$\leq33$&$\leq45$\T \B \\ 
&2021-09-22&NICER&4130010110&1.37&$\leq27$&$\leq30$&$\leq44$&$\leq48$&$\leq57$\T \B \\ 
&2021-09-23&NICER&4130010111$\dagger$&0.61&$\leq17$&$\leq18$&$\leq20$&$\leq34$&$\leq33$\T \B \\ 
&2021-09-23&NICER&4130010111$\dagger$&2.09&$\leq12$&$\leq16$&$\leq24$&$\leq24$&$\leq37$\T \B \\ 
&2021-09-24&NICER&4130010112$\dagger$&2.03&$\leq24$&$\leq29$&$\leq38$&$\leq46$&$\leq55$\T \B \\ 
&2021-09-24&NICER&4130010112$\dagger$&1.87&$\leq23$&$\leq31$&$\leq38$&$\leq43$&$\leq62$\T \B \\ 
&2021-09-26&NICER&4130010114$\dagger$&1.48&$\leq16$&$\leq19$&$\leq27$&$\leq25$&$\leq44$\T \B \\ 
&2021-09-26&NICER&4130010114$\dagger$&0.55&$\leq17$&$\leq16$&$\leq32$&$\leq33$&$\leq32$\T \B \\ 
&2021-09-27&NICER&4130010115&1.48&$\leq9$&$\leq9$&$\leq20$&$\leq19$&$\leq18$\T \B \\ 
&2021-09-28&NICER&4130010116$\dagger$&1.1&$\leq9$&$\leq8$&$\leq19$&$\leq18$&$\leq22$\T \B \\ 
&2021-09-28&NICER&4130010116$\dagger$&1.18&$\leq14$&$\leq17$&$\leq24$&$\leq27$&$\leq32$\T \B \\ 
&2021-09-29&NICER&4130010118&3.08&$\leq7$&$\leq7$&$\leq9$&$\leq10$&$\leq17$\T \B \\ 
&2021-10-01&NICER&4130010119$\dagger$&1.5&$\leq14$&$\leq14$&$\leq23$&$\leq23$&$\leq34$\T \B \\ 
&2021-10-01&NICER&4130010119$\dagger$&2.73&$\leq13$&$\leq13$&$\leq22$&$\leq23$&$\leq30$\T \B \\ 
&2021-10-11&NICER&4130010121&1.58&$\leq17$&$\leq17$&$\leq31$&$\leq32$&$\leq42$\T \B \\ 
&2021-10-12&NICER&4130010122&0.79&$\leq24$&$\leq26$&$\leq40$&$\leq44$&$\leq61$\T \B \\ 
&2021-10-13&NICER&4130010123&0.69&$\leq13$&$\leq19$&$\leq23$&$\leq34$&$\leq36$\T \B \\ 
&2021-10-14&NICER&4130010124&1.86&$\leq17$&$\leq17$&$\leq31$&$\leq37$&$\leq35$\T \B \\ 
&2021-10-21&NICER&4130010126&0.76&$\leq24$&$\leq30$&$\leq42$&$\leq52$&$\leq44$\T \B \\ 
&2021-10-23&NICER&4130010127&0.91&$\leq28$&$\leq31$&$\leq50$&$\leq55$&$\leq64$\T \B \\ 
&2021-10-25&NICER&4130010128&1.32&$\leq21$&$\leq21$&$\leq20$&$\leq29$&$\leq52$\T \B \\ 
&2021-10-28&NICER&4130010130&1.01&$\leq22$&$\leq22$&$\leq26$&$\leq49$&$\leq53$\T \B \\ 
&2022-02-19&NICER&4130010131&0.3&/&/&/&/&/\T \B \\ 
\hline\multirow{75}{*}{2022/2024}&2022-07-30&NICER&5130010101&2.14&\textbf{18}$_{-9}^{+8}$&\textbf{39}$_{-9}^{+10}$&$\leq58$&$\leq76$&/\T \B \\ 
&2022-07-31&NICER&5130010102&0.35&$\leq63$&\textbf{57}$_{-21}^{+30}$&/&/&/\T \B \\ 
&2022-08-06&NICER&5130010103&3.32&\textbf{23}$_{-5}^{+3}$&\textbf{41}$_{-8}^{+1}$&/&/&/\T \B \\ 
&2022-08-07&NICER&5130010104&3.55&\textbf{22}$_{-8}^{+3}$&\textbf{39}$_{-6}^{+5}$&/&/&/\T \B \\ 
&2022-08-08&NICER&5130010105&3.48&$\textbf{21}\pm4$&\textbf{37}$_{-4}^{+5}$&$\leq25$&/&/\T \B \\ 
&2022-08-09&NICER&5130010106&0.35&\textbf{28}$_{-12}^{+18}$&\textbf{46}$_{-12}^{+21}$&/&/&/\T \B \\ 
&2022-08-11&NICER&5130010108&0.06&/&/&/&/&/\T \B \\ 
&2022-08-13&NICER&5130010109&0.21&$\leq47$&$\leq100$&/&/&/\T \B \\ 
&2022-08-14&NICER&5130010110&3.36&$\textbf{23}\pm4$&\textbf{41}$_{-4}^{+5}$&/&/&/\T \B \\ 
&2022-08-15&NICER&5130010111&0.95&\textbf{19}$_{-6}^{+8}$&$\textbf{35}\pm8$&/&\textbf{41}$_{-15}^{+17}$&/\T \B \\ 
&2022-08-18&NICER&5130010112&0.4&$\leq40$&$\textbf{36}\pm11$&$\leq64$&$\leq76$&/\T \B \\ 
&2022-08-22&NICER&5501010101&1.91&$\textbf{18}\pm5$&$\textbf{49}\pm5$&$\textbf{26}\pm8$&/&/\T \B \\ 
&2022-08-23&NICER&5501010102&0.58&\textbf{20}$_{-8}^{+9}$&\textbf{32}$_{-9}^{+10}$&$\leq59$&/&/\T \B \\ 
&2022-08-25&NICER&5501010104&3.92&\textbf{19}$_{-4}^{+3}$&\textbf{47}$_{-4}^{+2}$&/&$\textbf{18}\pm7$&/\T \B \\ 
&2022-08-26&NICER&5501010105&3.06&\textbf{25}$_{-3}^{+4}$&$\textbf{47}\pm4$&\textbf{25}$_{-5}^{+8}$&\textbf{28}$_{-7}^{+9}$&\textbf{27}$_{-11}^{+10}$\T \B \\ 
&2022-08-27&NICER&5501010106&2.02&\textbf{22}$_{-4}^{+5}$&\textbf{47}$_{-4}^{+6}$&\textbf{20}$_{-8}^{+7}$&\textbf{35}$_{-8}^{+12}$&\textbf{29}$_{-13}^{+16}$\T \B \\ 
&2022-08-28&NICER&5501010107&2.8&\textbf{26}$_{-4}^{+5}$&\textbf{48}$_{-3}^{+5}$&$\textbf{21}\pm7$&$\textbf{20}\pm9$&/\T \B \\ 
&2022-08-29&NICER&5501010108&2.31&\textbf{26}$_{-4}^{+5}$&\textbf{44}$_{-3}^{+5}$&\textbf{19}$_{-5}^{+10}$&\textbf{36}$_{-8}^{+9}$&/\T \B \\ 
&2022-08-30&NICER&5501010109&2.27&\textbf{22}$_{-4}^{+5}$&$\textbf{44}\pm5$&\textbf{20}$_{-7}^{+8}$&\textbf{32}$_{-7}^{+10}$&\textbf{31}$_{-9}^{+15}$\T \B \\ 
&2022-08-31&NICER&5501010110&4.45&$\textbf{24}\pm3$&$\textbf{44}\pm3$&$\textbf{27}\pm7$&\textbf{34}$_{-6}^{+8}$&\textbf{25}$_{-9}^{+11}$\T \B \\ 
&2022-09-01&NICER&5501010111&0.48&\textbf{29}$_{-9}^{+10}$&\textbf{61}$_{-9}^{+13}$&/&\textbf{62}$_{-18}^{+23}$&/\T \B \\ 
&2022-09-18&NICER&5130010114&3.78&\textbf{31}$_{-3}^{+4}$&\textbf{51}$_{-3}^{+5}$&\textbf{23}$_{-7}^{+6}$&\textbf{24}$_{-11}^{+10}$&/\T \B \\ 
&2022-09-20&NICER&5130010116&3.32&\textbf{33}$_{-5}^{+3}$&\textbf{45}$_{-7}^{+4}$&/&/&/\T \B \\ 
&2022-09-27&NICER&5130010118&3.87&$\textbf{28}\pm4$&$\textbf{57}\pm4$&\textbf{37}$_{-7}^{+8}$&\textbf{36}$_{-9}^{+10}$&\textbf{31}$_{-11}^{+15}$\T \B \\ 
&2022-09-30&NICER&5130010119&0.23&$\leq60$&\textbf{60}$_{-16}^{+19}$&$\leq90$&/&/\T \B \\ 
&2022-10-06&NICER&5130010120&1.38&\textbf{38}$_{-4}^{+6}$&\textbf{57}$_{-6}^{+7}$&\textbf{38}$_{-11}^{+12}$&\textbf{34}$_{-12}^{+16}$&/\T \B \\ 
&2022-10-09&NICER&5130010121&2.18&\textbf{37}$_{-6}^{+5}$&$\textbf{53}\pm6$&\textbf{28}$_{-8}^{+10}$&\textbf{31}$_{-10}^{+12}$&/\T \B \\ 
&2022-10-12&NICER&5130010123&2.1&$\textbf{33}\pm6$&$\textbf{51}\pm6$&\textbf{46}$_{-7}^{+10}$&\textbf{38}$_{-10}^{+17}$&\textbf{44}$_{-16}^{+23}$\T \B \\ 
&2023-01-13&NICER&5665010101&0.49&$\leq63$&$\leq72$&/&/&/\T \B \\ 
&2023-01-28&NICER&5665010201&0.92&$\leq20$&\textbf{31}$_{-8}^{+10}$&$\leq31$&$\leq56$&$\leq49$\T \B \\ 
&2023-02-10&NICER&5665010301&0.48&\textbf{23}$_{-9}^{+7}$&\textbf{62}$_{-9}^{+12}$&/&\textbf{42}$_{-14}^{+22}$&/\T \B \\ 
&2023-02-23&NICER&5665010401&1.44&$\leq11$&$\leq12$&$\leq14$&$\leq17$&$\leq26$\T \B \\ 
&2023-02-24&NICER&5665010402&0.07&$\leq27$&$\leq45$&$\leq51$&$\leq88$&$\leq87$\T \B \\ 
&2023-02-28&NICER&5130010124&0.37&$\leq25$&$\leq33$&$\leq51$&$\leq50$&$\leq78$\T \B \\ 
&2023-03-02&NICER&6130010101&0.58&$\leq11$&$\leq12$&$\leq32$&$\leq30$&$\leq28$\T \B \\ 
&2023-03-03&NICER&6130010102&1.18&$\leq21$&$\leq21$&$\leq30$&$\leq34$&$\leq40$\T \B \\ 
&2023-03-05&NICER&6130010103&1.37&$\leq6$&$\leq7$&$\leq12$&$\leq15$&$\leq17$\T \B \\ 
&2023-03-06&NICER&6130010104&2.92&$\leq11$&$\leq14$&$\leq19$&$\leq21$&$\leq20$\T \B \\ 
&2023-03-08&NICER&6130010106&3.19&$\leq5$&$\leq6$&$\leq10$&$\leq11$&$\leq16$\T \B \\ 
&2023-03-09&NuSTAR&80801327002&12.16&$\leq8$&$\textbf{11}\pm2$&$\leq4$&$\leq5$&$\leq4$\T \B \\ 
&2023-03-10&NICER&6130010107&0.95&$\leq16$&$\leq16$&$\leq16$&$\leq28$&$\leq34$\T \B \\ 
&2023-03-10&NICER&5665010403&9.45&$\leq6$&$\leq7$&$\leq11$&$\leq12$&$\leq17$\T \B \\ 
&2023-03-10&NuSTAR&80902312002&10.8&\textbf{13}$_{-3}^{+5}$&$\textbf{8}\pm3$&$\leq6$&$\leq5$&$\leq6$\T \B \\ 
&2023-03-11&NICER&6557010201&12.67&$\leq4$&$\leq4$&$\leq6$&/&$\leq12$\T \B \\ 
&2023-03-11&NuSTAR&80902312004&7.94&$\leq7$&\textbf{9}$_{-4}^{+1}$&$\leq3$&$\leq4$&$\leq4$\T \B \\ 
&2023-03-12&NICER&6557010301&14.97&$\leq4$&$\leq4$&/&$\leq7$&$\leq12$\T \B \\ 
&2023-03-13&NuSTAR&80902312006&9.62&$\leq9$&\textbf{12}$_{-1}^{+4}$&$\leq3$&$\leq4$&$\leq3$\T \B \\ 
&2023-03-13&NICER&6557010302&1.28&$\leq12$&$\leq14$&$\leq19$&$\leq22$&$\leq27$\T \B \\ 
&2023-03-24&NICER&5665010404&2.61&$\leq12$&$\leq14$&$\leq17$&$\leq20$&$\leq29$\T \B \\ 
&2023-03-27&NICER&6130010109&1.11&$\leq17$&$\leq20$&$\leq27$&$\leq31$&$\leq37$\T \B \\ 
&2023-04-05&NICER&5665010406&0.87&$\leq21$&$\leq22$&$\leq34$&$\leq40$&$\leq50$\T \B \\ 
&2023-04-07&NICER&5665010407&4.34&$\leq4$&$\textbf{10}\pm3$&$\leq13$&$\leq12$&$\leq17$\T \B \\ 
&2023-04-08&NICER&5665010408&0.97&$\leq10$&$\textbf{15}\pm6$&$\leq19$&$\leq26$&$\leq33$\T \B \\ 
&2023-04-16&NICER&6130010110&0.73&$\leq12$&$\leq14$&$\leq33$&$\leq37$&$\leq47$\T \B \\ 
&2023-04-21&NICER&5665010409&1.33&$\leq10$&$\leq21$&$\leq18$&$\leq20$&$\leq38$\T \B \\ 
&2023-04-22&NICER&6130010111&3.51&$\leq8$&\textbf{23}$_{-5}^{+3}$&$\leq12$&$\leq22$&$\leq20$\T \B \\ 
&2023-05-01&NICER&6130010113&2.92&$\leq10$&\textbf{20}$_{-4}^{+5}$&$\leq23$&$\leq37$&$\leq28$\T \B \\ 
&2023-05-05&NICER&5665010410&1.95&$\leq11$&$\leq13$&$\leq23$&$\leq33$&$\leq27$\T \B \\ 
&2023-05-06&NICER&6130010114&1.98&$\leq8$&$\leq9$&$\leq18$&$\leq27$&$\leq30$\T \B \\ 
&2023-05-11&NICER&6130010115&0.72&$\leq18$&\textbf{16}$_{-7}^{+9}$&$\leq25$&$\leq38$&$\leq41$\T \B \\ 
&2023-05-19&NICER&5665010411&2.0&$\leq14$&$\textbf{25}\pm5$&$\leq18$&$\leq42$&/\T \B \\ 
&2023-05-21&NICER&6130010117&1.63&$\leq10$&$\leq22$&$\leq19$&$\leq24$&$\leq35$\T \B \\ 
&2023-05-28&NICER&6130010118$\dagger$&1.18&$\leq37$&$\leq43$&/&/&/\T \B \\ 
&2023-05-28&NICER&6130010118$\dagger$&1.17&$\leq25$&\textbf{32}$_{-8}^{+10}$&$\leq41$&$\leq45$&/\T \B \\ 
&2023-06-08&NICER&6130010119&2.21&\textbf{22}$_{-4}^{+5}$&$\textbf{36}\pm6$&/&\textbf{31}$_{-9}^{+11}$&/\T \B \\ 
&2023-06-16&NICER&6130010120$\dagger$&0.47&\textbf{24}$_{-9}^{+13}$&\textbf{30}$_{-11}^{+18}$&$\leq68$&$\leq73$&/\T \B \\ 
&2023-06-16&NICER&6130010120$\dagger$&0.47&\textbf{29}$_{-8}^{+13}$&\textbf{38}$_{-11}^{+15}$&$\leq62$&/&/\T \B \\ 
&2023-06-18&NICER&6130010121&2.0&\textbf{40}$_{-4}^{+5}$&\textbf{50}$_{-5}^{+7}$&$\textbf{32}\pm9$&\textbf{30}$_{-11}^{+13}$&/\T \B \\ 
&2023-06-25&NICER&6130010122&0.59&$\leq37$&\textbf{28}$_{-11}^{+14}$&$\leq60$&$\leq69$&/\T \B \\ 
&2023-07-02&NICER&6130010123&1.27&\textbf{21}$_{-7}^{+5}$&\textbf{39}$_{-5}^{+9}$&/&/&/\T \B \\ 
&2023-08-27&NICER&6588010111&2.94&$\leq8$&$\leq8$&$\leq13$&$\leq17$&$\leq22$\T \B \\ 
&2023-08-31&NICER&6588010112&2.46&$\leq19$&$\leq15$&$\leq28$&$\leq33$&$\leq44$\T \B \\ 
&2023-09-01&NICER&6588010113&0.08&$\leq68$&$\leq67$&/&/&/\T \B \\ 
&2023-10-08&NICER&6588010116&3.57&$\leq30$&$\leq35$&$\leq61$&$\leq89$&/\T \B \\ 
&2023-10-12&NICER&6588010117&3.79&$\leq51$&$\leq68$&/&/&/\T \B \\ 

\end{longtable}

}

\medskip
Table B.2: Main characteristics of significant K$\alpha$ line detections from the sample. Uncertainties regarding luminosity are not quoted, as they were negligible.
{
\vspace*{-0.5em}

\tiny

\centering
\begin{longtable}[h]{c c c || c c || c  c  c | c  c  c }
\hline
\hline
\multirow{2}{*}{Source}
& \multirow{2}{*}{Date}
& \multirow{2}{*}{ObsID}
& \multirow{2}{*}{HR$_{[6-10]/[3-10]}$}
& \multirow{2}{*}{$L_{[3-10]}/L_{Edd}$}
& \multicolumn{3}{c}{\FeKav{}}
& \multicolumn{3}{c}{\FeKavi{}}
\T \B \\

&
&
&
& $\times 10^{-2}$
& EW
& blueshift
& width
& EW
& blueshift
& width
\T \B \\
\hline
\hline
\endfirsthead
\caption{continued}\\
\hline
\hline
\multirow{2}{*}{Source}
& \multirow{2}{*}{Date}
& \multirow{2}{*}{ObsID}
& \multirow{2}{*}{HR$_{[6-10]/[3-10]}$}
& \multirow{2}{*}{$L_{[3-10]}/L_{Edd}$}
& \multicolumn{3}{c}{\FeKav{}}
& \multicolumn{3}{c}{\FeKavi{}}
\T \B \\

&
&
&
& $\times 10^{-2}$
& EW
& blueshift
& width
& EW
& blueshift
& width
\T \B \\
\hline
\hline
\endhead

\hline\multirow{1}{*}{2002/2004}&2004-08-04&4568&$0.351_{-0.003}^{+0.003}$&$5.6$&/&/&/&\textbf{11}$_{-4}^{+3}$&\textbf{-300}$_{-500}^{+500}$&\textbf{0}$^{+4200}$\T \B \\ 
\hline\multirow{6}{*}{2005/2006}&2006-02-08&400010010&$0.306_{-0.002}^{+0.001}$&$5.1$&$\textbf{9}\pm3$&\textbf{-1600}$_{-2000}^{+1500}$&/&$\textbf{36}\pm3$&\textbf{-1200}$_{-500}^{+400}$&/\T \B \\ 
&2006-02-15&400010020&$0.287_{-0.002}^{+0.001}$&$3.7$&\textbf{19}$_{-4}^{+3}$&\textbf{-1400}$_{-1400}^{+1500}$&/&\textbf{37}$_{-3}^{+4}$&\textbf{-1000}$_{-900}^{+700}$&/\T \B \\ 
&2006-02-28&400010030&$0.249_{-0.002}^{+0.001}$&$3.4$&$\textbf{20}\pm4$&\textbf{-1200}$_{-1500}^{+1400}$&/&$\textbf{34}\pm4$&\textbf{-700}$_{-900}^{+900}$&/\T \B \\ 
&2006-03-08&400010040&$0.239_{-0.002}^{+0.001}$&$3.0$&\textbf{25}$_{-5}^{+4}$&\textbf{-1300}$_{-1600}^{+1800}$&/&\textbf{36}$_{-4}^{+5}$&\textbf{-400}$_{-1300}^{+1100}$&/\T \B \\ 
&2006-03-15&400010050&$0.229_{-0.001}^{+0.001}$&$2.8$&$\textbf{22}\pm3$&\textbf{-800}$_{-1200}^{+1200}$&/&$\textbf{36}\pm3$&\textbf{-300}$_{-700}^{+800}$&/\T \B \\ 
&2006-03-23&400010060&$0.213_{-0.001}^{+0.001}$&$2.5$&\textbf{25}$_{-3}^{+4}$&\textbf{300}$_{-1300}^{+1100}$&/&$\textbf{38}\pm4$&\textbf{200}$_{-700}^{+800}$&/\T \B \\ 
\hline\multirow{14}{*}{2011/2013}&2012-01-17&13714&$0.362_{-0.003}^{+0.003}$&$4.7$&$\textbf{32}\pm4$&\textbf{0}$_{-100}^{+200}$&\textbf{1900}$_{-500}^{+500}$&$\textbf{57}\pm5$&\textbf{-300}$_{-100}^{+100}$&\textbf{2700}$_{-400}^{+400}$\T \B \\ 
&2012-01-20&13715&$0.344_{-0.002}^{+0.002}$&$4.6$&\textbf{34}$_{-5}^{+3}$&\textbf{100}$_{-100}^{+200}$&\textbf{2300}$_{-400}^{+600}$&\textbf{49}$_{-5}^{+4}$&\textbf{-300}$_{-100}^{+100}$&\textbf{2200}$_{-300}^{+600}$\T \B \\ 
&2012-01-26&13716&$0.347_{-0.003}^{+0.002}$&$4.4$&\textbf{47}$_{-2}^{+3}$&\textbf{500}$_{-200}^{+200}$&\textbf{3000}$_{-500}^{+400}$&\textbf{52}$_{-3}^{+1}$&\textbf{-300}$_{-100}^{+0}$&\textbf{2200}$_{-400}^{+700}$\T \B \\ 
&2012-01-30&13717&$0.389_{-0.003}^{+0.003}$&$5.1$&$\textbf{30}\pm3$&\textbf{200}$_{-300}^{+200}$&\textbf{2000}$_{-700}^{+800}$&$\textbf{48}\pm4$&\textbf{-200}$_{-200}^{+200}$&\textbf{1800}$_{-700}^{+700}$\T \B \\ 
&2012-02-13&906008010&$0.292_{-0.001}^{+0.001}$&$4.8$&\textbf{39}$_{-4}^{+3}$&\textbf{-800}$_{-900}^{+800}$&/&$\textbf{56}\pm4$&\textbf{-1700}$_{-500}^{+700}$&/\T \B \\ 
&2012-03-04&0670671501\_S003&$0.366_{-0.002}^{+0.002}$&$4.8$&$\textbf{35}\pm7$&\textbf{-5000}$_{-1900}^{+2200}$&/&\textbf{55}$_{-7}^{+9}$&\textbf{-5800}$_{-1200}^{+1600}$&/\T \B \\ 
&2012-03-04&0670671501\_U014&$0.347_{-0.0}^{+0.0}$&$5.3$&$\textbf{31}\pm2$&\textbf{-5200}$_{-300}^{+200}$&/&\textbf{48}$_{-2}^{+1}$&\textbf{-5200}$_{-200}^{+100}$&/\T \B \\ 
&2012-03-20&0670671301\_S003&$0.36_{-0.001}^{+0.001}$&$6.2$&$\textbf{21}\pm3$&\textbf{-3900}$_{-800}^{+900}$&/&\textbf{46}$_{-2}^{+4}$&\textbf{-4300}$_{-400}^{+400}$&/\T \B \\ 
&2012-03-25&0670672901\_S003&$0.401_{-0.0}^{+0.0}$&$5.8$&$\textbf{20}\pm1$&\textbf{-6000}$_{-500}^{+500}$&/&\textbf{45}$_{-2}^{+1}$&\textbf{-5900}$_{-300}^{+200}$&/\T \B \\ 
&2012-09-09&0670673001\_S003&$0.413_{-0.001}^{+0.001}$&$8.0$&\textbf{9}$_{-2}^{+3}$&\textbf{-4600}$_{-3000}^{+2600}$&/&\textbf{31}$_{-2}^{+3}$&\textbf{-4300}$_{-800}^{+1000}$&/\T \B \\ 
&2012-09-10&0670673001\_U002&$0.432_{-0.002}^{+0.002}$&$7.2$&/&/&/&\textbf{25}$_{-5}^{+6}$&\textbf{-3500}$_{-3000}^{+2700}$&/\T \B \\ 
&2012-09-11&0670673101\_S003&$0.467_{-0.002}^{+0.002}$&$9.4$&/&/&/&\textbf{9}$_{-4}^{+5}$&\textbf{-1200}$_{-6400}^{+5800}$&/\T \B \\ 
&2013-02-20&40014008002&$0.33_{-0.001}^{+0.001}$&$6.2$&/&/&/&\textbf{30}$_{-2}^{+3}$&\textbf{500}$_{-800}^{+700}$&/\T \B \\ 
&2013-02-21&40014009001&$0.331_{-0.001}^{+0.001}$&$6.4$&/&/&/&\textbf{37}$_{-1}^{+0}$&\textbf{900}$_{-600}^{+400}$&/\T \B \\ 
\hline\multirow{3}{*}{2015}&2015-02-20&409007010&$0.363_{-0.002}^{+0.002}$&$7.3$&/&/&/&\textbf{17}$_{-3}^{+4}$&\textbf{-1600}$_{-900}^{+1200}$&/\T \B \\ 
&2015-02-24&409007020&$0.327_{-0.002}^{+0.002}$&$5.8$&$\textbf{18}\pm3$&\textbf{-600}$_{-1100}^{+1200}$&/&$\textbf{44}\pm5$&\textbf{-1000}$_{-500}^{+600}$&/\T \B \\ 
&2015-02-27&409007030&$0.33_{-0.002}^{+0.002}$&$6.0$&/&/&/&\textbf{38}$_{-3}^{+4}$&\textbf{-1200}$_{-1000}^{+1100}$&/\T \B \\ 
\hline\multirow{1}{*}{2016/2017}&2016-10-21&19904&$0.311_{-0.002}^{+0.002}$&$5.6$&\textbf{23}$_{-5}^{+4}$&\textbf{-300}$_{-300}^{+300}$&\textbf{1800}$_{-1500}^{+1400}$&\textbf{45}$_{-7}^{+4}$&\textbf{-200}$_{-300}^{+300}$&\textbf{2400}$_{-800}^{+1000}$\T \B \\ 
\hline\multirow{24}{*}{2018/2019}&2018-06-11&1130010104&$0.199_{-0.002}^{+0.002}$&$2.4$&\textbf{31}$_{-7}^{+12}$&\textbf{0}$_{-1600}^{+1200}$&/&\textbf{52}$_{-8}^{+12}$&\textbf{200}$_{-800}^{+600}$&/\T \B \\ 
&2018-07-02&1130010105&$0.334_{-0.002}^{+0.002}$&$5.8$&\textbf{19}$_{-5}^{+6}$&\textbf{100}$_{-1100}^{+1000}$&/&\textbf{41}$_{-6}^{+5}$&\textbf{0}$_{-600}^{+600}$&/\T \B \\ 
&2018-07-04&1130010106&$0.337_{-0.003}^{+0.003}$&$5.8$&\textbf{18}$_{-7}^{+10}$&\textbf{-1500}$_{-2200}^{+1800}$&/&\textbf{46}$_{-8}^{+9}$&\textbf{400}$_{-800}^{+700}$&/\T \B \\ 
&2018-07-07&1130010109&$0.341_{-0.005}^{+0.004}$&$6.0$&\textbf{21}$_{-8}^{+12}$&\textbf{-600}$_{-2200}^{+1500}$&/&\textbf{42}$_{-10}^{+13}$&\textbf{-800}$_{-900}^{+1000}$&/\T \B \\ 
&2018-07-10&1130010111&$0.351_{-0.002}^{+0.002}$&$6.4$&\textbf{21}$_{-4}^{+6}$&\textbf{-900}$_{-800}^{+800}$&/&$\textbf{42}\pm5$&\textbf{-600}$_{-500}^{+500}$&/\T \B \\ 
&2018-07-13&1130010113&$0.335_{-0.002}^{+0.002}$&$6.0$&\textbf{23}$_{-4}^{+6}$&\textbf{-900}$_{-900}^{+900}$&/&\textbf{51}$_{-5}^{+6}$&\textbf{-300}$_{-400}^{+500}$&/\T \B \\ 
&2018-07-14&1130010114&$0.335_{-0.002}^{+0.002}$&$6.1$&$\textbf{20}\pm4$&\textbf{-500}$_{-1500}^{+1400}$&/&$\textbf{50}\pm5$&\textbf{-100}$_{-600}^{+600}$&/\T \B \\ 
&2018-07-16&1130010116&$0.328_{-0.003}^{+0.002}$&$6.0$&\textbf{25}$_{-6}^{+9}$&\textbf{-200}$_{-1600}^{+2200}$&/&\textbf{51}$_{-3}^{+10}$&\textbf{-200}$_{-800}^{+800}$&/\T \B \\ 
&2018-07-17&1130010117&$0.335_{-0.002}^{+0.002}$&$6.3$&$\textbf{25}\pm5$&\textbf{-900}$_{-1600}^{+1300}$&/&\textbf{51}$_{-6}^{+7}$&\textbf{-400}$_{-700}^{+700}$&/\T \B \\ 
&2018-07-21&1130010119&$0.335_{-0.002}^{+0.002}$&$5.9$&\textbf{22}$_{-4}^{+5}$&\textbf{-1400}$_{-1900}^{+1700}$&/&$\textbf{43}\pm5$&\textbf{-700}$_{-1100}^{+1000}$&/\T \B \\ 
&2018-07-22&1130010120&$0.333_{-0.002}^{+0.002}$&$6.0$&\textbf{19}$_{-5}^{+7}$&\textbf{-1400}$_{-5800}^{+3700}$&/&\textbf{50}$_{-6}^{+7}$&\textbf{-700}$_{-1000}^{+900}$&/\T \B \\ 
&2018-07-23&1130010121&$0.338_{-0.001}^{+0.002}$&$6.0$&\textbf{20}$_{-5}^{+4}$&\textbf{-900}$_{-800}^{+600}$&/&\textbf{48}$_{-5}^{+4}$&\textbf{-300}$_{-400}^{+300}$&/\T \B \\ 
&2018-07-25&1130010123&$0.337_{-0.002}^{+0.002}$&$6.0$&\textbf{21}$_{-5}^{+6}$&\textbf{-900}$_{-2100}^{+1900}$&/&$\textbf{49}\pm6$&\textbf{-200}$_{-1000}^{+900}$&/\T \B \\ 
&2018-07-26&1130010124&$0.335_{-0.002}^{+0.002}$&$6.0$&$\textbf{18}\pm6$&\textbf{-800}$_{-3200}^{+2500}$&/&\textbf{44}$_{-5}^{+7}$&\textbf{-600}$_{-1000}^{+1000}$&/\T \B \\ 
&2018-07-28&1130010125&$0.34_{-0.003}^{+0.003}$&$6.3$&\textbf{18}$_{-6}^{+8}$&\textbf{-700}$_{-3200}^{+2900}$&/&\textbf{38}$_{-8}^{+7}$&\textbf{-200}$_{-1600}^{+1300}$&/\T \B \\ 
&2018-08-02&1130010126&$0.359_{-0.002}^{+0.002}$&$7.3$&/&/&/&\textbf{38}$_{-5}^{+4}$&\textbf{-1300}$_{-900}^{+900}$&/\T \B \\ 
&2018-08-04&1130010128&$0.343_{-0.002}^{+0.002}$&$6.4$&/&/&/&\textbf{44}$_{-5}^{+4}$&\textbf{-500}$_{-600}^{+700}$&/\T \B \\ 
&2018-08-06&1130010130&$0.346_{-0.002}^{+0.002}$&$6.6$&/&/&/&\textbf{40}$_{-3}^{+4}$&\textbf{-300}$_{-800}^{+900}$&/\T \B \\ 
&2018-08-08&1130010132&$0.327_{-0.001}^{+0.001}$&$6.2$&\textbf{15}$_{-2}^{+1}$&\textbf{400}$_{-800}^{+800}$&/&$\textbf{44}\pm2$&\textbf{300}$_{-200}^{+300}$&/\T \B \\ 
&2018-08-10&1130010134&$0.333_{-0.002}^{+0.002}$&$6.0$&$\textbf{17}\pm4$&\textbf{-100}$_{-1000}^{+700}$&/&\textbf{47}$_{-3}^{+4}$&\textbf{0}$_{-400}^{+300}$&/\T \B \\ 
&2018-08-11&1130010135&$0.325_{-0.002}^{+0.001}$&$5.8$&$\textbf{19}\pm3$&\textbf{-600}$_{-1100}^{+1000}$&/&\textbf{45}$_{-2}^{+3}$&\textbf{-300}$_{-500}^{+400}$&/\T \B \\ 
&2018-08-20&1130010139&$0.318_{-0.001}^{+0.001}$&$5.5$&$\textbf{22}\pm2$&\textbf{300}$_{-400}^{+300}$&/&\textbf{47}$_{-2}^{+3}$&\textbf{100}$_{-200}^{+300}$&/\T \B \\ 
&2018-08-22&1130010141&$0.299_{-0.002}^{+0.002}$&$5.3$&\textbf{26}$_{-6}^{+4}$&\textbf{-100}$_{-700}^{+700}$&/&$\textbf{46}\pm5$&\textbf{-100}$_{-400}^{+400}$&/\T \B \\ 
&2018-10-05&1130010148&$0.195_{-0.001}^{+0.002}$&$2.4$&/&/&/&\textbf{28}$_{-6}^{+9}$&\textbf{1800}$_{-1100}^{+1200}$&/\T \B \\ 
\hline\multirow{7}{*}{2020}&2020-04-01&3130010106&$0.339_{-0.002}^{+0.002}$&$7.4$&/&/&/&\textbf{10}$_{-4}^{+5}$&\textbf{-1700}$_{-1600}^{+1800}$&/\T \B \\ 
&2020-04-03&3130010108&$0.318_{-0.002}^{+0.002}$&$6.1$&/&/&/&\textbf{24}$_{-4}^{+5}$&\textbf{600}$_{-800}^{+700}$&/\T \B \\ 
&2020-04-06&3130010110&$0.33_{-0.003}^{+0.003}$&$6.1$&/&/&/&\textbf{29}$_{-6}^{+9}$&\textbf{-300}$_{-1400}^{+1300}$&/\T \B \\ 
&2020-04-07&3130010111&$0.325_{-0.002}^{+0.002}$&$5.8$&/&/&/&\textbf{22}$_{-4}^{+5}$&\textbf{-200}$_{-700}^{+800}$&/\T \B \\ 
&2020-04-09&3130010112&$0.335_{-0.002}^{+0.002}$&$6.1$&/&/&/&\textbf{28}$_{-5}^{+4}$&\textbf{-400}$_{-600}^{+700}$&/\T \B \\ 
&2020-04-13&3130010116&$0.334_{-0.003}^{+0.003}$&$6.1$&/&/&/&\textbf{27}$_{-5}^{+8}$&\textbf{200}$_{-800}^{+1000}$&/\T \B \\ 
&2020-04-15&3130010118&$0.339_{-0.003}^{+0.003}$&$6.7$&/&/&/&$\textbf{24}\pm7$&\textbf{-400}$_{-1600}^{+1000}$&/\T \B \\ 
\hline\multirow{45}{*}{2022/2024}&2022-07-30&5130010101&$0.219_{-0.004}^{+0.003}$&$2.4$&\textbf{18}$_{-9}^{+8}$&\textbf{1200}$_{-1300}^{+1800}$&/&\textbf{39}$_{-9}^{+10}$&\textbf{0}$_{-800}^{+900}$&/\T \B \\ 
&2022-07-31&5130010102&$0.208_{-0.005}^{+0.005}$&$2.4$&/&/&/&\textbf{57}$_{-21}^{+30}$&\textbf{-2000}$_{-3000}^{+1600}$&/\T \B \\ 
&2022-08-06&5130010103&$0.25_{-0.001}^{+0.001}$&$3.6$&\textbf{23}$_{-5}^{+3}$&\textbf{-900}$_{-900}^{+900}$&/&\textbf{41}$_{-8}^{+1}$&\textbf{100}$_{-500}^{+600}$&/\T \B \\ 
&2022-08-07&5130010104&$0.259_{-0.003}^{+0.002}$&$3.5$&\textbf{22}$_{-8}^{+3}$&\textbf{-1200}$_{-900}^{+900}$&/&\textbf{39}$_{-6}^{+5}$&\textbf{-900}$_{-400}^{+700}$&/\T \B \\ 
&2022-08-08&5130010105&$0.273_{-0.002}^{+0.002}$&$3.8$&$\textbf{21}\pm4$&\textbf{-300}$_{-1000}^{+1000}$&/&\textbf{37}$_{-4}^{+5}$&\textbf{-600}$_{-500}^{+600}$&/\T \B \\ 
&2022-08-09&5130010106&$0.274_{-0.005}^{+0.004}$&$4.2$&\textbf{28}$_{-12}^{+18}$&\textbf{-2800}$_{-4300}^{+2700}$&/&\textbf{46}$_{-12}^{+21}$&\textbf{200}$_{-2400}^{+1700}$&/\T \B \\ 
&2022-08-14&5130010110&$0.298_{-0.002}^{+0.002}$&$4.9$&$\textbf{23}\pm4$&\textbf{100}$_{-800}^{+800}$&/&\textbf{41}$_{-4}^{+5}$&\textbf{-300}$_{-400}^{+300}$&/\T \B \\ 
&2022-08-15&5130010111&$0.297_{-0.003}^{+0.002}$&$4.9$&\textbf{19}$_{-6}^{+8}$&\textbf{-1300}$_{-1700}^{+1800}$&/&$\textbf{35}\pm8$&\textbf{-1200}$_{-1000}^{+800}$&/\T \B \\ 
&2022-08-18&5130010112&$0.307_{-0.004}^{+0.004}$&$5.3$&/&/&/&$\textbf{36}\pm11$&\textbf{800}$_{-4600}^{+4000}$&/\T \B \\ 
&2022-08-22&5501010101&$0.327_{-0.003}^{+0.003}$&$5.6$&$\textbf{18}\pm5$&\textbf{-200}$_{-1300}^{+1100}$&/&$\textbf{49}\pm5$&\textbf{-600}$_{-400}^{+400}$&/\T \B \\ 
&2022-08-23&5501010102&$0.331_{-0.004}^{+0.004}$&$6.0$&\textbf{20}$_{-8}^{+9}$&\textbf{-1300}$_{-2100}^{+1500}$&/&\textbf{32}$_{-9}^{+10}$&\textbf{-1200}$_{-1600}^{+1300}$&/\T \B \\ 
&2022-08-25&5501010104&$0.331_{-0.002}^{+0.001}$&$6.0$&\textbf{19}$_{-4}^{+3}$&\textbf{600}$_{-600}^{+700}$&/&\textbf{47}$_{-4}^{+2}$&\textbf{0}$_{-300}^{+300}$&/\T \B \\ 
&2022-08-26&5501010105&$0.328_{-0.002}^{+0.002}$&$5.6$&\textbf{25}$_{-3}^{+4}$&\textbf{100}$_{-1100}^{+1200}$&/&$\textbf{47}\pm4$&\textbf{-100}$_{-600}^{+700}$&/\T \B \\ 
&2022-08-27&5501010106&$0.344_{-0.002}^{+0.002}$&$5.7$&\textbf{22}$_{-4}^{+5}$&\textbf{-100}$_{-1400}^{+1500}$&/&\textbf{47}$_{-4}^{+6}$&\textbf{0}$_{-800}^{+900}$&/\T \B \\ 
&2022-08-28&5501010107&$0.331_{-0.003}^{+0.003}$&$5.7$&\textbf{26}$_{-4}^{+5}$&\textbf{-400}$_{-1300}^{+900}$&/&\textbf{48}$_{-3}^{+5}$&\textbf{-400}$_{-500}^{+600}$&/\T \B \\ 
&2022-08-29&5501010108&$0.333_{-0.002}^{+0.002}$&$5.5$&\textbf{26}$_{-4}^{+5}$&\textbf{-1100}$_{-1100}^{+1300}$&/&\textbf{44}$_{-3}^{+5}$&\textbf{-700}$_{-700}^{+900}$&/\T \B \\ 
&2022-08-30&5501010109&$0.337_{-0.002}^{+0.002}$&$5.8$&\textbf{22}$_{-4}^{+5}$&\textbf{-800}$_{-1400}^{+1300}$&/&$\textbf{44}\pm5$&\textbf{-300}$_{-800}^{+900}$&/\T \B \\ 
&2022-08-31&5501010110&$0.33_{-0.002}^{+0.002}$&$5.5$&$\textbf{24}\pm3$&\textbf{-800}$_{-1000}^{+1100}$&/&$\textbf{44}\pm3$&\textbf{-1000}$_{-700}^{+700}$&/\T \B \\ 
&2022-09-01&5501010111&$0.316_{-0.004}^{+0.003}$&$5.3$&\textbf{29}$_{-9}^{+10}$&\textbf{-500}$_{-2500}^{+2400}$&/&\textbf{61}$_{-9}^{+13}$&\textbf{-200}$_{-1100}^{+1200}$&/\T \B \\ 
&2022-09-18&5130010114&$0.332_{-0.003}^{+0.002}$&$5.2$&\textbf{31}$_{-3}^{+4}$&\textbf{-1300}$_{-500}^{+500}$&/&\textbf{51}$_{-3}^{+5}$&\textbf{-700}$_{-300}^{+400}$&/\T \B \\ 
&2022-09-20&5130010116&$0.296_{-0.002}^{+0.002}$&$4.7$&\textbf{33}$_{-5}^{+3}$&\textbf{-700}$_{-900}^{+800}$&/&\textbf{45}$_{-7}^{+4}$&\textbf{-1800}$_{-500}^{+800}$&/\T \B \\ 
&2022-09-27&5130010118&$0.341_{-0.003}^{+0.003}$&$5.1$&$\textbf{28}\pm4$&\textbf{-600}$_{-1100}^{+1100}$&/&$\textbf{57}\pm4$&\textbf{-1400}$_{-600}^{+700}$&/\T \B \\ 
&2022-09-30&5130010119&$0.325_{-0.007}^{+0.006}$&$5.1$&/&/&/&\textbf{60}$_{-16}^{+19}$&\textbf{-2600}$_{-2500}^{+2800}$&/\T \B \\ 
&2022-10-06&5130010120&$0.346_{-0.003}^{+0.003}$&$5.1$&\textbf{38}$_{-4}^{+6}$&\textbf{-800}$_{-1400}^{+1200}$&/&\textbf{57}$_{-6}^{+7}$&\textbf{-1300}$_{-800}^{+800}$&/\T \B \\ 
&2022-10-09&5130010121&$0.319_{-0.003}^{+0.003}$&$4.8$&\textbf{37}$_{-6}^{+5}$&\textbf{-400}$_{-1000}^{+900}$&/&$\textbf{53}\pm6$&\textbf{-1100}$_{-700}^{+800}$&/\T \B \\ 
&2022-10-12&5130010123&$0.31_{-0.002}^{+0.002}$&$4.4$&$\textbf{33}\pm6$&\textbf{-400}$_{-500}^{+600}$&/&$\textbf{51}\pm6$&\textbf{-1500}$_{-500}^{+400}$&/\T \B \\ 
&2023-01-28&5665010201&$0.392_{-0.004}^{+0.004}$&$7.3$&/&/&/&\textbf{31}$_{-8}^{+10}$&\textbf{-900}$_{-2400}^{+2700}$&/\T \B \\ 
&2023-02-10&5665010301&$0.421_{-0.008}^{+0.007}$&$6.5$&\textbf{23}$_{-9}^{+7}$&\textbf{-700}$_{-4500}^{+3300}$&/&\textbf{62}$_{-9}^{+12}$&\textbf{-800}$_{-1300}^{+1100}$&/\T \B \\ 
&2023-03-09&80801327002&$0.484_{-0.003}^{+0.0}$&$12.7$&/&/&/&$\textbf{11}\pm2$&\textbf{-5700}$_{-1400}^{+1400}$&/\T \B \\ 
&2023-03-10&80902312002&$0.478_{-0.001}^{+0.001}$&$12.2$&\textbf{13}$_{-3}^{+5}$&\textbf{4600}$_{-1500}^{+300}$&/&$\textbf{8}\pm3$&\textbf{-6700}$_{-2900}^{+2700}$&/\T \B \\ 
&2023-03-11&80902312004&$0.685_{-0.003}^{+0.001}$&$18.4_{-0.1}$&/&/&/&\textbf{9}$_{-4}^{+1}$&\textbf{-5000}$_{-1500}^{+1700}$&/\T \B \\ 
&2023-03-13&80902312006&$0.641_{-0.001}^{+0.001}$&$18.1$&/&/&/&\textbf{12}$_{-1}^{+4}$&\textbf{-4400}$_{-1200}^{+1700}$&/\T \B \\ 
&2023-04-07&5665010407&$0.391_{-0.002}^{+0.002}$&$10.2$&/&/&/&$\textbf{10}\pm3$&\textbf{-1100}$_{-1400}^{+1300}$&/\T \B \\ 
&2023-04-08&5665010408&$0.379_{-0.002}^{+0.002}$&$8.7$&/&/&/&$\textbf{15}\pm6$&\textbf{-1300}$_{-1400}^{+1200}$&/\T \B \\ 
&2023-04-22&6130010111&$0.335_{-0.001}^{+0.001}$&$6.6$&/&/&/&\textbf{23}$_{-5}^{+3}$&\textbf{-400}$_{-600}^{+800}$&/\T \B \\ 
&2023-05-01&6130010113&$0.323_{-0.002}^{+0.002}$&$6.0$&/&/&/&\textbf{20}$_{-4}^{+5}$&\textbf{-100}$_{-1200}^{+900}$&/\T \B \\ 
&2023-05-11&6130010115&$0.359_{-0.004}^{+0.004}$&$7.1$&/&/&/&\textbf{16}$_{-7}^{+9}$&\textbf{-700}$_{-2700}^{+2200}$&/\T \B \\ 
&2023-05-19&5665010411&$0.311_{-0.002}^{+0.002}$&$5.5$&/&/&/&$\textbf{25}\pm5$&\textbf{-1100}$_{-1000}^{+900}$&/\T \B \\ 
&2023-05-28&6130010118&$0.31_{-0.003}^{+0.003}$&$5.4$&/&/&/&\textbf{32}$_{-8}^{+10}$&\textbf{-1000}$_{-1200}^{+1300}$&/\T \B \\ 
&2023-06-08&6130010119&$0.296_{-0.002}^{+0.002}$&$4.7$&\textbf{22}$_{-4}^{+5}$&\textbf{-900}$_{-1100}^{+1000}$&/&$\textbf{36}\pm6$&\textbf{-500}$_{-500}^{+700}$&/\T \B \\ 
&2023-06-16&6130010120&$0.323_{-0.004}^{+0.004}$&$4.9$&\textbf{24}$_{-9}^{+13}$&\textbf{500}$_{-6500}^{+4000}$&/&\textbf{30}$_{-11}^{+18}$&\textbf{-2700}$_{-5300}^{+6600}$&/\T \B \\ 
&2023-06-16&6130010120&$0.301_{-0.004}^{+0.004}$&$4.8$&\textbf{29}$_{-8}^{+13}$&\textbf{200}$_{-2900}^{+3000}$&/&\textbf{38}$_{-11}^{+15}$&\textbf{200}$_{-2400}^{+2400}$&/\T \B \\ 
&2023-06-18&6130010121&$0.302_{-0.002}^{+0.002}$&$4.6$&\textbf{40}$_{-4}^{+5}$&\textbf{-300}$_{-1100}^{+900}$&/&\textbf{50}$_{-5}^{+7}$&\textbf{-600}$_{-1000}^{+800}$&/\T \B \\ 
&2023-06-25&6130010122&$0.314_{-0.005}^{+0.005}$&$5.2$&/&/&/&\textbf{28}$_{-11}^{+14}$&\textbf{0}$_{-3400}^{+4900}$&/\T \B \\ 
&2023-07-02&6130010123&$0.304_{-0.003}^{+0.003}$&$5.4$&\textbf{21}$_{-7}^{+5}$&\textbf{-500}$_{-2300}^{+3600}$&/&\textbf{39}$_{-5}^{+9}$&\textbf{-2200}$_{-1500}^{+1500}$&/\T \B \\ 

\end{longtable}

}
    
\end{onecolumn}
\end{document}